\documentclass[aps,twocolumn,nofootinbib,preprintnumbers,eqsecnum]{revtex4-1}
\pdfoutput=1

\usepackage{color}

\newcommand{\nb}[1]{\color{blue}}

\usepackage[
	  pagebackref=false,
	  colorlinks=true,
      linkcolor=blue,
      urlcolor=blue,
      filecolor=black,
      citecolor=red,
      pdfstartview=FitV,
      pdftitle={},
        pdfauthor={Thomas Faulkner, Hong Liu, Mukund Rangamani},
        pdfsubject={},
        pdfkeywords={},
        pdfpagemode=None,
        bookmarksopen=true
      ]{hyperref}

\usepackage[normalem]{ulem}
\usepackage{amsmath}
\usepackage{enumerate}
\usepackage{amsfonts}
\usepackage{yfonts}

\usepackage{subfigure}
\usepackage{psfrag}

\usepackage{epsfig}
\usepackage[latin1]{inputenc}
\usepackage{float}
\usepackage{graphicx}
\usepackage{cancel}
\usepackage{mathrsfs}
\usepackage{amssymb}
\usepackage{amsfonts}
\usepackage{amsmath}
\usepackage{slashed}

\usepackage{placeins}

\usepackage{graphicx}
\usepackage{bm}

\def\({\left(}
\def\){\right)}
\def\[{\left[}
\def\]{\right]}
\def\<{\langle}
\def\>{\rangle}

\newcommand\half{{\ensuremath{\frac{1}{2}}}}
\newcommand\p{\ensuremath{\partial}}

\newcommand\field[1]{{\ensuremath{\mathbb{{#1}}}}}

\newcommand\vev[1]{{\ensuremath{\left\langle{#1}\right\rangle}}}

\newcommand{\RR}{\field{R}}

\newcommand{\be}{\begin{equation}}
\newcommand{\ee}{\end{equation}}
\newcommand{\bea}{\begin{eqnarray}}
\newcommand{\eea}{\end{eqnarray}}
\newcommand{\bwt}{\begin{widetext}}
\newcommand{\ewt}{\end{widetext}}

\newcommand{\bi}{\begin{itemize}}
\newcommand{\ei}{\end{itemize}}
\newcommand{\ben}{\begin{enumerate}}
\newcommand{\een}{\end{enumerate}}
\newcommand{\bca}{\begin{cases}}
\newcommand{\eca}{\end{cases}}
\newcommand{\bln}{\begin{align}}
\newcommand{\eln}{\end{align}}
\newcommand{\bst}{\begin{split}}
\newcommand{\est}{\end{split}}

\newcommand\al{{\alpha}}
\newcommand\ep{\epsilon}

\newcommand\Sig{\Sigma}
\newcommand\lam{\lambda}
\newcommand\Lam{\Lambda}
\newcommand\om{\omega}

\newcommand\ga{{\ensuremath{{\gamma}}}}
\newcommand\Ga{{\ensuremath{{\Gamma}}}}
\newcommand\de{{\ensuremath{{\delta}}}}
\newcommand\De{{\ensuremath{{\Delta}}}}
\newcommand\vp{\varphi}

\def\th{{\theta}}

\newcommand\ov{\over}
\newcommand\ha{{\half}}

\def\le{\left}
\def\ri{\right}

\newcommand\sG{{\ensuremath{{\mathcal G}}}}

\newcommand\sL{{\ensuremath{{\mathcal L}}}}
\newcommand\sM{{\ensuremath{{\mathcal M}}}}

\newcommand\sO{{\ensuremath{{\mathcal O}}}}
\newcommand\sR{{\ensuremath{{\mathcal R}}}}

\renewcommand{\Im}{\textrm{Im}\,}

\newcommand{\vk}{{\vec k}}
\newcommand{\ka}{{\kappa}}

\newcommand{\ta}{\tilde a}

\newcommand{\oir}{{\Phi}}
\newcommand{\chis}{{\chi_*}}

\newcommand{\ircft}{{SLQL}}

\newcommand{\Slql}{{SLQL}}

\begin{document}

\title {
Quantum phase transitions in semi-local quantum liquids}

\preprint{MIT-CTP 4283}

\author{Nabil Iqbal}
\affiliation{Center for Theoretical Physics, Massachusetts Institute of Technology,
Cambridge, MA 02139 }
\author{ Hong Liu}
\affiliation{Center for Theoretical Physics,
Massachusetts
Institute of Technology,
Cambridge, MA 02139 }
\author{M\'ark Mezei}
\affiliation{Center for Theoretical Physics, Massachusetts Institute of Technology,
Cambridge, MA 02139 }

\begin{abstract}

We consider several types of quantum critical phenomena from finite-density 
gauge-gravity duality which to different degrees lie outside the Landau-Ginsburg-Wilson paradigm. These include: (1) 
 a ``bifurcating'' critical point, for which the order parameter remains gapped at the critical point,  and thus is {\it not} driven by soft order parameter fluctuations. 
 Rather it appears to be driven by ``confinement'' which arises when two fixed points annihilate and lose conformality.
On the condensed side, there is an infinite tower of condensed states and  the nonlinear response of the tower   exhibits an infinite spiral structure; (2)  a ``hybridized'' critical point which can be described by a standard Landau-Ginsburg sector of order parameter fluctuations hybridized with a strongly coupled sector; (3) a ``marginal'' critical 
point which is obtained by tuning the above two critical points to occur together and whose bosonic fluctuation spectrum coincides with that postulated to underly the ``Marginal Fermi Liquid'' description of the optimally doped cuprates.

\end{abstract}


\maketitle

\tableofcontents

\section{Introduction and summary}

In a strongly correlated many-body system, small changes of  external control parameters 
can lead to qualitative changes in the ground state of the system, resulting in a quantum 
phase transition. The quantum criticality associated with continuous quantum phase transitions 
give rise to some of the most interesting phenomena in condensed matter physics, especially in itinerant electronic systems \cite{Gegenwart.08,Lohneysen.07,Natphys.08}.
Among these are the breakdown of Fermi liquid theory and the emergence
of unconventional superconductivity.

Quantum criticality is traditionally formulated within the Landau paradigm
of phase transitions~\cite{Hertz.76,Sachdev_book,Sondhietal}. The critical theory can be understood in terms of the fluctuations of the
order parameter, a coarse-grained variable manifesting the breaking
of a global symmetry. This critical theory lives in $d+z$ dimensions \cite{Hertz.76}, where $d$ is the spatial dimension, and $z$ the dynamic exponent. 

More recent experimental and theoretical developments \cite{Gegenwart.08,Lohneysen.07,Natphys.08,Si.01,Senthil.04}, however, have pointed to
new types of quantum critical points. New modes, which are
inherently quantum and are beyond order-parameter fluctuations,
emerge as part of the quantum critical excitations.
For example, continuous quantum phase transitions observed in various antiferromagnetic heavy fermion compounds, involve a nontrivial interplay between local and extended degrees of 
freedom. While the extended degrees of freedom can be described by an antiferromagnetic order parameter, the Kondo breakdown and the interplay between Kondo breakdown and antiferromagnetic fluctuations cannot be captured in the standard Landau-Ginsburg-Wilson formulation.

It is thus of great interest to identify other examples of strongly correlated quantum critical points that do not fit easily into the standard formalism. In this paper we will discuss a set of such phase transitions using holographic duality \cite{AdS/CFT}. We will be studying a $d$-dimensional field theory that is conformal\footnote{Choosing a theory which is conformal in the UV is solely based on technical convenience and 
our discussion is not sensitive to this.} in the UV and has a $U(1)$ global symmetry. Consider turning on a nonzero chemical potential $\mu$ for the $U(1)$ charge. This finite charge density system has a disordered phase described in the bulk by a charged black hole in AdS$_{d+1}$~\cite{Romans:1991nq,Chamblin:1999tk}.  The conserved current $J_\mu$ of the boundary global $U(1)$ is mapped to a bulk $U(1)$ gauge field $A_M$, under which the black hole is charged. Various examples exist of boundary gauge theories with such a gravity description.

Now consider a scalar operator $\sO$ dual to a bulk scalar field $\phi$, which at a finite chemical potential could exhibit various instabilities towards the condensation of $\sO$. If we tune  parameters an instability can be made to vanish, even at zero temperature, with a critical point  separating an ordered phase characterized by a nonzero expectation value $\langle \sO \rangle$ from a disordered phase in which $\langle \sO \rangle$ vanishes.  The physical interpretation of the condensed phase depends on the quantum numbers carried by $\sO$: for example, if it is charged under some $U(1)$ it can be interpreted as a superconducting phase, whereas if it transforms as a triplet under an $SU(2)$ denoting spin it can be interpreted as an antiferromagnetic order parameter. If it is charged under a $Z_2$ symmetry, the condensed phase can be used to model, for example, a Ising-Nematic phase from  a Pomeranchuck instability. 
The critical point is largely independent of the precise interpretation of the condensed phase.

We will essentially discuss two different kinds of critical phenomena, which we refer to as a ``{\it bifurcating}'' and a ``{\it hybridized}'' quantum critical point.  A ``bifurcating'' quantum phase transition happens when a bulk scalar dips below the Breitenlohner-Freedman bound~\cite{bf} in the deep interior of the spacetime.  It was shown previously \cite{Iqbal:2010eh,Jensen:2010ga} that the thermodynamical behavior of this system has an exponentially generated scale reminiscent  of Berezinskii-Kosterlitz-Thouless transition and there is an infinite tower of geometrically separated condensed states analogous to the Efimov effect~\cite{efimov} in  the formation of three-body bound states.\footnote{See also~\cite{Jensen:2010vx,Kutasov:2011fr,Evans:2011zd}.}  Here we study the dynamical critical behavior in detail. We find that at a bifurcating critical point, the static susceptibility for the order parameter does not diverge (i.e. the order parameter remains {\it gapped} at the critical point), but rather develops a branch point singularity. When extended beyond the critical point into the (unstable) disordered phase, the susceptibility attempts to bifurcate into the complex plane. As the order parameter remains gapped at the critical point,  the quantum phase transition is {\it not} driven by soft order parameter fluctuations as in the Landau-Ginsburg-Wilson paradigm. 
Rather it appears to be driven by ``confinement'' which leads to the formation of a tower of bound states, which then Bose condense, i.e. it can be interpreted as a {\it quantum confinement/deconfinement} critical point.\footnote{As will be elaborated later here we use the term ``confinement'' in a somewhat loose sense, as in our story the ``confined'' state still has gapless degrees of freedom left and thus the  ``confinement'' only removes part of the deconfined spectrum. }
On the condensed side, we find the nonlinear response of the tower of condensed states  exhibits an infinite spiral structure that is shown in~Fig. \ref{fig:spiral1} in Sec.~\ref{sec:COND}. 

The instability corresponding to a ``hybridized'' phase transition occurs when the bulk extremal black hole geometry allows for certain kinds of scalar hair~\cite{Faulkner09}. One can approach the critical point for onset of the instability by a double-trace deformation in the field theory~\cite{Faulkner:2010gj}. Here we review and extend 
the results of~\cite{Faulkner:2010gj}. At a hybridized critical point the static susceptibility does diverge, but the small frequency and temperature behavior near the critical point does not follow the standard Landau-Ginzburg-Wilson formulation due to presence of some soft degrees of freedom other than the order parameter fluctuations. In particular, in some parameter range the dynamical susceptibility exhibits the local quantum critical behavior observed in quantum phase transitions of certain heavy fermion materials. 
 
Finally, one can tune the parameters of the system such that both types of critical point happen at the same time, resulting in yet another kind of critical point, which we call a ``marginal critical point'', as it is driven by 
a marginally relevant operator. Intriguingly, the critical fluctuations at such a point are precisely the same as the bosonic fluctuation spectrum postulated to underly the ``Marginal Fermi Liquid'' \cite{Varma89} description of the optimally doped cuprates (see also~\cite{SY,bgg}).  

Underlying the various sorts of novel quantum critical behavior described above is the  ``semi-local quantum liquid'' (or \Slql\ for short) nature of the disordered phase.  SLQL is a quantum phase dual to gravity in AdS$_2 \times \RR^{d-1}$ which is the near-horizon geometry of a zero-temperature charged black hole. It has a finite spatial correlation length, but a scaling symmetry in the time direction, and has gapless excitations at generic finite momenta; its properties have been discussed in detail recently in~\cite{Iqbal:2011in}~(and are also reviewed below in Sec.~\ref{sec:ads2}). A hybridized QCP can be described by a standard Landau-Ginsburg sector of order parameter 
hybridized with degrees of freedom from \Slql.  A bifurcating QCP can be understood as the transition of \Slql\ to a confining phase, as a consequence of two fixed points describing \Slql\ annihilate. 
The infinite tower of condensed states and the associated infinite spiral can be understood as consequences of  a spontaneously broken discrete scaling symmetry in the time direction.

The plan of the paper is as follows. In the next section, we discuss various aspects of the disordered phase 
and in particular the notion of  semi-local quantum liquid from the point of view taken in \cite{Iqbal:2011in}. In Section \ref{sec:insta} we discuss various instabilities of a generic AdS spacetime, and in Section \ref{sec:finin} we discuss how these instabilities manifest themselves in the AdS$_2$ factor in the disordered phase, resulting in quantum phase transitions. In Section \ref{sec:LEFT} we attempt to illuminate the nature of these quantum phase transitions by providing a low-energy effective theory for them. In Section \ref{sec:cond} we discuss various aspects of the condensed phase. In Sections \ref{sec:biI} and \ref{sec:hybrid} we provide a description of the critical behavior around the bifurcating and hybridized critical points respectively. In Section \ref{sec:doucri} we discuss the ``marginal'' critical point that is found if parameters are tuned so that the hybridized and bifurcating critical points collide. Finally in Section \ref{sec:disc} we conclude with a discussion of the interpretation of the \Slql\ as an intermediate-energy phase and the implications for our results. 

Due to the length of this paper various details and most derivations have been relegated to the appendices. We do not summarize all appendices here, but we do point out that in an attempt to make this paper more modular an index of important symbols (including brief descriptions and the location of their first definition) is provided in Appendix~\ref{app:symindex}. 

{\it Note:} While this paper was in preparation, we become aware of related work by Kristan Jensen~\cite{jensen}.

\section{Disordered phase and semi-local quantum liquids} \label{sec:ads2}

We will be interested in instabilities to the condensation of a scalar operator for a holographic system at a finite density, and in particular, the quantum critical behavior near a critical point for the onset of an instability.
An important set of observables for diagonalizing possible instabilities and 
characterizing the dynamical nature of a critical point 
are susceptibilities of the  order parameter. Suppose the order parameter is given by the expectation value of some bosonic operator $\sO$, then the corresponding susceptibility $\chi (\om, \vec k)$ are given by the retarded function  for $\sO$, which captures the linear responses of the system to an infinitesimal source\footnote{For example if $\sO$ is the magnetization of the system, then the corresponding source is the magnetic field.} conjugate to $\sO$.

In a stable phase in which $\sO$ is uncondensed, turning on an infinitesimal source will result in an 
expectation value for $\sO$ which is proportional to the source with the proportional constant given by the  susceptibility. However, if the system has an instability to the condensation of $\sO$, turning on an infinitesimal source will lead to modes exponentially growing with time. Such growing modes are reflected in the presence of singularities of  $\chi (\om, \vk)$ in the {\it upper} complex $\om$-plane. Similarly, at the onset of an instability (i.e. a critical point, both thermal and quantum), the static susceptibility typically diverges, reflecting that the 
tendency of the system to develop an expectation value of $\sO$ even in the absence of an external source. The divergence is characterized by a critical exponent $\ga$ (see Appendix~\ref{app:critexp} for a review of definitions of other critical exponents)
\be
\chi(k=0,\om=0) \sim \le| g-g_c  \ri|^{-\gamma} 
\ee
where $g$ is the tuning parameter (which is temperature for a thermal transition) with $g_c$ the critical point.

In this section we first review the charged black hole geometry describing the disordered phase
and the retarded response function for a scalar operator in this phase. We also elaborate on the semi-local 
behavior of the system, which is a central theme of our paper. 

While the qualitative features of our discussion apply to any field theory spacetime dimension\footnote{Explicit examples of the duality are only known for  $d=3,4,6$.} $d \geq 3$, for definiteness we will restrict our quantitiative discussion to $d=3$.

 \subsection{AdS$_2$ and Infrared (IR) behavior}

At zero temperature a boundary CFT$_3$ with a chemical potential $\mu$ is described by an 
extremal AdS charged black hole, which has a metric and background gauge field given by
\be
{ds^2 } = \frac{R^2}{z^2}(-f dt^2 + d\vec{x}^2) + \frac{R^2}{z^2}\frac{dz^2}{f}  \label{RNmetric} \
\ee 
with 
\be
A_t = \mu (1 - \mu_* z) , \quad f = 1 + {3 \mu_*^4 z^4 } - {4 \mu_*^3 z^3}, \quad  \mu_* \equiv {\mu \ov \sqrt{3} g_F}  \label{gaugeexp}
\ee
where $R$ is the curvature radius of AdS$_4$ and $g_F$ is a dimensionless constant which determines the unit of charge\footnote{It is equal to the bulk gauge coupling in appropriate units.}. Note that the chemical potential $\mu$ is the only scale of the system and provides the basic energy 
unit. For convenience we introduce the appropriately rescaled $\mu_*$, which will be used often below as it avoids having the factor $\sqrt{3} g_F$ flying around.   
 $f$ has a double zero at the horizon $z=z_* \equiv {1 \ov \mu_*}$, with  
 \be
 f (z) \approx 6 {(z_* -z)^2 \ov z_*^2} + \dots, \; z \to z_* 
 \ee 
As a result the near-horizon geometry factorizes into AdS$_2 \times \mathbb{R}^2$:
\be \label{ads2M}
ds^2 = \frac{R_2^2}{\zeta^2}(-dt^2 + d\zeta^2) +\mu_*^2 R^2 d\vec{x}^2 \qquad A = {g_F \ov \sqrt{12} \zeta} dt.
\ee
Here we have defined a new radial coordinate $\zeta$ and $R_2$ is the curvature radius of AdS$_2$,
\be
\zeta \equiv  {z_*^2 \ov 6 (z_*-z)}, \qquad R_2 \equiv {R \ov \sqrt{6}}  \ .
\label{defZ}
\ee
The metric~\eqref{ads2M} applies to the region ${z_*-z \ov z_*} \ll 1$ which translates into $\mu \zeta \gg 1$. 
Also note that the metric~\eqref{RNmetric} has a finite horizon size and thus has a nonzero entropy density.


 \begin{figure}[h]
\begin{center}
\includegraphics[scale=0.35]{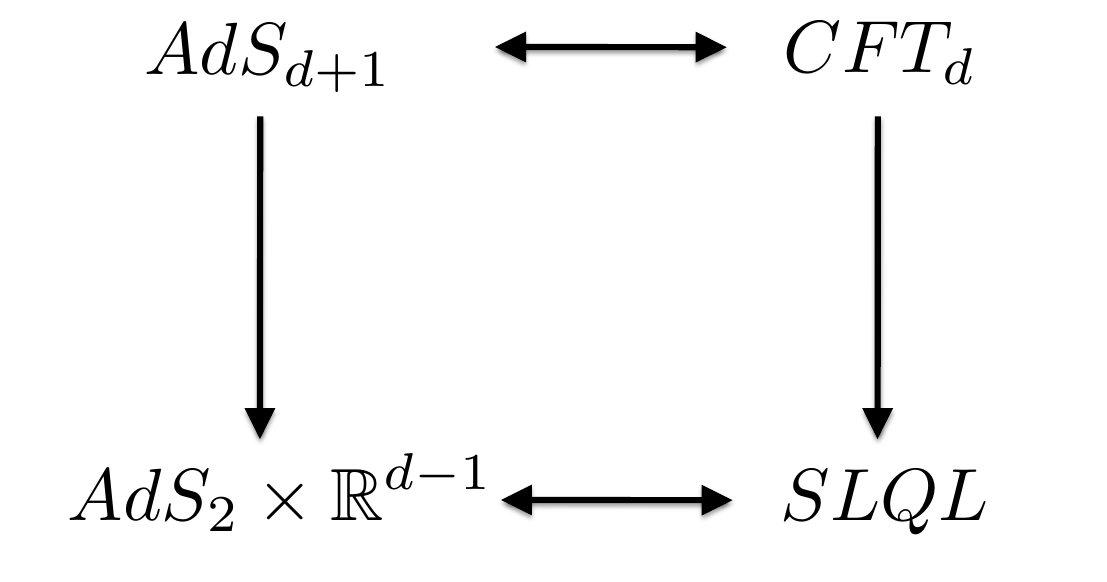}
\end{center}
\caption{At a finite chemical potential, a CFT$_d$ flows in the IR to \Slql. On the gravity side this is realized geometrically via the flow of the AdS$_{d+1}$ near the boundary to AdS$_2 \times \mathbb{R}^{d-1}$ near the horizon.
} 
\label{fig:flow}
\end{figure}

As discussed in~\cite{Faulkner09} the black hole geometry predicts that 
at a finite chemical potential the system is flowing to a nontrivial IR fixed point dual to
 AdS$_2 \times  \mathbb{R}^{2}$~\eqref{ads2M}. See Fig.~\ref{fig:flow}. Note that the metric~\eqref{ads2M} has a scaling symmetry 
\be 
t \to \lam t, \qquad \zeta \to \lam \zeta , \qquad \vec x \to \vec x 
\ee
under which only the time coordinate scales. Thus the IR fixed point has nontrivial scaling behavior  only in the time direction with the $\mathbb{R}^2$ directions becoming spectators. Thus we expect that it should be described by a conformal quantum mechanics, to which we will refer as  ``eCFT$_1$'' with ``e'' standing for ``emergent.'' This conformal quantum mechanics is somewhat unusual due to the presence of the $\RR^2$ factor on the gravity side,  with scaling operators labeled by continuous momentum 
along $\RR^2$ direction (as we shall see below).  As emphasized 
in~\cite{Iqbal:2011in},  the  quantum phase described by such an eCFT$_1$ has some interesting properties in terms of the dependence on spatial directions, and a more descriptive name {\it semi-local quantum liquid} (or \Slql\;for short) was given (see Sec.~\ref{sec:semi} for further elaboration).  Below the terms eCFT$_1$ and \Slql\ can be used interchangeably.  \Slql\ will be more often used to emphasize the IR fixed point as a quantum phase. 
 
Let us now consider a scalar operator $\sO(t, \vec x)$ corresponding to a bulk scalar field $\phi$ of mass $m^2$ and charge $q$.
Its conformal dimension $\De$ in the vacuum of the CFT$_3$ is related to $m$ by
\be \label{uvdim}
\De = {3 \ov 2} + \nu_U, \qquad \nu_U \equiv \sqrt{m^2 R^2 + {9 \ov 4}} \ . 
\ee
At a finite chemical potential,  in the IR its Fourier transform  $\sO_\vk (t)$ along the spatial directions, with momentum $\vk$,  should match onto some operator $\oir_\vk (t)$ in the \ircft. The conformal dimension of $\oir_\vk$ in the \ircft\ can be found from asymptotic behavior of classical solutions of $\phi$ in the AdS$_2 \times  \mathbb{R}^2$ geometry~\eqref{ads2M} and is given by\footnote{Note that depending on the value of $\nu_k$ there may be an alternative choice for $\delta_{k} = \frac{1}{2} - \nu_k$, by imposing a Neumann boundary condition for $\phi$ at the AdS$_2$ boundary~\cite{Klebanov:1999tb}. We will review this in more detail below when needed.}~\cite{Faulkner09} 
\be
\delta_k = \frac{1}{2} + \nu_k   \  \label{nudef}
\ee
with
\be \label{opep}
 \nu_k  =\sqrt{{m^2 R^2_2}- q_*^2 + {1 \ov 4}  +  {k^2 \ov 6 \mu_*^2} }, \quad 
  k = |\vec k|, \quad q_* = {q g_F \ov \sqrt{12}} \ .
 \ee
Equation~\eqref{opep} has some interesting features. Firstly, the IR dimension 
increases with momentum $k$, as a result operators with larger $k$ become less important in the IR. Note, however, this increase with momentum only becomes significant as $k \sim \mu$. For $k \ll \mu$, we can approximately treat $\de_k$  as momentum independent. Secondly, $\nu_k$ decreases with $q$, i.e. an operator with larger $q$ will have more significant IR fluctuations (given the same 
vacuum dimension $\De$). 

In the low frequency limit $\om \ll \mu$ the susceptibility (i.e. retarded Green function) $\chi (\om, \vk)$ for $\sO$ in the full CFT$_3$ can be written as~\cite{Faulkner09} 
\be \label{roep21}
 \chi (\om, \vk) = \mu_*^{2\nu_U} {b_+ (k,\om) + b_- (k,\om) \sG_k (\om) \mu_*^{-2 \nu_k}  \ov a_+ (k,\om) + a_- (k,\om) \sG_k (\om) \mu_*^{-2 \nu_k}}, 
\ee
where $\sG_k$ is the retarded function for $\oir_\vk$ in the \ircft\  and can be computed 
exactly by solving the equation of motion for $\phi$ in~\eqref{ads2M}. It is given by~\cite{Faulkner09}  
\be \label{iiRc}
\sG_k(\omega) = \frac{\Ga(-2\nu_k)}{\Ga(2 \nu_k)} {\Ga(\ha + \nu_k- iq_*) \ov \Ga (\ha -\nu_k - iq_*)}
(-2i \om)^{2 \nu_{k}} \ . 
\ee
$a_\pm (k,\om)$ and $b_\pm (k,\om)$ in~\eqref{roep21} are real (dimensionless) functions which can be extracted (numerically) by solving the equation of motion of $\phi$ in the full black hole geometry~\eqref{RNmetric}.  For the reader's convenience  we review the analytic properties of $a_\pm, b_\pm$ and outline derivation of~\eqref{roep21} in Appendix~\ref{app:rev}. $a_\pm, b_\pm$ are analytic in $\om$ and can be expanded for small $\omega$ as 
\be \label{wexp}
a_+ (k, \om) = a_+^{(0)} (k) + \om a_+^{(1)} (k) + \dots
\ee
$a_\pm (k, \om), b_\pm (k, \om)$ are also analytic functions of $\nu_k$ and $k^2$.  
Note that for a neutral scalar the linear term in $\om$ vanishes and the first nontrivial order starts with $\om^2$. 
A relation which will be useful below is (see Appendix~\ref{app:rev} for a derivation)
\be \label{oneu}
a_+^{(0)} (k) b_-^{(0)} (k) - a_-^{(0)} (k)  b_+^{(0)} (k) = {\nu_k \ov \nu_U}  \ .
\ee

We also introduce the uniform and static susceptibilities, given by 
\begin{align} \label{stsus}
\chi \equiv \chi (\om=0, \vk=0) & =  \mu_*^{2\nu_U} {b_+^{(0)} (0)  \ov a_+^{(0)} (0) } \nonumber \\
\qquad \chi (\vk) \equiv \chi (\om =0, \vk) & =  \mu_*^{2\nu_U}  {b_+^{(0)} (k)  \ov a_+^{(0)} (k) } \ .
\end{align}
Note that for notational simplicity, we distinguish $\chi, \chi (\vk)$ and $\chi (\om, \vk)$ only by their arguments. 


\subsection{Finite temperature scaling}

The previous considerations were all at precisely zero temperature; at finite temperature the  factor $f(z)$ in \eqref{RNmetric} develops a single zero at a horizon radius $z_0 < z_*$. For ${z_* - z_0 \ov z_*} \ll 1$,  the near-horizon region is now obtained by replacing the AdS$_2$ part of~\eqref{ads2M} by a Schwarzschild black hole metric in AdS$_2$, i.e. 
\be
ds^2 = \frac{R_2^2}{\zeta^2}\le(-\le(1-\frac{\zeta^2}{\zeta_0^2}\ri) dt^2 + \frac{d\zeta^2}{1-\frac{\zeta^2}{\zeta_0^2}}\ri) + {\mu_*^2}{R^2} d\vec{x}^2  \ \label{AdS2bh}
\ee
where $\zeta_0 \equiv {z_*^2 \ov 6 (z_* - z_0)}$.  The inverse Hawking temperature is given to leading order in ${z_* - z_0 \ov z_*}$ by
\be
T  
=  {1 \ov 2 \pi \zeta_0}  \ .
\ee
The metric~\eqref{AdS2bh} applies to the region ${z_*-z \ov z_*} \ll 1$ with the condition ${z_*-z_0 \ov z_*} \ll1$ which translates into $\mu \zeta \gg1 $ with the condition $\mu \zeta_0 \sim {\mu \ov T} \gg 1$. 

Thus at a temperature $T \ll \mu$,  one essentially heats up the \ircft\ and equation~\eqref{roep21} can be generalized to
\be \label{roep22}
\chi (\om, \vk, T) =  \mu_*^{2\nu_U} {b_+ (k,\om,T) + b_- (k,\om,T) \sG_k^{(T)} (\om) \mu_*^{-2 \nu_k}  \ov a_+ (k,\om,T) + a_- (k,\om,T) \sG_k^{(T)} (\om) \mu_*^{-2 \nu_k}}, 
\ee
where $\sG_k^{(T)}$ is the retarded function for $\oir_\vk$ in the \ircft\ at temperature $T$ and is given by \cite{Faulkner09,Faulkner:2011tm}\bwt
\be
\sG_k^{(T)}(\om) = (4 \pi T)^{2\nu_k} \frac{\Ga(-2\nu_k)}{\Ga(2 \nu_k)} {\Ga(\ha + \nu_k- iq_*) \ov \Ga (\ha -\nu_k - iq_*)}  \frac{\Ga\le(\frac{1}{2} + \nu_k - i\frac{\om}{2\pi T} + i q_* \ri)}{\Ga\le(\frac{1}{2} - \nu_k - i\frac{\om}{2\pi T}  + i q_* \ri) } \label{finiteTch1} \ .
\ee
\ewt
Note that at finite $T$, $a_\pm, b_\pm$ also receive {\it analytic} corrections in $T$ as indicated in~\eqref{roep22}. The retarded function $\sG_k^{(T)}$ in the \ircft\ has a scaling form in terms of $\om/T$ as expected from the scaling symmetry at the zero temperature. Note that there is no scaling in the spatial momentum  and analytic dependence on $T$ and $\om$ in~\eqref{roep22}.  

In both~\eqref{iiRc} and~\eqref{finiteTch1} the $k$ dependence solely arises from $\nu_k$, which in turn depends on $k$ through $k^2/\mu^2$. This implies that for $k \ll \mu$, $\sG_k$ is approximately $k$-independent.  

For most of this paper we will be considering a neutral scalar, for which~\eqref{iiRc} and~\eqref{finiteTch1} simplify to 
\be
\sG_k(\om) =  \le(-\frac{i \om}{2}\ri)^{2\nu_k} 
\frac{\Ga(-\nu_k)}{\Ga(\nu_k)} \  , \label{iiRc1}
\ee 
and
\begin{align}
\sG^{(T)}_k(\om) & =  (\pi T)^{2\nu_k}\frac{\Ga(-\nu_k)}{\Ga(\nu_k)}\frac{\Ga\le(\frac{1}{2} + \nu_k - i\frac{\om}{2\pi T}\ri)}{\Ga\le(\frac{1}{2} - \nu_k - i\frac{\om}{2\pi T}\ri)} \nonumber \\
& \equiv T^{2 \nu_k} g \le({\om \ov T} , \nu_k\ri) \ , \label{finiteTgf21}
\end{align}
with $g(x,\nu_k)$ a universal scaling function. 

\subsection{Semi-local quantum liquids} \label{sec:semi}

We expect that the leading low frequency behavior of the spectral function of $\sO$ should be given by that of the IR fixed point, i.e. that of $\oir_\vk$ in the \ircft. Indeed, suppose $a_+^{(0)} (k) \neq 0$ and $\nu_k$ is real, we can expand~\eqref{roep21} at small frequency as 
\be \label{gre}
\chi (\omega, k) = \chi (k) + \chi_2 (k) {\cal G}_k (\omega) + {\rm real \; analytic \; in \; \omega}
\ee
where
\be
\chi (k) =  \mu_*^{2\nu_U} {b_+^{(0)}  (k) \over a_+^{(0)} (k)} , \qquad 
\chi_2 (k) =  \mu_*^{2\nu_U} {\nu_k \over \nu_U}{1 \over (a_+^{(0)})^2}  \ 
\ee
are real analytic functions of $\nu_k$ and $k^2$. Note we have used~\eqref{oneu} in the above expression for $\chi_2$.  The spectral function is then given by
 \be 
 \Im \chi (\om, k) = \chi_2 (k) \Im \sG_k(\omega)  + \dots \ .  \label{spectrDens}
 \ee
 The factor $\chi_2 (k)$ can be interpreted as a wave function renormalization of the operator. The $\dots$~in~\eqref{spectrDens} denote higher order corrections which can be interpreted as coming from irrelevant perturbations to the \Slql. 
 
In subsequent sections we will describe situations in which~\eqref{gre} and~\eqref{spectrDens}  break down 
and give rise to instabilities. Below we briefly review the semi-local behavior of the IR fixed point discussed in~\cite{Iqbal:2011in} to provide some physical intuition as to the nature of the disordered phase. 

An important feature of the \Slql\ is that the spectral weight, which is defined by the imaginary part of the retarded function~\eqref{iiRc}, scales with $\om$ as a power for any momentum $k$, which indicates the presence of low energy excitations for all momenta (although at larger momenta, with a larger scaling dimension the weight will be more suppressed).

Another interesting feature of the \Slql, which is a manifestation of the disparity between the spatial and time directions of the spacetime metric~\eqref{ads2M}, is that the system has an infinite correlation time, but a finite correlation length in the spatial directions~(where the scale is provided by the nonzero chemical potential).  This is intuitively clear from the presence of (and lack of) scaling symmetry in the time (and spatial) directions in the near horizon region. The correlation length $\xi$ in spatial directions can be read from the branch point $k= i \xi^{-1}$ in $\nu_k$.  More explicitly, $\nu_k$ in~\eqref{opep} can be rewritten as 
\be \label{defnk}
\nu_k = {1 \ov \sqrt{6} \mu_*} \sqrt{k^2 + {1 \ov \xi^2}}
\ee
with
\be \label{corL}
\xi  \equiv {1 \ov \mu_*} {1 \ov \sqrt {{m^2 R^2}- 6 q_*^2 + {3 \ov 2}}} 
= {1 \ov \sqrt{6} \nu_{0} \mu_*}  \ .
\ee
By  Fourier transforming~\eqref{spectrDens} to coordinate space one obtains Euclidean correlation function $G_E (\tau = i t , \vec x )$ with the following behavior (For details of the Fourier transform see Appendix of~\cite{TASI}. See also~\cite{Iqbal:2011in} for arguments based on geodesic approximation):
\ben 
\item  For $x \equiv |\vec x|  \ll \xi$ (but not so small that the vacuum behavior takes over), 
\be \label{timde}
G_E (\tau, x) \sim {1 \ov \tau^{2 \de_{k=0}}} \ .
\ee

\item   For $x \gg \xi$, the correlation function decays at least 
exponentially as 
\be \label{spaco}
G_E (\tau, x) \sim e^{- {x \ov \xi}} \ .
\ee 

\een

 \begin{figure}[h]
\begin{center}
\includegraphics[scale=0.55]{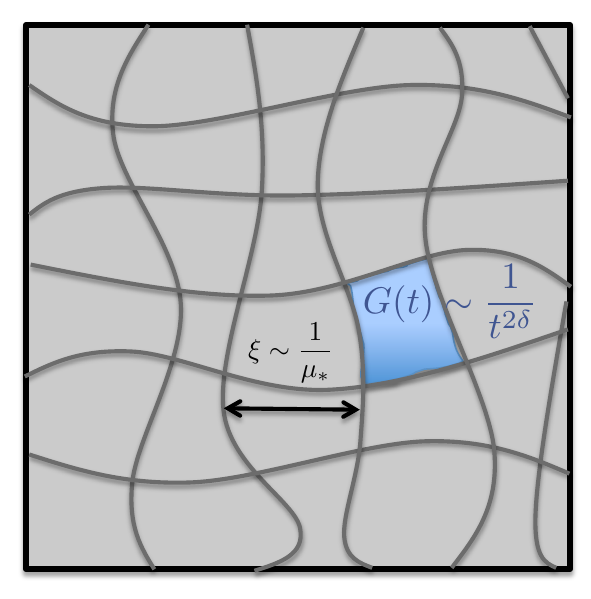}
\end{center}
\caption{A cartoon picture: the system separates into domains of size $\xi \sim {1 \ov \mu}$. 
Within each domain a conformal quantum mechanics governs dynamics in the time direction with a power law correlation (i.e. infinite relaxation time).} 
\label{fig:local}
\end{figure}

From the above we see that the system separates into domains of size $\xi \sim {1 \ov \mu}$. 
Within each domain a conformal quantum mechanics governs dynamics in the time direction with a power law correlation (i.e. infinite relaxation time)~\eqref{timde}. Domains separated by distances larger than $\xi$ are uncorrelated with one another. See Fig.~\ref{fig:local} for a cartoon picture.

This behavior is reminiscent of the {\it local quantum critical} behavior discussed in~\cite{Si.01} as proposed for heavy fermion quantum critical points and also that exhibited in the electron spectral function for the strange metal phase for cuprates~\cite{Varma89}. We note that there are also some important differences. Firstly, here the behavior happens to be a {\it phase}, rather than a critical point. Secondly, while there is nontrivial scaling only  in the time direction, the local AdS$_2$ correlation functions depend nontrivially on $k$. From~\eqref{defnk} it is precisely this dependence of $\nu_k$ on $k$ that gives the spatial correlation length of the system. Also while at a generic point in parameter space, the dependence of $\nu_k$ and $\sG_k$ on $k$ is analytic and only through $k/\mu$ (and thus can be approximated as $k$-independent for $k \ll \mu$), 
as we will see in Sec.~\ref{sec:biI}, near a bifurcating quantum critical point, the dependence becomes nonanalytic at $k=0$ and is important for understanding the behavior around the critical point. For these reasons, such a phase was named as a semi-local quantum liquid (\Slql) in~\cite{Iqbal:2011in}. As also discussed there,  \Slql\ should be interpreted as a universal intermediate energy phase rather than as a zero temperature phase. This will have important implications for the interpretation of quantum critical behavior 
to be discussed in later sections, a point to which we will return in the conclusion section. For now we will treat it as a zero-temperature phase. 

\section{Scalar instabilities of an AdS spacetime} \label{sec:insta}

In preparation for the discussion of instabilities and quantum phase transitions for the finite density system we introduced in last section,  here we review the scalar instabilities of a pure AdS$_{d+1}$ spacetime. As we will see in the later sections, the instabilities and critical behavior of our finite density system are closely related to those of the near horizon AdS$_2$ region. Below we will first consider general $d$ and then point out some features specific to AdS$_2$. We will mainly state the results; for details see Appendix~\ref{app:dt}.

Consider a  scalar field $\phi$ in AdS$_{d+1}$, which is dual to an operator~$\Phi$ 
in some boundary CFT$_d$. The conformal dimension of $\Phi$  is given by
\be \label{dim1}
\De_\pm = {d \ov 2} \pm \nu  , \qquad \nu = \sqrt{M^2 R^2 + {d^2 \ov 4}}
\ee
where $M^2$ is the mass square for $\phi$. For $\nu \geq 1$, only the $+$ sign in~\eqref{dim1} is allowed. For $\nu \in (0,1)$, there are two ways to quantize $\phi$ by imposing Dirichlet or Neumann conditions at the AdS boundary,  which are often called standard and alternative quantizations respectively, and lead to two different CFTs. 
We will call the CFT$_d$ in which $\Phi$ has dimension $\De_+ = {d \ov 2} + \nu$ the CFT$_d^{\rm IR}$ and the corresponding operator $\Phi_+$. The one in which 
$\Phi$ has dimension $\De_- = {d \ov 2} - \nu$ will be denoted as the CFT$_d^{\rm UV}$ and the 
operator $\Phi_-$. The range of dimensions in the CFT$_d^{\rm UV}$  is $\De_- \in ({d \ov 2}-1, {d \ov 2})$ with the lower limit (corresponding to $\nu \to 1$) approaching that of a free particle in $d$ spacetime dimension. 

\begin{figure}[h]
\begin{center}
\includegraphics[scale=0.6]{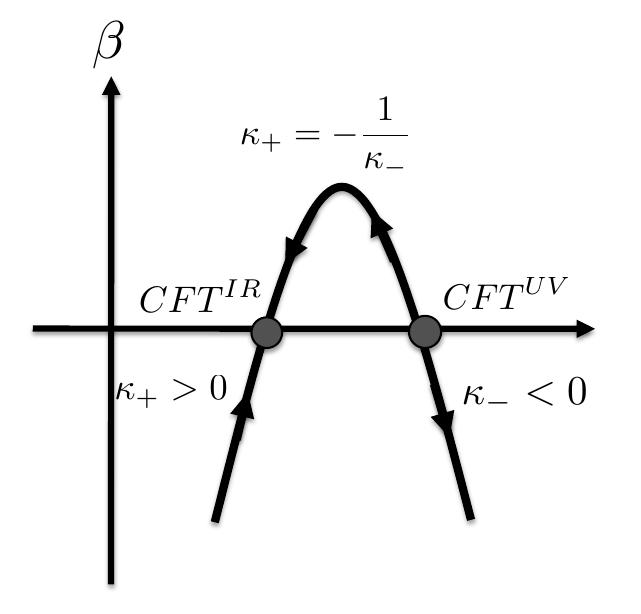}
\end{center}
\caption{Flow from CFT$_d^{\rm UV}$ to CFT$_d^{\rm IR}$ (the arrows denote flows to the IR). In the region between two fixed points, one can describe the system using either fixed point. To the left of the fixed point  
corresponding to CFT$_d^{\rm IR}$, the system develops a UV instability. To the right of CFT$_d^{\rm UV}$, the system develops an IR instability.} 
\label{fig:phs}
\end{figure}

Let us consider first $\nu \in (0,1)$.  In the CFT$_d^{\rm UV}$ the double trace operator $\Phi_-^2$  is relevant (as $2 \De_- < d$). Turning on a double trace deformation 
\be \label{douir}
{\kappa_-  \mu_*^{2\nu} \ov 2}  \int \, \Phi_-^2 
\ee
with a positive $\kappa_-$, the theory will flow in the IR to the CFT$_d^{\rm IR}$~\cite{Witten:2001ua} (and thus their respective names). Turning on~\eqref{douir} with a $\kappa_- < 0$ will instead lead to an instability in the IR, and $\Phi_-$ will condense (see Appendix~\ref{app:dt} for explanation).\footnote{As discussed recently in~\cite{Faulkner:2010gj} this instability can be used to generate a new type of holographic superconductors.}
Thus $\ka_- = 0$, i.e. the CFT$_d^{\rm UV}$, is a {\it quantum critical point} for the onset of instability for condensing the scalar operator.  The double trace deformation 
 \be \label{douv}
{\kappa_+  \mu_*^{-2\nu} \ov 2}  \int \, \Phi_+^2 
\ee
in the CFT$_d^{\rm IR}$ is an irrelevant perturbation and the theory flows in the UV to the CFT$_d^{\rm UV}$ for  negative $\kappa_+$. Note that  CFT$_d^{\rm IR}$ deformed by~\eqref{douv} with $\ka_+ < 0$ is equivalent to CFT$_d^{\rm UV}$ deformed by~\eqref{douir} with the relation\footnote{For a recent discussion of these issues see Appendix of~\cite{Faulkner:2010jy}. \label{foot:1}}
\be \label{euive}
\ka_+ = -{1 \ov \ka_-} \ .
\ee
 Thus the alternative quantization corresponds to the limit $\ka_+ \to -\infty$.   For positive $\kappa_+$ the system develops  a {\it UV instability}  (see Appendix~\ref{app:dt}). See Fig.~\ref{fig:phs} for a summary. 
 
For $\nu> 1$, there is only CFT$_d^{\rm IR}$ corresponding to the standard quantization and the double trace deformation is always irrelevant. There is a UV instability for $\ka_+ < 0$ ($\ka_+ > 0$) for $\nu \in (n, n+1)$ 
for $n$ an odd (even) integer (see Appendix~\ref{app:dt}). For example for $\nu \in (1,2)$ there is a UV instability for $\ka_+ < 0$.

 As $\nu \to 0$, i.e. $M^2 \to -{d^2 \ov 4}$, the two CFT$_d$'s merge into one at $\nu =0$.  When $M^2$ drops below $M_c^2 \equiv -{d^2 \ov 4}$, the so-called Breitenlohner-Freedman bound, the fixed points become complex and 
 the conformal symmetry is broken. Relatedly  $\nu$ becomes complex and 
 $\Phi$ develops exponentially growing modes~\cite{bf}. The system becomes unstable to the 
 condensation of $\Phi$ modes. Introducing a UV cutoff $\Lam$, then there is an emergent IR energy scale $\Lam_{IR}$ below which the condensate sets in~\cite{Kaplan:2009kr} 
  \be \label{irsca0}
\Lambda_{IR} \sim \Lam \exp\le(-\frac{\pi}{\sqrt{M^2_c - M^2}}\ri)  \ .
\ee
 Thus $\nu=0$ is another critical point; for all $M^2 < M_c^2$ an instability occurs. 

Now consider being precisely at $\nu=0$, i.e. fix $M^2 = M_c^2$. Our system still has another control knob: the double trace deformation 
\be 
{\kappa \ov 2} \int  \Phi^2
\ee
is marginal. One can show that it is marginally relevant for $\ka < 0$ and  irrelevant for $\ka > 0$~\cite{Witten:2001ua}~(see Appendix~\ref{app:A} for a derivation). Thus for $\ka > 0$ the system is stable in the IR, but for $\ka < 0$ there is an exponentially generated IR scale ($\Lam$ is a UV cutoff)
\be \label{newir}
\Lam_{IR} = \Lam \exp \le({1 \ov \ka} \ri) \ 
\ee
below which the operator will again condense. As it requires tuning two parameters, $\nu=0$ and $\ka =0$ is a multi-critical point. See Fig.~\ref{fig:MFT} in Appendix \ref{app:A}.

 The above discussions apply to any $d$ including $d=1$.  There are some new elements for $d=1$, i.e. AdS$_2$,\footnote{We will assume the AdS$_2$ has a constant radial electric field as in most applications with~\eqref{ads2M} as one such example.}  which does not happen for $d \geq 2$.  Firstly, from~\eqref{opep}, the dimension of an operator also depends on its charge, i.e. the second equation in~\eqref{dim1} is modified to\footnote{Note that in~\eqref{opep}, $k^2$ term comes from dimensional reduction on $\RR^2$ and should be considered as part of the AdS$_2$ mass square, i.e. the AdS$_2$ mass square is $M^2 R_2^2 = m^2 R_2^2 +  {k^2 \ov 6 \mu_*^2}$.}
   \be \label{ads2E}
 \nu = \sqrt{M^2 R_2^2 - q_*^2 + {1 \ov 4}}  \ .
 \ee
 Thus $\nu$ can become imaginary when charge is sufficiently large even for a positive $M^2$.  Secondly, for a charged scalar in AdS$_2$, the range in which both quantizations exists becomes $\nu \in (0, \ha)$ (see Appendix~\ref{app:ad2i} for details). Both features have 
to do with that the gauge potential in~\eqref{ads2M} blows up at the infinity and 
thus affects the boundary conditions (including normalizability)
  of a charged scalar.

In our discussion below, double trace deformations of the eCFT$_1$ describing the \Slql\ will play 
an important role. In particular, for an operator $\Phi$ in which alternative quantization exists we should 
also distinguish eCFT$_1^{\rm{IR}}$ and eCFT$_1^{\rm{UV}}$, where $\Phi$ has dimension $\ha \pm \nu$ respectively.

\section{Instabilities and quantum critical points at finite density} \label{sec:finin}

We now go back to the $(2+1)$-dimensional system at a finite density introduced in sec.~\ref{sec:ads2}. We will slightly generalize the discussion there by also including double trace deformations in the dual CFT$_3$. We will mostly work in the standard quantization 
so that our discussion also applies to $\nu_U  > 1$ and the results for the alternative quantization can be obtained from those for standard quantization using~\eqref{euive}. 
 We will work in the parameter region that the vacuum theory is stable in the IR,  i.e. 
 \be \label{vacsta}
m^2 R^2 \geq - {9 \ov 4} ,
\quad {\rm and} \quad \ka_+ < 0 \ .
\ee
Turning on a finite chemical potential can lead to new IR instabilities and  quantum phase transitions. In this section we discuss these instabilities and 
the corresponding quantum critical points and in next section we give an effective theory description. The subsequent sections will be devoted to a detailed  
study of critical behavior around these quantum critical points. 

\subsection{Finite density instabilities}

Potentially instabilities due to the condensation of a scalar operator $\sO$  can be diagonalized 
by examining the retarded function~\eqref{roep21}, which can be generalized to include double trace deformations~\eqref{douv} as \cite{Klebanov:1999tb}
\be \label{roep23}
\chi (\om, \vk) = \mu_*^{2\nu_U} {b_+ (k,\om) + b_- (k,\om) \sG_k (\om) \mu_*^{-2 \nu_k}  \ov \tilde a_+ (k,\om) + \tilde a_- (k,\om) \sG_k (\om) \mu_*^{-2 \nu_k}} \  
\ee
 where we have used~\eqref{douGF} and 
 \begin{align}
 \tilde a_\pm (\om, k) &= a_\pm (\om,k) + \ka_+ b_\pm (\om,k) \nonumber \\
 &= a_\pm^{(0)} (k) + \ka_+ b_\pm^{(0)} (k) + O(\om) \ .
 \end{align}
Instabilities will manifest themselves as poles of~\eqref{roep23} in the {\it upper} half complex-$\om$-plane, which gives rise to
exponentially growing modes and thus leads to condensation of $\phi$.

In~\cite{Faulkner09}, it was found  that when one of the following two conditions happens,~\eqref{gre}--\eqref{spectrDens} do not apply and Eq.~\eqref{roep23} always has poles in the upper $\om$-plane\footnote{Ref.~\cite{Faulkner09} considered only the standard and alternative quantization. The argument there generalizes immediately to~\eqref{roep23} with double trace deformations.}, implying instabilities:
\ben

\item $\nu_k$ becomes imaginary for some $k$, for which there are an infinite number of poles 
in the upper half $\om$-plane\footnote{For example, see the right plot of Fig. 1 of~\cite{Faulkner09}.}.  Writing~\eqref{opep} as  
\be \label{uude}
\nu_k = \sqrt{u + {k^2 \ov 6 \mu_*^2}}, \qquad u \equiv m^2 R_2^2 + {1 \ov 4} - q_*^2  \ 
\ee
$\nu_k$ becomes complex for sufficiently small $k$  whenever $u < 0$. 
For a given $m$, this always occurs for a sufficiently 
large $q$. For a neutral operator $q=0$, $u$ can be negative for $m^2R^2$ lying in the window 
\be \label{winmag}
-{9 \ov 4} < m^2 R^2 < -{3 \ov 2}
\ee
where the lower limit comes from the stability of vacuum theory~\eqref{vacsta} and the upper limit comes from the condition $u <0$ after using the relation~\eqref{defZ}. 
Interpreting $m^2 R_2^2  - q_*^2$ as an effective AdS$_2$ mass square (at $k=0$), on the gravity side the instability can be interpreted as violating the AdS$_2$ BF bound~\cite{Gubser:2005ih,hartlon,Gubser:2008pf,Denef:2009tp}. For a charged scalar the instability is also related to pair production of charged particles from the black hole and superradiance~\cite{Faulkner09}.
On the field theory side, the instability can be interpreted as due to formation of bound states in \Slql~\cite{Iqbal:2011in} (see also discussion in Sec.~\ref{sec:bifsum}).

\item $\tilde a_+^{(0)} (k)$ can become zero for some special values of momentum $k_F$.  At $k=k_F$ it is clear from~\eqref{roep23} that since $\tilde a_+^{(0)} =0$, $\chi$ has a singularity at $\om=0$. Furthermore since $a_+^{(0)}$ changes sign near $k=k_F$, it was shown in~\cite{Faulkner09} (see sec.~VI B), the phase of~\eqref{iiRc} is such that a pole moves from the upper half $\om$-plane (for $k< k_F$) to the lower half $\om$-plane (for $k>k_F)$ through $\om =0$. 
In Appendix~\ref{app:phase} Fig.~\ref{fig:neu},~\ref{fig:neu2} we show some examples of a  neutral scalar field for which 
$\ta_+^{(0)}$ has a zero at some momentum. On the gravity side a zero of $\tilde a_+^{(0)} (k)$ 
corresponds to the existence of a normalizable solution of scalar equation in the black hole geometry, i.e. a scalar hair~\cite{Faulkner09}. Such a normalizable mode implies in the boundary field theory the
existence of some soft degrees of freedom and as we shall see in Sec.~\ref{sec:Heff} the instability can be captured by a standard Landau-Ginsburg model.

\een

 In the parameter range (say for $m, q, \ka_+$) where either (or both) instability appears, the system is unstable to the condensation of the operator $\sO$ (or in bulk language condensation of $\phi$). For a charged scalar the condensed phase corresponds to a  holographic superconductor~\cite{herzogr,Horowitz:2010gk} and the first instability underlies that of~\cite{holographicsc,hartnolletal}  as was first pointed out in~\cite{Denef:2009tp}, while holographic superconductors due to the second type instability has been discussed recently in~\cite{Faulkner:2010gj}. For a neutral scalar, the first type of instability was first pointed out in~\cite{hartlon}, and as discussed in~\cite{Iqbal:2010eh} the condensed phase can be used as a model for antiferromagnetism when the scalar operator is embedded as part of a triplet transforming under a global $SU(2)$ symmetry corresponding to spin. For a single real scalar field with a $Z_2$ symmetry, the condensed phase can be considered as a model for an Ising-nematic phase. 

 Both types of instabilities can be cured by going to sufficiently high temperature; there exists a critical temperature $T_c$, beyond which these instabilities no longer exist and at which the system undergoes a continuous  superconducting (for a charged scalar) or antiferromagnetic (for a neutral scalar) 
phase transition. As has been discussed extensively in the literature  such finite temperature phase transitions are of the mean field type, as the boundary conditions of the finite-temperature black hole horizon are analytic (see e.g.~\cite{Iqbal:2010eh,Maeda:2009wv,Herzog:2010vz}). Alternatively one can continuously dial external parameters of the system at zero temperature
to get rid of the instabilities.  The critical values of the parameters at which the instabilities disappear then correspond to quantum critical points (QCP) where quantum phase transitions into a superconducting or an antiferromagnetic phase occur.

\subsection{Bifurcating quantum critical point}

For the first type of instability a quantum critical point occurs 
when the effective AdS$_2$ mass becomes zero for $k=0$~\cite{Iqbal:2010eh,Jensen:2010ga}, i.e. 
from~\eqref{uude}, at 
\be \label{biy}
u = u_c =0 \ .
\ee
For example for a neutral scalar field (with $q=0$) this happens at 
\be \label{crimas}
m^2_c R^2 = -{ 3\ov 2} \ .
\ee
Note that while in AdS/CFT models the mass square $m^2$ for the vacuum theory is typically not an externally tunable parameter, the effective AdS$_2$ mass square can often be tuned. For example, in the set-up of~\cite{Jensen:2010ga,Jensen:2010vx}, the effective AdS$_2$ mass square can be tuned by dialing an external magnetic field and so is the the example discussed in~\cite{Iqbal:2010eh} when considering a holographic superconductor in a magnetic field. See also~\cite{Iqbal:2010eh} for a phenomenological model.  In this paper we will not worry about the detailed mechanism to realize the $u_c =0$ critical point and will just treat $u$ as a dialable parameter (or just imagine dialing the mass square for the vacuum theory). Our main purpose is to identify and understand the critical behavior around the critical point which is independent of 
the specific mechanism to realize it. As will be discussed in subsequent sections, as we approach $u_c=0$ from the uncondensed side ($u > 0$), the static susceptibility remains finite, but develops a cusp at $u=0$ and if we naively continue it to $u < 0$ the susceptibility becomes complex. Below we will refer to this critical point as a {\it bifurcating} QCP. 

\subsection{Hybridized quantum critical point}

The second type of instability results in an intricate phase structure in the $u-\ka_+$ plane. For illustration, we restrict our attention here to $0<u<{1\ov24}$ (i.e. $-\frac32<m^2 R^2<-\frac54$) where the story is relatively simple, and relegate the discussion of the $u>{1\ov24}$ regime to Appendix~\ref{app:phase}.

For $0<u<{1\ov24}$, one can readily check numerically that $\ta_+^{(0)}$ is a monotonically increasing function of $k$ for negative $\ka_+$.  See Fig.~\ref{fig:neu}. Thus to diagnose possible instability we need only  to examine 
the sign of $\ta_+^{(0)}  (k=0)$ with the stable region having $\ta_+^{(0)}  (k=0) > 0$. 
This implies that the system is stable for $\ka_+$ satisfying\footnote{Note that for $m^2 R^2 < 0$, both $b_+^{(0)} (k=0)$ and $a_+^{(0)} (k=0)$ are  positive, see Fig.~\ref{fig:neu} in Appendix~\ref{app:phase}.} 
\be \label{stare}
0 > \ka_+ \geq \ka_c \equiv -{a_+^{(0)} (k=0) \ov b_+^{(0)} (k=0)}
\ee 
where the upper limit is required by~\eqref{vacsta}. For $\ka_+ > 0$ there is a UV instability already present in the vacuum, and this instability is unaffected by the introduction of finite density.

We will focus on the the critical point $\ka_c$ in~\eqref{stare} below. 
 Note that at the critical point $\ka_+ = \ka_c$
\be \label{hybq}
\ta_+^{(0)} (k=0, \ka_c)= 0
\ee
and as a result the uniform susceptibility $\chi$ in~\eqref{stsus} diverges. Such a quantum critical point has been discussed recently in~\cite{Faulkner:2010gj}. As already mentioned in~\cite{Faulkner:2010gj} and will be elaborated more in Sec.~\ref{sec:hybrid}, the presence of the strongly coupled IR sector described by AdS$_2$  gives rise to a variety of new phenomena which cannot be captured by the standard Laudau-Ginsburg-Wilson paradigm. For reasons to be clear in sec.~\ref{sec:hybrid}, below we will refer to such a critical point as a {\it hybridized} QCP.

Note that it is rather interesting that despite that $\ka_+$ being an irrelevant coupling, tuning it 
could nevertheless result in an IR instability due to finite density effect. 
In the $u$-range we are working in $\nu_U < 1$, and this phenomenon can be understood more intuitively 
through the description in terms of alternative quantization. From~\eqref{euive}, the stable region~\eqref{stare} translates into 
\be 
\ka_- \geq  -{1 \ov \ka_c} \ 
\ee
with the alternative quantization itself ($\ka_- =0$) falling into the unstable region. Note that turning on a double trace deformation in the alternative quantization $\ha \ka_- \mu_*^{2\nu_U} \int \sO^2$ translates in the bulk description  into turning on a bulk boundary action $\ha f_- \int \phi^2$ where $f_- \propto \ka_-$ and $\phi$ is the bulk field dual to $\sO$.  Thus we see that at finite density 
one needs to turn on a nonzero ``boundary mass'' to stabilize the alternative quantization.

\subsection{A marginal quantum critical point}

We can also tune $\ka_+$ and $u$ together to have a doubly tuned critical point at $u = 0, \ka_+=\ka_+^*$, where the susceptibility both diverges and bifurcates. The value of $\ka_+^*$ can be obtained from $u \to 0$ limit of the expression for $\kappa_c$ given in \eqref{stare}, leading to 
\be  \label{mcpv1}
\ka_+^* = -{\al \ov \beta}  \  
\ee
where $\al$ and $\beta$ are constants defined in Eq. ~\eqref{req}. For the specific example~\eqref{crimas} of tuning the AdS$_4$ mass of a neutral scalar to reach $u=0$, the values of $\al, \beta$ are given in~\eqref{pp1}--\eqref{pp2} which gives $\ka_* = - 2.10$.

As we will show in sec.~\ref{sec:doucri}, the dynamical susceptibility around such a critical point coincide with that of the bosonic fluctuations underlying the ``Marginal Fermi Liquid'' postulated in~\cite{Varma89} for describing the strange metal region of the high $T_c$ cuprates.\footnote{It has also been pointed out by David Vegh~\cite{veghun} that the retarded function for a scalar operator with $\nu=0$ in AdS$_2$ gives the bosonic fluctuations of  the ``Marginal Fermi Liquid''.}

\begin{figure}[h]
\begin{center}
\includegraphics[scale=0.7]{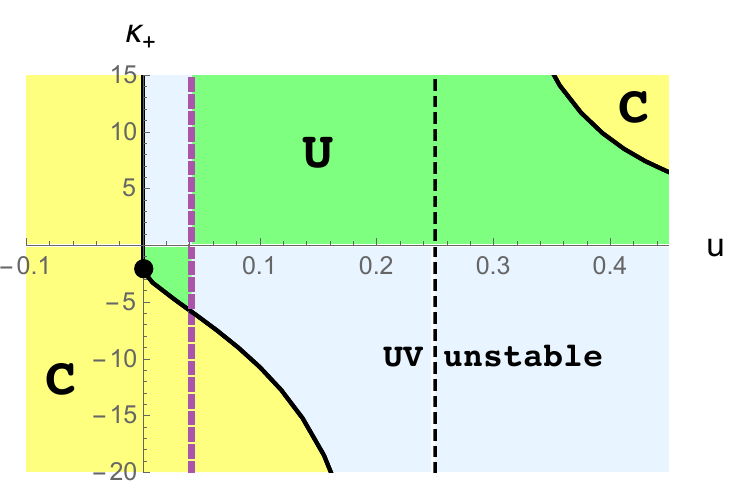} 
\includegraphics[scale=0.7]{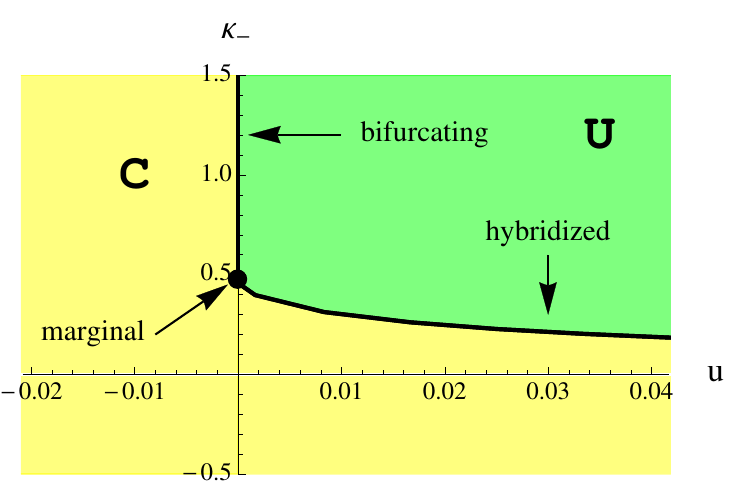}
\end{center}
\caption{The full phase diagram of the system for a neutral scalar. $C$ ($U$) denotes regions with (without)
IR instabilities; $C$ stands for condensed, $U$ for uncondensed phase.  The region with UV instability is filled with light blue.\\
 {\bf Top plot}: phase diagram for the standard quantization.
For $u < 0$, i.e. $m^2 R^2 < -{3 \ov 2}$ the system is always unstable in the IR with $u=0$ the critical line for a bifurcating QCP. The vertical purple dashed line is at $u = {1 \ov 24}$ corresponding to $m^2 R^2 = -{5 \ov 4}$. 
There is no alternative quantization to the right of this line. The vertical black dashed line is at $u={1 \ov 4}$ corresponding to $m^2 = 0$. The curve separating $C$ and $U$ approaches infinity when approaching this line. \\
{\bf Bottom plot}: phase diagram for the alternative quantization (for $\nu_{U} \in (0,1)$, hence the limited range in $u$ compared to the top plot, $u<{1\ov 24}$). The $\ka_->0$ part of the phase diagram can be obtained from the $\ka_+<0$ part of the standard quantization  phase diagram by using the relation~\eqref{euive}.
 In the vacuum, the system has an IR instability for $\ka_- < 0$, i.e. with $\ka_- =0$ the critical line. At a finite density the critical line is pushed into the region $\ka_- > 0$. } 
\label{fig:fullphase}
\end{figure}

The full phase diagram for a neutral scalar operator is given in fig.~\ref{fig:fullphase}.\footnote{A similar phase diagram for AdS$_5$ was determined in~\cite{Ren:2012hg} for a wider range of $u$.} Additional details about the construction of the phase diagram can be found in Appendix~\ref{app:phase}.


\section{Effective theory description of the critical points $(\mu_*=1)$} \label{sec:LEFT}

In this section we illuminate the nature of various quantum critical points discussed in the last section by giving a low energy effective boundary theory description for them. 
For a hybridized QCP, the discussion below slightly generalizes an earlier discussion of~\cite{Faulkner:2010gj}. 

For definiteness, for the rest of the paper we will restrict our discussion to a neutral scalar field with $q=0$. Almost all qualitative features of our discussion apply to the charged case except for some small differences which we will mention along the way. To avoid clutter we set $\mu_*=1$ in this section. 



On general ground we expect that the low energy effective action of the system can be written as
\be \label{eff1}
S_{eff} = S_{\rm eCFT_1} + S_{UV} 
\ee
where $S_{\rm eCFT_1}$ is the action for the IR fixed point \Slql, for which we do not have an explicit Lagrangian description, but (as discussed in Sec.~\ref{sec:ads2}) whose operator dimensions and correlation functions are known from  from gravity in AdS$_2 \times \RR^{2}$.  $S_{UV}$ arises from integrating out higher energy degrees of freedom, and can be expanded in terms of scaling operators in $S_{\rm eCFT_1}$. 
 The part relevant for $\sO$ can be written as
  \be \label{uvact}
S_{UV}=  
{1 \over 2} \int \chi (k) J_{\vec k} J_{-\vec k}  - {1 \over 2} \int  \, \xi_k
\Phi_{\vec k} \Phi_{-\vec k} + \int \eta_{k} \Phi_{\vec k} J_{-\vec k} 
  + \dots
\ee
where $\Phi_\vk$ is the scaling operator at the IR fixed point to which $\sO_\vk$ matches. We have written the action in momentum space since the dimension of  $\Phi_{\vec k} (t)$ is momentum-dependent, and the integral signs should be understood as 
$\int = \int dt {d \vec k }$. We have introduced a source $J_{\vec k}$ for $\sO_{\vec k} $ and $\dots$ denotes higher powers of $\Phi_{\vec k}$ and $J$.  Since we are only interested in two-point functions it is enough to keep $S_{UV}$ to quadratic order in $\Phi$ and $J$.  We have also only kept the lowest order terms in the expansion in time derivatives.  The ``UV data'' $\chi (k), \eta_k$ and $\xi_k$ 
can be found from by integrating out the bulk geometry all the way to the boundary of the near-horizon AdS$_2$ region~\cite{Faulkner:2010jy};
$\chi (k)$ is the static 
susceptibility and other coefficients can be expressed in terms of functions $a_\pm, b_\pm$  we introduced earlier as 
\begin{align}  \label{coffG}
\chi (k) = {b_+^{(0)} (k) \ov \tilde a_+^{(0)} (k)} , \qquad \xi_k = {\tilde a_-^{(0)} (k) \ov \tilde  a_+^{(0)} (k)}, \nonumber \\ \eta_k  = {\sqrt{W} \ov  \tilde a_+^{(0)} (k)}, \qquad 
W \equiv a_+^{(0)} b_-^{(0)} - a_-^{(0)} b_+^{(0)} \ .
\end{align}

In~\eqref{eff1} we are working with the standard quantization of eCFT$_1$; in terms of the notation introduced at the beginning of Sec.~\ref{sec:insta},  it corresponds to ${\rm eCFT}_1^{\rm IR}$ 
and $\Phi_k$ corresponds to $\Phi_+$ with dimension $\ha + \nu_k$. 
Since the full action~\eqref{eff1} is essentially given by  eCFT$_1$ with (irrelevant) double trace deformations, the full correlation function following 
from~\eqref{eff1} can be readily obtained using~\eqref{douGF},  
\be \label{g1}
\chi (\om, k) = \chi(k)  + {\eta^2_k \ov \sG_k^{-1} + \xi_k} \ .
\ee
It can be readily checked that~\eqref{g1} agrees with the lowest order $\om$ expansion of~\eqref{roep23}
with the substitution of~\eqref{coffG}. Alternatively, one can obtain~\eqref{coffG} by requiring~\eqref{g1} to match~\eqref{roep23} ~\cite{Faulkner:2010jy}. \footnote{This is the approach taken by \cite{Faulkner:2010tq}.}

 When $\nu_k$ for $\Phi_\vk$ lies in the range $\nu_k \in (0,1)$ (or $(0,\ha)$ for a charged operator), it is also useful to write the low energy theory in terms of the operator in the alternative quantization, i.e. in terms of eCFT$_1^{\rm UV}$. Again following the procedure of~\cite{Faulkner:2010jy} we find  
\be \label{eff4}
S_{eff} =  
S_{\rm eCFT_{1}^{UV}}   - {1 \over 2} \int  \, \xi_-
\Phi_{-}^2 + \int \eta_{-} \Phi_{-} J  +
{1 \over 2} \int \chi_- J^2
\ee
with
\begin{align} \label{verex}
\chi_- = {b_-^{(0)} (k) \ov \tilde a_-^{(0)} (k)} , & \qquad \xi_-  = - {\tilde a_+^{(0)} (k) \ov \tilde  a_-^{(0)} (k)} = - {1 \ov \xi_k}  \nonumber \\ & \eta_-   = {\sqrt{W} \ov  \tilde a_-^{(0)} (k)}  \ .
\end{align}
In~\eqref{eff4} to distinguish from~\eqref{eff1} we have reinstated the subscript $(-)$ and suppressed $k$-dependence. 

\subsection{Hybridized QCP} \label{sec:Heff}

Near a hybridized QCP~\eqref{hybq}, the effective action~\eqref{eff1}--\eqref{uvact} breaks down as all the coefficient functions in~\eqref{uvact} diverge at $k=0$. 
For example, near $\ka_c$ at  small $k$,  the static susceptibility $\chi (k)$ has the form 
\be 
\chi(\vk) \approx \frac{1}{(\kappa_+ - \kappa_c) + h_k \vk^2  } 
, \quad  
h_k \equiv {\p_{k^2} \tilde a^{(0)}_+ (k) \ov b^{(0)}_+ (k)} \biggr|_{k=0, \ka_+ = \ka_c}  \ 
 \label{hybridizedG0}
\ee
which is the standard mean field behavior with the spatial correlation length scaling as 
\be  \label{colen}
\xi \sim (\kappa_+ - \kappa_c)^{- \nu_{crit}}, \qquad \nu_{crit}=\ha \ .
\ee
The reason for these divergences is not difficult to identify; we must have integrated out some gapless modes, which should be put back to the low energy effective action. 
Indeed as discussed in~\cite{Faulkner09,Faulkner:2010jy}, when $\tilde a_+^{(0)} (k)$ becomes zero at some values of $k$, the bulk equation of motion develops a normalizable mode with $\om =0$, which will give rise to gapless excitations in the boundary theory. Thus near a hybridized QCP, we should introduce a new field $\vp$ in the low energy theory. Clearly there is no unique way of doing this\footnote{We can for example make a field redefinition in $\vp$ as 
$\vp \to Z_1 \vp + Z_2 \Phi$.} 
 and the simplest choice is
 \begin{align} \label{eff2}
S_{eff} = S_{\rm eCFT_1} - {1 \over 2} \int c_k \Phi^2_\vk +\int \, {\lambda_k } \Phi_{-\vk} \vp_\vk 
\nonumber \\ - {1 \over 2} \int \, \vp_{-\vk} \, \chi^{-1} \, \vp_{\vk} + 
  \int \vp J
\end{align}
where 
\begin{align} \label{newpa}
\chi (k) = {b_+^{(0)} (k) \ov \tilde a_+^{(0)} (k)} , \qquad \lam_k = {\eta_k \ov \chi_k} = {\sqrt{W} \ov b_+^{(0)} (k)}, \nonumber \\  
 c_k = \xi_k + \chi_k \lam_k^2 = {b_-^{(0)} (k) \ov b_+^{(0)} (k)}  \ .
\end{align}
Now all the coefficient functions are well defined near a hybridized QCP~\eqref{hybq}, where $\vp$ becomes gapless.  

 It is worth reemphasizing that both $\vp$ and $\Phi$ should be considered as 
low energy degrees of freedom, representing different physics. Given that in~\eqref{eff2} only $\vp$ couples to the source $J$ for the operator $\sO$,  $\vp$ can be considered as the standard Laudau-Ginsburg order parameter (i.e. essentially $\sO$ written as an effective field) 
representing {\it extended} correlations. In particular the phase transition is signaled by that it becomes gapless. The last two terms in~\eqref{eff2} then corresponds to the standard Landau-Ginsburg action for the order parameter. In contrast, as we discussed in Sec.~\ref{sec:semi}, field $\Phi$ from \ircft\ can be considered as representing some strongly coupled {\it semi-local} degrees of freedom whose effective action is given by the first two terms in~\eqref{eff2}.  
The key element in~\eqref{eff2} is that the Laudau-Ginsburg order parameter $\vp$ is now hybridized  with (through the mixing term $\int \, {\lambda_k } \Phi_\vk \vp_\vk$)  some degrees of freedom in \Slql , which are not present in conventional phase transitions. This is the origin of the name  ``hybridized QCP.''  

In summary, the action~\eqref{eff2} can be written as 
\be \label{thep}
S_{eff} = \tilde S_{\rm eCFT_1} [\Phi] + \lam \int  \, \Phi \vp  + S_{LG} [\vp]+ \int \vp J
\ee
where $\tilde S_{\rm eCFT_1}$ is given by the first two terms in~\eqref{eff2} and
\be \label{LGQ}
S_{LG} =  -  {1 \over 2} \int  \vp_{-\vk} (\ka_+ - \ka_c  + h_k  k^2)  \vp_{\vk} \ .
\ee
In this coupled theory, there is then an interesting interplay between  semi-local and extended degrees of freedom. As shown in \cite{Faulkner:2010gj} and as we will see in Sec.~\ref{sec:hybrid} this leads to a variety of novel critical behavior.  When $\vp$ is massive, i.e. away from the critical point, one can integrate out $\vp$ and obtain a low energy effective theory solely in terms of $\Phi$ as in~\eqref{eff1}.

In the range of $\nu_k$ for which the alternative quantization for the \ircft\ applies, the low energy theory can also be described using~\eqref{eff4}. It is interesting that in this formulation all coefficients functions in~\eqref{eff4} are well defined near a hybridized QCP. Thus there is no need to introduce $\vp$ any more and~\eqref{eff4} is the full low energy effective theory. Then how does the SLQL sector of~\eqref{eff4} know that we are dialing the effective mass of $\vp$  in~\eqref{eff2} to drive a quantum phase transition through a hybridized QCP? What happens is that through hybridization between $\Phi$ and $\vp$, when one drives the effective mass for $\vp$ to zero, the double trace coupling in~\eqref{eff1}, $\xi_k$ is driven to infinity and when expressed in terms of alternative quantization, the corresponding double trace coupling $\xi_-$ is driven through $\xi_- =0$ (see~\eqref{verex}), which is precisely a quantum critical point of the 
eCFT$_1$ itself (see discussion of Sec.~\ref{sec:insta} and Appendix~\ref{app:douins}). 
Thus in this formulation, dialing the external parameter directly drives to a 
critical point  of the SLQL sector. 
 
It is also important to emphasize that in the formulation of~\eqref{eff4}, while eCFT$_1$ involves only the time direction, this theory can nevertheless describe the quantum phase transition of the full system including that the spatial correlation length goes to infinity near the critical point, since spatial correlations are encoded in the various $k$-dependent coefficient functions (including the cosmological constant). But this is achieved by some level of conspiracy among various coefficient functions in~\eqref{eff4}, which will not work using generic coefficients as one would normally do in writing a general low energy effective theory. In this sense the effective action~\eqref{eff2} in terms of two sectors is a more ``authentic'' low energy theory.

\subsection{Bifurcating QCP}
 
Let us now consider a bifurcating QCP~\eqref{biy}. Since $\ka_+$ does not play a role here, for notational simplicity we  will set it to zero, i.e. $\ta_\pm$ become $a_\pm$.  

At $u=0$,  the fixed points corresponding to the standard and alternative quantization for $\Phi_{k=0}$ merge into a single one, and for $u < 0$, the \ircft\ becomes unstable as $\Phi$ will develop exponentially growing modes as discussed in Sec.~\ref{sec:insta} and  Appendix~\ref{app:A}.

At a bifurcating QCP, all the coefficient functions in~\eqref{uvact} remain finite. For example, in the $u \to 0$ limit, the susceptibility $\chi (k=0)$  can be written as 
\be \label{unsce}
\chi (k=0) = {\beta \ov \al} - {\sqrt{u} \ov 2 \nu_U \al^2}  + \dots 
\ee
where $\al, \beta$ are some numerical constants and we have used~\eqref{req},~\eqref{W1}. Thus there is no need to introduce the Landau-Ginsburg field $\vp$ as for a hybridized QCP. In other words, in~\eqref{eff2}, at a bifurcating QCP, $\vp$ remains gapped and we can integrate it out.  
Nevertheless, various coefficient functions in~\eqref{uvact} do become singular at $u=0$, with a {\it branch point} singularity, as can be seen from the second term in~\eqref{unsce}.  If we naively extending~\eqref{unsce} and $\eta_k, \xi_k$ to $u < 0$, they become complex.\footnote{This complexity is of course unphysical as for $u < 0$ the disordered phase based on which~\eqref{unsce}  is calculated is unstable. As we will see in Sec.~\ref{sec:biI} the susceptibility for the condensed
phase is indeed real.} 
Also note that from equation~\eqref{corL} the spatial correlation length of \Slql\ does diverge when $u \to 0$ as 
\be 
\xi = {1 \ov \sqrt{6} \sqrt{u}}  \ .
\ee

Let us now focus on the homogenous mode (i.e. $k=0$) and consider the $u \to 0$ limit 
of~\eqref{eff1}--\eqref{coffG}. Note as $u \to 0$, 
\be
\xi_{k=0} = 1 - 2 \sqrt{u} {\tilde \al \ov \al} + \dots , \qquad \eta_{k=0}^2 = {\sqrt{u} \ov \nu_U \al^2} + \dots 
\ee
where we have used~\eqref{req} and~\eqref{oneu}.  Also note from~\eqref{iiRc1}
\be \label{eiro}
\sG_{k=0}  (\om) \to - 1 + 2 \sqrt{u} \sG_0 (\om), \qquad u \to 0 
\ee
with $\sG_0 (\om)$ the retarded function at $\nu=0$, given by\footnote{$\sG_0$ is obtained by solving directly the bulk equation of motion at $\nu=0$.}  
\be \label{eurp}
\sG_{0} (\om) = - \log \le(- {i \om \ov 2 \mu_*} \ri)  -\gamma_E \ 
\ee
where $\mu_*$ is a UV regulator (it is convenient to chose to use the same $\mu_*$~\eqref{gaugeexp} that is supplied by the full theory). \footnote{Contrary to other parts of the section we reintroduced $\mu_*$ for~\eqref{eurp} only.}

Since $S_{\rm eCFT_1}$ in~\eqref{eff1} is defined to be the theory which gives $\sG_{k}$ of~\eqref{eiro}, we see that the $u \to 0$ is a bit subtle as a straightforward limit does not gives an action whose retarded function is~\eqref{eurp}. An efficient way to proceed is to write down the general action
\be \label{effmarg}
S_{eff} = S_{\rm eCFT_1}^{(\nu=0)} + {\tilde \chi_0 \over 2} \int dt \, J^2  - {\tilde \xi_0 \over 2} \int dt  \, \Phi^2
 + \tilde \eta \int dt \, \Phi J 
  + \dots
\ee
 where $J, \Phi$ are shorthand for $J_{k=0}$ and $\Phi_{k=0}$ and $S_{\rm eCFT_1}^{(\nu=0)}$ denotes the action for the IR fixed point in which the retarded function for $\Phi$ is given by~\eqref{eurp}. 
Various coefficients in~\eqref{effmarg} can then be deduced by matching the retarded function from~\eqref{effmarg} with the  $u=0$ limit of~\eqref{roep21}, which is 
\be  \label{sus12}
\chi^{(u=0)} (\om, k=0) = {\beta \sG_0 (\om) + \tilde \beta \ov \al \sG_0 (\om) + \tilde \al} \ .
\ee 
We find that\footnote{There is a systematic procedure to derive these coefficients directly from the $u \to 0$ limit of~\eqref{eff1}, which is not needed here.}  
\be \label{newff1}
\tilde \chi_0 = {\tilde \beta \ov \tilde \al}, \qquad \tilde \xi_0 = {\al \ov \tilde \al}, \qquad \tilde \eta_0^2 = {1 \ov 2 \nu_U \tilde \al^2} \ .
\ee 
For the specific example of tuning to $u=0$ by dialing the mass for the bulk scalar field~\eqref{crimas}, the numerical values of $\al, \beta, \tilde \al, \tilde \beta, \nu_U$ are given in equations~\eqref{pp1}--\eqref{pp2} in Appendix~\ref{app:rev}.

\subsection{Marginal critical point} \label{sec:mcp}
 
In~\eqref{effmarg}, $\Phi$ has dimension $\ha$ and thus the double trace term $\Phi^2$ is marginal. 
As discussed in Sec.~\ref{sec:insta} and Appendix~\ref{app:A}, it is marginally irrelevant when its coupling $\tilde \xi_0$ is positive  and marginally relevant when $\tilde \xi_0 $ is negative (leading to a condensed phase), with $\tilde \xi =0$ being a multi-critical point.  When $\ka_+ =0$, the value of $\tilde \xi_0$ is given by~\eqref{newff1}, which
for the specific example of~\eqref{crimas} has a positive value and thus the system is IR stable. Turning on a nonzero $\ka_+$, $\tilde \xi_0$ generalizes to  
\be 
\tilde \xi_0^{(\ka)} = {\al + \ka_+ \beta \ov \tilde \al + \ka_+ \tilde \beta}  
\ee
and the susceptibility~\eqref{sus12} to
\be  \label{sus13}
\chi^{(u=0,\ka_+)} (\om, k=0) = {\beta \sG_0 (\om) + \tilde \beta \ov (\al + \ka_+ \beta) \sG_0 (\om) + (\tilde \al + \ka_+ \tilde \beta)} \ .
\ee 
There is thus a critical point at 
\be  \label{mcpv}
\ka_+^* = -{\al \ov \beta} 
\ee
which agrees with~\eqref{mcpv1} obtained from directly taking the $u \to 0$ limit of~\eqref{stare} . 
At the critical point, there is a divergent static susceptibility 
\be 
\chi^{(u=0,\ka_+)} (\om=0, k=0) = {\beta \ov \al + \ka_+ \beta } \ .
\ee 
For $\ka_+ < \ka_+^*$, $\tilde \xi^{(\ka)}_0 < 0$, and the system is unstable to the condensation of $\sO$. 
In this case, the condensation is driven by a marginally relevant operator (thus for the name marginal critical point) which generates an exponential IR scale~\eqref{newir} 
\be 
\Lam_{IR} \sim \mu\exp \le({1 \ov \tilde \xi^{(\ka)}_0} \ri) \ 
\ee
just as in the BCS instability for superconductivity.

\section{Aspects of the condensed phase} \label{sec:cond}

In this section we discuss some qualitative features of the spacetime geometry corresponding to the condensed state of a {\it neutral} scalar. In particular, we show that in the IR, the solution again asymptotes 
to AdS$_2 \times \RR^{d-1}$, but with a different curvature radius and transverse size compared with the uncondensed solution.  The discussion applies to both types of instabilities discussed in Sec.~\ref{sec:finin}.

Consider the Einstein-Maxwell action coupled to a {\it neutral} scalar field $\phi$ 
\be \label{pact}
S = \frac{1}{2\kappa^2} \int d^{d+1}x \sqrt{-g} \le(\sR_{d+1} + \frac{d(d-1)}{R^2} - \frac{R^2}{g_F^2}F^2 \ri) + S_{\phi}
\ee 
where $F = dA$ and
\be
S_{\phi} =  \frac{1}{2\kappa^2 g}  \int d^{d+1} x \, \sqrt{-g} \, \le(-\ha (\p \phi)^2 - V (\phi) \ri) \ ,
\ee
where $g$ is the coupling constant for the matter field. 

In the absence of any charged matter the equation of motion for $A_t$ is simply Gauss's law
\be
\p_r \le(\frac{1}{g_F^2} \sqrt{-g}g^{rr}g^{tt}\partial_r A_t \ri) = 0
\ee
Note we work in a gauge in which $A_r = 0$. This equation is nothing but the electric flux conservation  
\be \label{fluxc}
{ E_{\rm prop} \ov g_F^2} A = {\rm const} \ 
\ee
with $E_{\rm prop} \equiv \sqrt{g^{rr}g^{tt}}\p_r A_t$ the electric field in a local proper frame and 
$A \equiv \sqrt{-g \ov  g_{tt} g_{rr}} $ the transverse area. The boundary charge density $\rho_B$ is the canonical momentum with respect to $A_t$ at infinity, which can be written as 
\be \label{chden}
\rho_B =  \frac{2 R^2}{\kappa^2} { E_{\rm prop} \ov g_F^2} A \biggr|_{\infty} =  \frac{2 R^2}{\kappa^2} { E_{\rm prop} \ov g_F^2} A \biggr|_{r_h} \ 
\ee
where we have used~\eqref{fluxc}.
  Given that the entropy density $s$ is the area of the horizon, 
\be
s = \frac{2\pi}{\kappa^2}A (r = r_h)
\ee
 equation~\eqref{chden} also implies that
\be
\frac{\rho_B}{s}
= {R^2 \ov \pi} {E_{\rm prop} \ov g_F^2}\bigg|_{r = r_h}   \  .
\label{rhosrel}
\ee
This is a rather intriguing result which expresses the dimensionless ratio of charge density over the entropy density 
in terms of the local electric field at the horizon in units of the asymptotic AdS radius.

We now express this in terms of more geometric quantities. To do this, we assume that $\phi$ goes to zero (which is a local maximum of $V$) at asymptotic AdS$_{d+1}$ infinity and in the interior settles into a constant value $\phi = \phi_0$ which is a nearby local minimum. We choose the normalization of $V(\phi)$ so that $V(0) =0$ and thus $V(\phi_0) < 0$. At the IR fixed point, the effective cosmological constant is modified from the asymptotic value.  For convenience we define
\be \label{newr2}
{1 \ov \tilde R_2^2} =   {1 \ov R_2^2} - {V (\phi_0)\ov g} \ , \qquad {1 \ov R_2^2} = \frac{d(d-1)}{R^2} 
\ee 
where $R_2$ is the AdS$_2$ radius in the uncondensed phase. Since $V (\phi_0)<0$, we have $R_2 > \tilde R_2$.

Now if we require a nonsingular solution, i.e.  if the electric field is nonsingular,\footnote{If there is a horizon, this means nonsingular also at the horizon.} flux
conservation~\eqref{fluxc} tells us the area $A$ should be finite in the IR. We thus expect that the IR geometry factorizes into the form $\sM_2  \times \mathbb{R}^{d-1}$ with $\sM_2$ some two dimensional manifold involving $r,t$.  Near the horizon we can thus write the $d+1$ dimensional metric as
\be \label{mefc}
ds^2 = g_{MN} dx^M dx^N = g^{(2)}_{\mu\nu}dx^\mu dx^\nu + a^2 d\vec{x}_{d-1}^2
\ee
where $\mu, \nu$ run over the 2d space $(r,t)$ and $a$ is a constant. Now dimensionally reduce along all the spatial directions, the action becomes
\be 
S = \frac{1}{2\kappa^2} \int d^{2}x \sqrt{-g^{(2)}} a ^{d-1} \le(\sR_{2} + {1 \ov \tilde R_2^2} - \frac{R^2}{g_{F}^2}F^2 \ri) \ .
\ee 
Note that here we assume that all active fields do not couple in a special way to the transverse spatial components of the metric $a$, whose effect can thus be taken into account purely from the $a^{d-1}$ factor in the metric determinant; a nonzero magnetic field $F_{xy}$ for example would violate this assumption and introduce extra $a$-dependence into the action. 

Varying the 2d metric $g^{(2)}$ we find
\be
2 E_{\rm prop}^2 = - F^2 =   \frac{g_F^2  }{R^2 \tilde R_2^2} \  \label{Feqn}
\ee
which is simply a constraint on the electric field.  Varying with respect to $a$ and using \eqref{Feqn} in the resulting equation, we find
\be
\sR_2 = - {2 \ov \tilde R_2^2}  \ 
 \ee
 which implies that $\sM_2$ is given by an AdS$_2$ with radius $\tilde R_2$, i.e. the IR metric can be written as 
\be 
ds^2 = {\tilde R_2^2 \ov \xi^2} \le(-dt^2 + d \xi^2 \ri) + a^2 d \vec x^2_{d-1}  \ .
\ee
From~\eqref{Feqn} we also find that
\be 
A_t = {e_d \ov \xi}, \quad {\rm with} \quad e_d \equiv {g_F \tilde R_2 \ov \sqrt{2} R}  \ .
\ee
Given $\rho_B$ we can now also determine the value of $a$ from~\eqref{chden}
\be
a^{d-1} = {\rho_B \kappa^2 g_F \ov \sqrt{2}} {\tilde R_2 \ov R} \ .
\ee

Now using~\eqref{Feqn} in \eqref{rhosrel} we also find that
\be
\frac{\rho_B}{s} = \frac{1}{\tilde R_2}\le(\frac{R}{\sqrt{2}\pi g_F}\ri) \ .
 \label{srho}
\ee
Note the combination of $R$ and $g_F$ appearing in the brackets is the ratio of gauge to gravitational couplings. 
Our discussion leading to~\eqref{srho} only depends on the factorized form of~\eqref{mefc}, which of course also applies to the uncondensed phase with $\tilde R_2$ replaced by $R_2$. 
As we are now in a lower point in the bulk effective potential, we have $\tilde R_2< R_2$, and thus ${\rho_B \ov s}$ increases in the condensed phase. Keeping the charged density fixed, this implies that the entropy density $s$ is smaller in the condensed phase, i.e. 
the condensate appears to have gapped out some degrees of freedom. 
Note that~\eqref{srho}  also provides a boundary theory way to interpret the AdS$_2$ radius: it measures the number of degrees of freedom needed to store one quantum of charge. 

\begin{figure}[h]
\begin{center}
\includegraphics[scale=0.35]{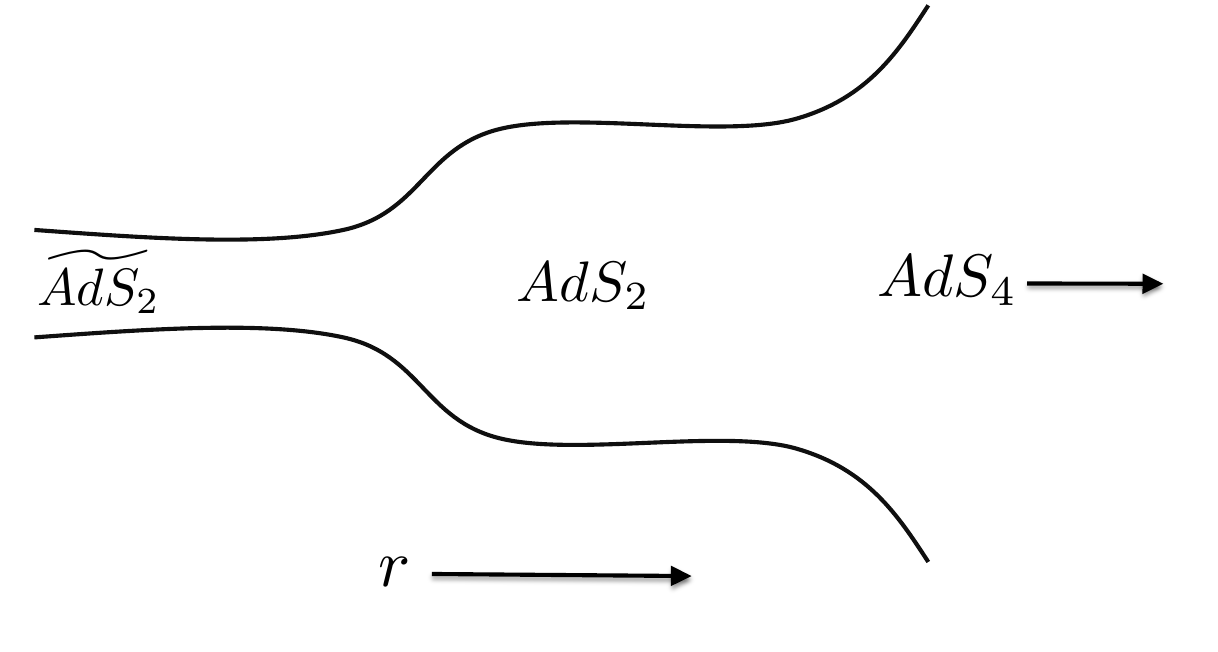}
\includegraphics[scale=0.35]{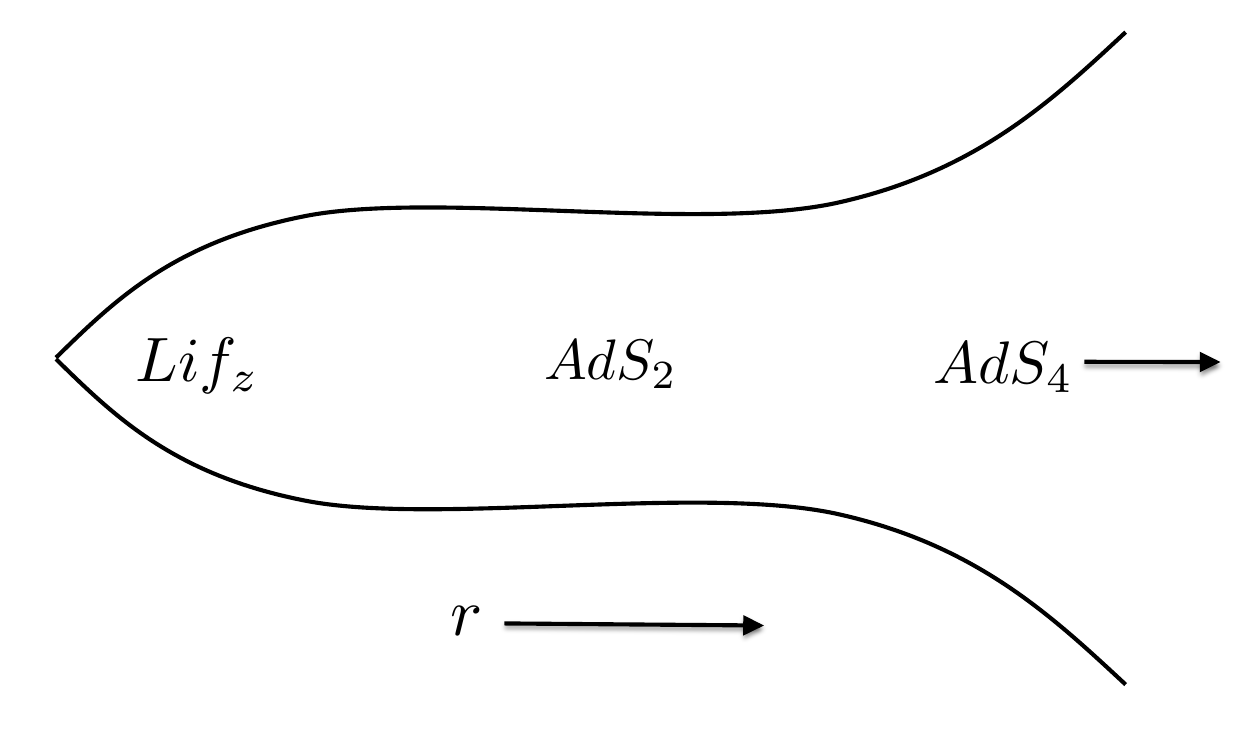}
\end{center}
\caption{Comparison of the spacetime geometries (close to the critical point) corresponding to the condensed state of a neutral (left) and charged scalar (right). The vertical direction in the plot denotes the $\RR^2$ 
directions. For AdS$_2$ the transverse directions has a constant size independent of radial coordinates, while for the Lifshitz geometry, the size of the transverse directions shrinks to zero in the interior. 
Close to the critical point, the IR scale at which the scalar condensate sets
in is much smaller than the chemical potential and we expect an intermediate spacetime region described by the AdS$_2$ of the original black hole geometry.   
} 
\label{fig:ads2t}
\end{figure}

We conclude this section by pointing out a difference between the geometries corresponding to the condensed states of a charged~(holographic superconductor) and a neutral (AFM-type state) scalar. 
As discussed in the above the infrared region of the bulk geometry for the condensate of a neutral scalar is still given by an AdS$_2 \times \RR^2$, but with a smaller curvature radius and entropy density than those of the uncondensed geometry. This implies that such a neutral condensate is not yet the stable ground state, and at even lower energy some other order has to take over~\cite{Iqbal:2011in}. We will return to this point in the conclusion section. 
In contrast, the geometry for a holographic superconductor at zero temperature is given by a Lifshitz geometry (which includes AdS$_4$ as a special example)~\cite{Gubser:2009cg,Horowitz:2009ij,g1} in the infrared. The black hole has disappeared and the system has zero entropy density. Such a solution may be stable and thus could describe the genuine ground state. Note, however, in both cases, the condensed state still has some gapless degrees of freedom left.
See Fig.~\ref{fig:ads2t} for a cartoon which contrasts the difference between the two cases. 

\section{Critical behavior of a bifurcating QCP} \label{sec:biI}

We now proceed to study the critical behavior of the various types of critical points identified in section~\ref{sec:finin}. In this section we study the bifurcating quantum critical point, including the static and finite frequency behavior at zero temperature and then thermal behavior. In this section we will set the double trace deformation to zero, i.e. $\ka_+ =0$, as the story for a nonzero $\ka_+$ 
is exactly the same.

\subsection{Zero temperature:  from uncondensed side}

For convenience we reproduce the expression for the zero-temperature susceptibility \eqref{roep21},
\be
\chi (\om, \vk) = \mu_*^{2\nu_U} {b_+ (k,\om) + b_- (k,\om) \sG_k (\om) \mu_*^{-2 \nu_k}  \ov a_+ (k,\om) + a_- (k,\om) \sG_k (\om) \mu_*^{-2 \nu_k}} \  \label{roeprpt}
\ee
with 
\be
\nu_k = \sqrt{u + {k^2 \ov 6 \mu_*^2}}, \qquad u \equiv m^2 R_2^2 + {1 \ov 4}  \ .
\ee
To study the behavior near the critical point $u=0$ we study the implications of taking $\nu_k \to 0$ in \eqref{roeprpt}, i.e. both $k^2/\mu^2$ and $u$ are small. 

\subsubsection{Static properties}

\begin{figure}[!h]
\begin{center}
\includegraphics[scale=0.8]{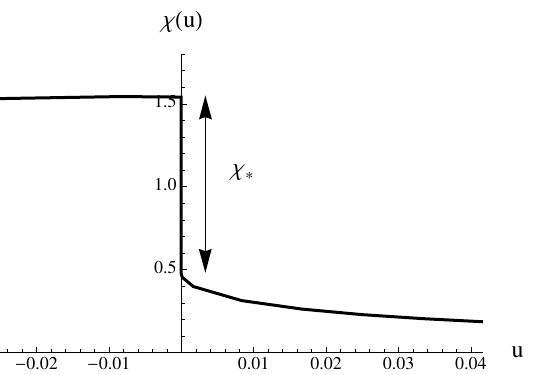}
\end{center}
\caption{A plot of $\chi (u)$ as a function of $u$ with $\mu_*=1$. We also include the behavior on the $u< 0$ side to be worked out in Sec.~\ref{sec:COND}. Note that while there is a cusp in $\chi$ approaching the critical point from the uncondensed side ($u>0$), there is no cusp approaching the critical point from the condensed side ($u<0$). From both sides the susceptibility is finite at the critical point, but there is a jump 
in their values.} 
\label{fig:cusp}
\end{figure}

We first study the critical behavior of the static susceptibility~\eqref{stsus} by setting $\om \to 0$ in \eqref{roeprpt} and taking $u \to 0$ from the uncondensed side $u > 0$. From equation~\eqref{req} we find that for small $\nu_k$, 
\be \label{finkc1}
\chi (k) =\mu_*^{2 \nu_U}  {\beta + \nu_k \tilde \beta \ov \al + \nu_k \tilde \al} + O(\nu^2_k, k^2)
\ee
where $\al, \beta, \tilde \al, \tilde \beta$ are numerical constants.  Setting $k=0$ we find  the zero momentum susceptibility is given by
\be \label{bifu}
\chi =\mu_*^{2 \nu_U} {\beta + \sqrt{u} \tilde \beta \ov \al + \sqrt{u} \tilde \al} \ . 
\ee
As already mentioned earlier, at the critical point the static susceptibility remains finite, given by 
 \be  \label{chiz}
\chi|_{u \to 0_+}=  \chi_0 \equiv \mu_*^{2 \nu_U}  {\beta \ov \al} \ .
 \ee 
which is in sharp contrast with the critical behavior from the Landau paradigm where one expects that the uniform susceptibility always  diverges approaching a critical point. Due to the square root appearing in~\eqref{bifu}, $\chi$ has a branch point at $u=0$ and  
bifurcates into the complex plane for $u < 0$. Of course, when $u < 0$, eq.~\eqref{bifu} can no longer be used, but the fact that it becomes complex can be considered an indication of instability. Furthermore, taking a derivative with respect to $u$  we find that
\be \label{nonan1}
\p_u \chi =\mu_*^{2 \nu_U}  {\al \tilde \beta - \beta \tilde \al \ov 2 \al^2}  {1 \ov  \sqrt{u}}
=  -   {\mu_*^{2 \nu_U} \ov 4 \nu_U \al^2}  {1 \ov \sqrt{u}} \to \infty, \quad (u \to 0) \ 
\ee
where we have used~\eqref{W1} in the second equality.
Thus even though $\chi (u)$ is finite at $u=0$, it develops a cusp there, as shown in Fig.~\ref{fig:cusp}. 
It will turn out convenient to introduce a quantity 
\be  \label{chiS}
\chis \equiv \mu_*^{2 \nu_U}  {1 \ov 4 \nu_U \al^2}  = \chi_0 {1 \ov 4 \nu_U \al \beta}
\ee
and then~\eqref{nonan1} becomes 
\be 
\p_u \chi = - {\chi_* \ov \sqrt{u}} \ .
\ee 
Similarly, 
taking derivative over $k^2$ in~\eqref{finkc1} and then setting $k=0$, we find that 
\be \label{nonan2}
 \p_{k^2} \chi (\vec k)\bigr|_{k=0}= - {\chis \ov 6 \mu^2_* \sqrt{u}}  , \qquad u \to 0 \ .
\ee
Note that this divergence is related to the fact for any $u > 0$, $\chi (\vk)$ is analytic in $k^2$, but not at $u=0$, where $\nu_k \propto k$.

The above non-analytic behavior at $k=0$ should have important consequences when we 
Fourier transform $\chi (\vk)$ to coordinate space. Indeed by comparing~\eqref{uude} with~\eqref{defnk}, we find that 
\be \label{zeroXi}
\xi = {1 \ov \sqrt{6} \mu_* \sqrt{u}} \ .
\ee
Thus as $u \to 0$, the correlation length $\xi$ diverges as $u^{-\ha}$ which is the same as that in a mean field theory.  More explicitly, Fourier transforming $\chi (k)$ to coordinate space 
we find 
asymptotically at large $x$, 
\be
G(x) \equiv \int \frac{d^2k}{(2\pi)^2} \chi(\vk) e^{ikx} \approx \frac{\chis \sqrt{u} }{\pi x^2} \exp\le(- {x \ov \xi} \ri) \ . \label{bifcorr}
\ee
Note however that there is additional suppression by factors of $\sqrt{u}$ in the numerator of this expression; this suggests that the actual power law falloff at the critical point is not the one found from setting $u \to 0$ above, but is rather faster. Indeed performing the integral at precisely $u= 0$ we find 
\be
G(x)\big|_{u=0} \sim 
 \frac{\mu_*^{2\Delta -1}}{(\mu_* x)^3} 
\label{bifcorrcrit}
\ee
with a different exponent $\sim x^{-3}$. 


\subsubsection{Dynamical properties} \label{sec:biow}


We now turn to the critical behavior of the susceptibility~\eqref{roep21} at a nonzero $\om$ near the critical point from uncondensed side $u >0$. 
We should be careful with the $\nu_k \to 0$ limit as the factor $\le(\frac{\om}{\mu_*}\ri)^{2\nu_k}$ in the AdS$_2$ Green function~\eqref{iiRc1} behaves differently depending 
on the order we take the $\nu_k \to 0$ and $\om \to 0$ limits. For example, the Taylor expansion of such a term in small $\nu_k$  involves terms of the form $\nu_k\log(\om/\mu_*)$, but in the small $\om$ limit, the resulting large logarithms may invalidate the small $\nu_k$ expansion. 
 
To proceed,  we note first that the expression~\eqref{roeprpt} together with the explicit expression for the AdS$_2$ Green's function \eqref{iiRc1} can be written
\bwt
\be \label{rew1}
\chi (\om, \vk) = \mu_*^{2\nu_U}\le(\frac{b_+ \Ga(\nu_k)\le({-i\om \ov 2\mu_*}\ri)^{-\nu_k} + b_-\Ga(-\nu_k)\le({-i\om \ov 2\mu_*}\ri)^{\nu_k}}{a_+ \Ga(\nu_k)\le({-i\om \ov 2\mu_*}\ri)^{-\nu_k} + a_-\Ga(-\nu_k)\le({-i\om \ov 2\mu_*}\ri)^{\nu_k}}\ri) \ .
\ee
\ewt
Now from the discussion at the beginning of Appendix~\ref{app:abpro}, we can write  
\be \label{rew2}
a_{\pm} = a(\pm\nu_k; k^2,\om), \quad b_{\pm} = b(\pm\nu_k; k^2,\om)
\ee
where $a(\nu; k^2, \om)$ and $b (\nu; k^2,\om)$ are some functions analytic in all its variables. 
Using~\eqref{rew2}, eq.~\eqref{rew1} can be further written as 
\be \label{nicex}
\chi (\om, \vk) = \mu_*^{2\nu_U}\frac{f_b(\nu_k) - f_b(-\nu_k)}{f_a(\nu_k) - f_a(-\nu_k)}
\ee
where\footnote{Note that $\Ga(\nu \to 0) \sim \frac{1}{\nu} - \ga + O(\nu)$, necessitating the extra factor of $\nu$ in the definition of $f_{a,b}(\nu)$ to obtain a nonsingular Taylor expansion.} 
\be 
f_{b}(\nu) \equiv b(\nu)\nu\Ga(\nu)\le(\frac{-i\om}{2\mu_*}\ri)^{-\nu}
\ee
and similarly for $f_a(\nu)$. 
The point of this rewriting is to illustrate that if $f_{a,b}(\nu)$ have nonsingular Taylor expansions in $\nu$ -- which is the case for any finite $\om$ -- then if we expand numerator and denominator in $\nu$ all the terms that are odd in $\nu$ will cancel, and thus $\chi (\om, \vk)$ contains only {\it even} powers of $\nu_k$, i.e. $\chi (\om, \vk) = \chi (\om, u, k^2)$ and for any nonzero $\om$, $\chi (\om, \vk)$ is {\it analytic} at $u=0$ and $k^2 =0$. There is no branch-point  singularity that was found in \eqref{finkc1}. In particular the expression for $\chi (\om, \vk)$ approaching $u=0$ for the condensed side can be simply obtained by analytically continuing~\eqref{nicex} to $u < 0$. 
This should be expected  since for a given $\om$, as we take $u \to 0_-$, it should always be the case that $\om$ is much larger than the scale where the physics of condensate sets in, which should go to zero with $u$. Thus the physics of the condensate is 
not visible at a given nonzero $\om$. We will see in next section that the same thing happens at finite temperature.    

Now expanding the Gamma function and  $a_\pm ,b_{\pm}$ in~\eqref{rew1} to leading order in $\nu_k$, but keeping the full dependence on $\om$, we find that 
\be
\chi (\om, \vk) = \chi_0 \frac{\sinh\le(\nu_k \log\le(\frac{-i\om}{\om_b}\ri)\ri)}{\sinh\le(\nu_k \log\le(\frac{-i\om}{\om_a}\ri)\ri)} + \dots  \label{finomsinh}
\ee
where the energy scales $\om_{a,b}$ are given by
\be \label{defwab}
\om_{a} = 2\mu_* \exp \le(\frac{\tilde{\al}}{\al} - \ga_E \ri), \quad 
\om_b = 2\mu_* \exp \le(\frac{\tilde{\beta}}{\beta} - \ga_E \ri),
\ee
 where $\ga_E$ is the Euler-Mascheroni constant, and $\chi_0$ is uniform susceptibility at the critical point given earlier in~\eqref{chiz}. For a charged scalar, equations~\eqref{nicex} and~\eqref{finomsinh} still apply with slightly different functions $f_a, f_b$ and $\om_a, \om_b$ becoming complex.

 Considering $\nu_k \to 0$ in~\eqref{finomsinh} with a fixed $\om$,  we then find
\be \label{zernu}
\chi(\om, \vk ) =\chi_0 \frac{\log\le(\frac{\om}{\om_b}\ri) - i\frac{\pi}{2}  }{\log\le(\frac{\om}{\om_a}\ri) - i\frac{\pi}{2} }  + O(u, k^2)\ 
\ee
whose leading term is simply~\eqref{sus12} and  the corrections are analytic in both $u$ and $k^2$.  Note that both above expression and~\eqref{finomsinh} have a pole at $\om = i \om_a$ in the upper half $\om$-plane.  But this should not concern us as our expressions are only valid for $\om \ll \mu_* \sim \om_a$.


Further taking the $\om \to 0$ limit in~\eqref{zernu} then gives 
\bea \label{zerwL}
\chi (\om, \vk) &=& \chi_0 \le(1 + {1 \ov 2 \nu_U \al \beta} {1 \ov \log \om} + {i \pi \ov 4\nu_U \al \beta}
{1 \ov (\log \om)^2} + \dots \ri) \cr
& = &  \chi_0 +   {2 \chis \ov \log \om}   +
{i \pi \chis \ov (\log \om)^2}   + \dots
\eea
where we have kept the leading nontrivial $\om$-dependence in both real and imaginary parts
and used~\eqref{chiS}. 

Equations~\eqref{zernu} and~\eqref{zerwL} give the leading order expression at nonzero $u$ (for both signs, as $\chi (\om,k)$ is analytic at $u=0$ at a nonzero $\om$) and $k^2$ as far as $\nu_k \log {\om \ov \om_{a,b}}$  remains small. They break down when $\om$ becomes exponentially small in $\frac{1}{\nu}$, 
\be \label{uepp}
\om \sim \Lam_{\rm CO}, \quad  
\Lam_{\rm CO} \sim \mu_* e^{-{\# \ov \sqrt{u}}}  \ 
\ee
where $\#$ denotes some $O(1)$ number. In the regime of~\eqref{uepp}, the susceptibility~\eqref{finomsinh} crosses over to 
\be \label{theoL}
\chi (\om \to 0, \vk) = \chi_0 - 2 \nu_k \chis - 4 \nu_k \chis \le({-i \om \ov 2 \mu_*}\ri)^{2 \nu_k} + \dots
\ee
which is the low energy behavior~\eqref{gre} for the uncondensed phase and also 
consistent with~\eqref{finkc1}. Note that $\dots$ in the above equation also includes perturbative corrections in $\om$.

\subsection{Zero temperature: from the condensed side} \label{sec:COND}

When $u < 0$, the IR scaling dimension of $\sO_\vk$ becomes complex for sufficiently small $k$ as  $\nu_k = \sqrt{u + {k^2 \ov 6 \mu_*^2}} = - i \lam_k$ is now pure imaginary.\footnote{Note that the choice of branch of the square root does not matter as~\eqref{finomsinh} is a function of $\nu_k^2$.} For a given nonzero $\om$ and $|u|$ sufficiently small, as discussed after~\eqref{nicex} the corresponding expression for $\chi (k, \om)$ can be obtained from~\eqref{finomsinh} by simply taking $u$ to be negative, after which we find
 \be 
  \chi (\om, \vk) = \chi_0 \frac{\sin\le(\lam_k \log\le(\frac{-i\om}{\om_b}\ri)\ri)}{\sin\le(\lam_k \log\le(\frac{-i\om}{\om_a}\ri)\ri)} + \dots  \ . \label{usmS}
\ee
While~\eqref{finomsinh} is valid to arbitrarily small $\om$, equation~\eqref{usmS} has poles in the {\it upper} half frequency plane (for $k=0$)  at\footnote{Note that~\eqref{usmS} also have poles 
for non-positive integer $n$. But at these values $\om$ is either of order or much larger than the chemical potential $\mu$ to which our analysis do not apply.} 
\be  \label{polemm}
\om_n = i \om_a \exp\le(-{n \pi \ov \sqrt{-u}} \ri) \equiv i \Lam_n , 
, \qquad n=1,2 \dots  \ 
\ee
with 
\be  \label{irsca2}
\Lam_n \sim \mu  \exp\le(-{n \pi \ov \sqrt{-u}} \ri) \ . 
\ee
In particular, we expect~\eqref{usmS} to break down for $\om \sim \Lam_1$, the largest among~\eqref{irsca2}, and at which scale the physics of the condensate should set in. 
This is indeed consistent with an earlier analysis of classical gravity solutions in~\cite{Iqbal:2010eh,Jensen:2010ga} where it was found that  $\sO$ develops an expectation value of order 
\be \label{exvev1}
{\vev{\sO} \ov \mu^\De} \sim \le({\Lam_{1} \ov \mu} \ri)^{\ha}   \ . 
\ee
The exponent $\ha$ in~\eqref{exvev1} is the scaling dimension of $\sO$ in the \Slql\
for $u=0$, while $\De$ is its UV scaling dimension in the vacuum. It was also found in~\cite{Iqbal:2010eh,Jensen:2010ga} there are an infinite number of excited condensed states with a dynamically generated scale given by $\Lam_n$  and $\vev{\sO} \sim \Lam^\ha_n$, respectively. Thus the pole series in~\eqref{polemm} in fact signal 
a geometric series of condensed states. This tower of condensed states with geometrically spaced expectation values is reminiscent of Efimov states~\cite{efimov}.\footnote{In fact the gravity analysis (from which~\eqref{usmS} arises) reduces to  the same quantum mechanics problem as that of the formation of three-body bound states in~\cite{efimov}.} 
The largest is in the first  state $n = 1$, which is the energetically favored vacuum (see the discussion of free energy below). 

\subsubsection{Static susceptibility}

\begin{figure*}[!ht]
    \begin{minipage}[l]{1.0\columnwidth}
       \begin{center}
\includegraphics[scale=0.5]{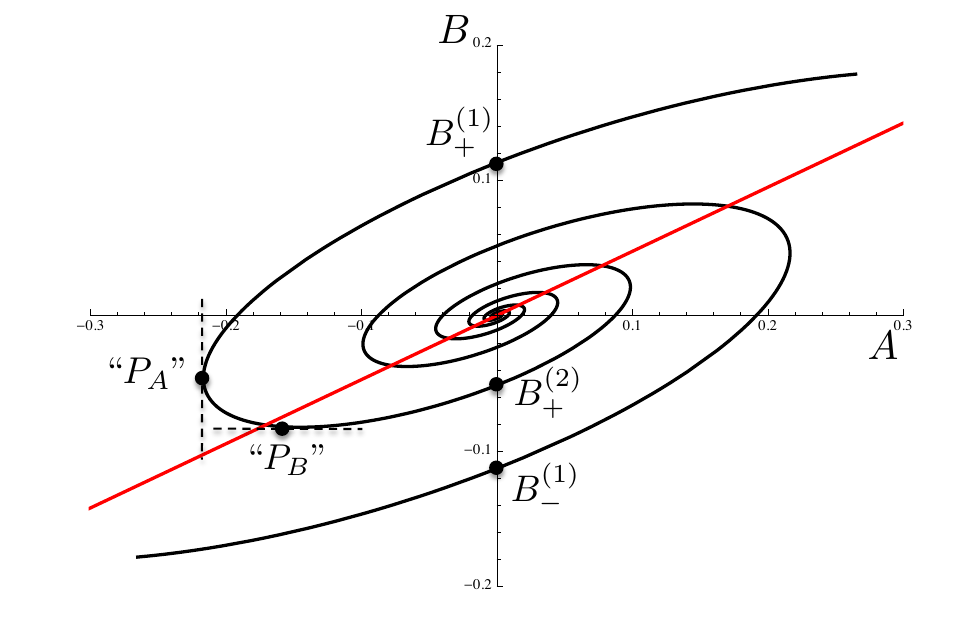} 
\end{center}
\caption{The spiral in the $B-A$ plane passing through  $B_+^{(1)}$ gives the solution described by~\eqref{ABexpr12} as $\zeta_*$ is varied. The spiral passing through $B_-^{(1)} = - B_+^{(1)}$ gives the mirror curve from $\sO \to - \sO$ reflection symmetry. 
The normalizable solutions in the standard quantization are given by the intersections of the spiral with respect to the $B$-axis with $B_\pm^{(1)}$ the ground states and $B_\pm^{(2)}$ the first excited states and etc.  
The red straight line has slope given by~\eqref{limitslope}. As $\sqrt{-u} \to 0$ most part
of the spiral becomes parallel to it.  For ease of visulazation a nonlinear mapping has been performed along the the major and minor axes of the spiral; while the zeros of $A$ and $B$ are preserved by this mapping the location of divergences and zeros of $\frac{dB}{dA}$ are {\it not} (hence the quotation marks in the location of ``$P_A$'' and ``$P_B$'', which are only for illustrative purposes).
}
\label{fig:spiral1}
    \end{minipage}
    \hfill{}
    \begin{minipage}[r]{1.0\columnwidth}
        \begin{center}
\includegraphics[scale=0.5]{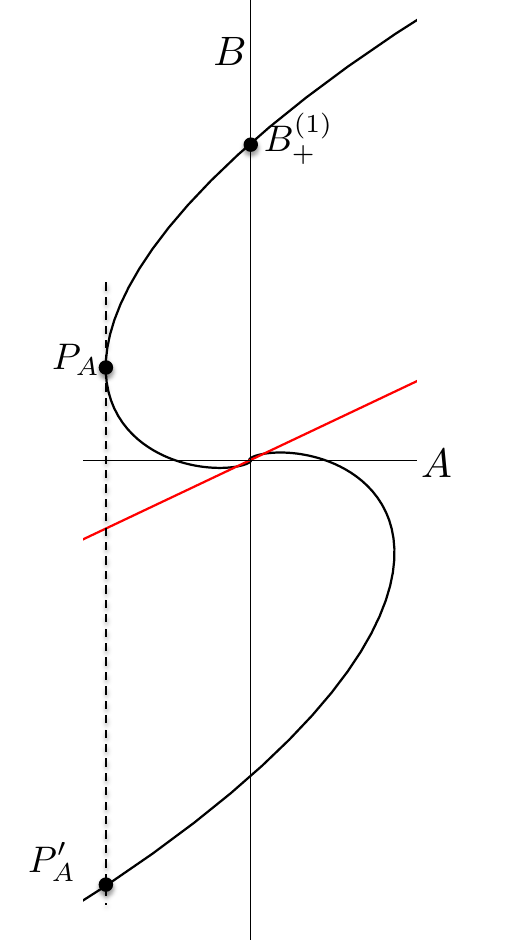}
\end{center}
\caption{A zoomed in version of the spiral, where there has been no nonlinear mapping and so the location of $P_A$ is faithfully reproduced. As described in the text, at $P_A$ the system becomes locally unstable and relaxes to $P_A'$. Appearances to the contrary, the spiral continues to wind around infinitely many times as it approaches the origin, a fact that is difficult to see without the nonlinear mapping due to the exponential spacing of the intersections.
  }
\label{fig:spiral2}
    \end{minipage}
\end{figure*}

In Appendix~\ref{app:COND}, we compute the response of the system to a static and uniform external source in this tower of ``Efimov'' states. The result 
is rather interesting and can be described as follows. One finds that the response in all the ``Efimov''  states can be read from a pair of continuous spiral curves described parametrically 
by (for $\sqrt{-u} \ll  1$)\footnote{Note that the following result applies to both neutral and charged cases.}
\bea \label{ABexpr12}
A  &=& z_*^{3-\De} {\ga \ov \sqrt{-u}} \al  \sqrt{z_{*} \ov \zeta_{*}}  \sin\le(\sqrt{-u} \log {\zeta_{*} \ov z_*}  + \sqrt{-u} {\tilde \al \ov \al} \ri), \nonumber \\
B &=&  z_*^{-\De} {\ga \ov \sqrt{-u}} \beta  \sqrt{z_{*} \ov \zeta_{*}}  \sin\le(\sqrt{-u} \log {\zeta_{*} \ov z_*} + \sqrt{-u} {\tilde \beta \ov \beta}\ri). \nonumber \\ 
\eea
where $A$ and $B$ denote the source and expectation value for $\sO$ respectively, $\ga$ is a $O(1)$ constant. $\zeta_{*}^{-1}$ is a dynamical energy scale which parametrizes movement through the solution space; as we vary $\zeta_*$, we trace out a {\it spiral} in the $(A,B)$ plane.\footnote{Infinite spirals in holography were previously  found by~\cite{Mateos:2007vn} in a different setting resulting in first order phase transitions, by~\cite{Moroz:2009kv} in nonrelativistic holography and very recently by~\cite{Evans:2011zd}.} See fig.~\ref{fig:spiral1}. Since we are considering a system with a $Z_2$ symmetry $\sO \to - \sO$, in~fig.~\ref{fig:spiral1} there is also a mirror spiral obtained from~\eqref{ABexpr12} by taking $(A,B) \to - (A,B)$.

The tower of ``Efimov'' states is obtained by setting the source $A=0$, which leads to 
\be \label{doew1}
\zeta_* = \zeta_n \equiv  z_{*} e^{{n \pi \ov \sqrt{-u}} - {\tilde \al \ov \al}} , \qquad n=1,2, \dots
\ee
which when plugged into the expression for $B$ in~\eqref{ABexpr12} gives
\be
\vev{\sO} \propto |B| =  \mu_*^\De {\ga \ov 2\nu_U\al}  e^{-{n \pi \ov 2\sqrt{-u}} + {\tilde \al \ov 2 \al}} \sim \mu_*^\De \exp\le(-n\frac{\pi}{2\sqrt{-u}}\ri) \label{efimovVEV1}
\ee
where we have used~\eqref{W1}. These are the values at which the spiral intersects with the 
vertical axis, with that for the $n=1$ state corresponding to the outermost intersection. Note that $\zeta_n \sim \Lam_n^{-1}$ and equation~\eqref{efimovVEV1} is consistent with the discussion 
below~\eqref{exvev1}. 

As $\sqrt{-u} \to 0$, from~\eqref{ABexpr12}, $A$ and $B$ are becoming in phase, and the spiral is being squeezed into a straight line, with limiting slope
\be
\frac{B}{A}\bigg|_{\sqrt{-u} \to 0} = \mu_*^{2\nu_U}\frac{\beta}{\al} = \chi_0 \ . \label{limitslope}
\ee
This slope agrees with the value found from linear response approaching the critical point from the other side \eqref{chiz}. This is however {\it not} the relevant slope for the susceptibility, which should be given by 
\be \label{lsus}
\chi_L = {dB \ov d A}\biggr|_{A=0}
\ee
which in the usual models of spontaneous symmetry breaking, corresponds to the longitudinal susceptibility. From~\eqref{ABexpr12} we find
 \be
\begin{split}
\chi_L 
&= \mu_*^{2 \nu_U} {\beta \ov \al} \le(1+ \frac{\tilde \al \beta-\al \tilde  \beta}{2 \al \beta} \ri)+O(u)= \chi_0 + \chis
 +O(u) \label{slopechange1}
\end{split}
\ee
where we have used~\eqref{W1} and~\eqref{chiS}. 
Essentially, even though the spiral is squished into a straight line as we approach the transition, each {\it intersection} of the spiral with the $A = 0$ axis has a different slope than the limiting slope of the entire spiral. 
Note that this result is independent of $n$ and in particular applies to $n=1$, the ground state. Since $\chi_0$ is the value at $u= 0$ from the uncondensed side, we thus find a jump in the value of uniform susceptibility  in crossing $u=0$ (see Fig.~\ref{fig:cusp}) and the difference is precisely the same coefficient as the divergent terms in~\eqref{nonan1}, which also appears in other places 
such as~\eqref{zerwL}.

We now elaborate a bit more on the interpretation of various parts of the spirals in Fig.~\ref{fig:spiral1}. Let us start with the ground state\footnote{Equivalently we can also start with its $Z_2$ image $B_-^{(1)}$.} $B_+^{(1)}$, and first follow the spiral to the right, i.e. we apply an external source $A$ in the {\it same} direction as the condensate. This will increase $B$ according to~\eqref{lsus} and~\eqref{slopechange1}. Note that near the critical point $B_+^{(1)}$ is exponentially small; thus as we  increase $A$ further, we will eventually reach a regime where the forced response is much larger than the condensate $B \gg B_+^{(1)}$ but is still much smaller than $1$, $B \ll 1$. One thus expects that here the system should not care about the (exponentially small) condensate and the response should simply be given by that at the critical point, i.e. the linear response line given by $\chi_0$. Thus the spiral will approach a straight line  parallel to the red straight line in the figure.  

Now consider applying $A$ in the opposite direction to the condensate. As the $Z_2$ symmetry was {\it spontaneously} broken, we now expect that 
$B_-^{(1)}$ should be the global minimum and $B_+^{(1)}$ should be only locally stable. Nevertheless, we can choose to stay in the ``super-cooled'' state given by $B_+^{(1)}$ and stay on the response curve given by 
following the spiral at $B_+^{(1)}$ to the left, where now the source acts to {\it reduce} $B$. 
The response curve in the region between $B_+^{(1)}$ and $P_A$ is nonlinear as the effect of the condensate is important. At $P_A$ the susceptibility ${d B \ov d A} \to + \infty$ ($P_A$ corresponds to an inflection point in the effective potential) and the state that we are on becomes genuinely (i.e. even {\it locally}) unstable and if we continue to increase $|A|$, then the system will relax to the point $P_A'$  on the other branch of the spiral
starting from $B_-^{(1)}$.  Note that 
\be 
{B(P_A) \ov B_+^{(1)}} \sim O(1)
\ee
where by $O(1)$ we mean that the ratio is independent of the small parameter $\sqrt{-u}$. 

To complete the story let us now consider starting from the first excited state $B_+^{(2)}$ and again apply the external source along the direction of the condensate, which now corresponds to following the spiral to the left. Near $B_+^{(2)}$, the response is again controlled by~\eqref{slopechange1}, but again when $1 \gg |B| \gg |B_+^{(2)}|$, the system will forget that 
it is in a condensed state and the response will again be controlled by $\chi_0$. The response curve will once again be parallel to the linear response line until we reach the region near $P_B$, where the response has now become exponentially large compared with the value at $B_+^{(2)}$, i.e. it is now comparable to the value of $|B_+^{(1)}|$:
\be 
 {|B(P_B)| \ov B_+^{(1)}} \sim O(1), \quad 
\le|{B(P_B) \ov B_+^{(2)}} \ri| \sim O\le(e^{\pi \ov \sqrt{-u}} \ri) \ . 
\ee
Near $P_B$ the nonlinear effects due to the condensate again become important. In the region between $P_A$ and $P_B$ the susceptibility has the wrong sign and thus the system becomes locally thermodynamically unstable.  Also note that even though $B_+^{(2)}$ is an excited state and so not a global minimum of the free energy, it does appear to be locally thermodynamically stable. 

The discussion above also gives a physical explanation as to why as $u \to 0_-$ the whole spiral is squished into a straight line with slope given by~\eqref{chiz}: the vast majority of the spiral (e.g. the exponentially large region between $B_+^{(2)}$ and $P_B$) must become parallel to such a straight line.

The existence of a tower of  ``Efimov'' states with geometrically spaced expectation values may be considered as a consequence of spontaneous breaking of the discrete scaling symmetry
of the system. With an imaginary scaling exponent,~\eqref{usmS} exhibits a  discrete scaling symmetry with (for $k=0$)
\be 
\om \to e^{ {2 \pi \ov \sqrt{-u}}}  \om \  
\ee
which is, however, broken by the condensate.\footnote{Note that for $n=1$ state, since the physics of the condensate sets in already at $\Lam_1$, the range of validity for~\eqref{usmS} is not wide enough for the discrete scaling symmetry to be manifest.} The tower of ``Efimov'' states may then be considered as the ``Goldstone orbit'' for this broken discrete symmetry. 

We would also like to point out that $\p_u \chi_L$ and $\p_{k^2} \chi_L$ do not diverge at the critical point unlike from the uncondensed side. Hence we do not get a cusp approaching the critical point from the condensed side. This is due to that small $u$ corrections to~\eqref{slopechange1} are all analytic, which can be checked by explicit calculations to next nontrivial order as already indicated in~\eqref{slopechange1}.

\subsubsection{Free energy across the quantum phase transition} \label{sec:bfenerg}

The fact the order parameter~\eqref{efimovVEV1} is continuous (to an infinite number of derivatives) across the transition implies that the free energy is also continuous (to an infinite number of derivatives). We outline the arguments here.  The free energy is simply the (appropriately renormalized) Euclidean action of the scalar field configuration. We can divide the radial integral into two parts, the UV part and the IR AdS$_2$ part. It is clear that the contribution from the UV portion of the geometry will scale like $\phi^2 \sim \vev{\sO}^2 \sim \exp \le[-{\pi \ov \sqrt{-u }} \ri]$, since the scalar is small there and so a quadratic approximation to the action is sufficient. 

To make a crude estimate of the IR contribution in which region $\phi \sim O(1)$, let us ignore backreaction and imagine that in the IR the scalar is simply a domain wall: for $\zeta > \Lam_{IR}^{-1}$ it sits at the bottom of its potential $\phi(\zeta) = \phi_0$ and that for $\zeta < \Lam_{IR}^{-1}$ it is simply $0$. Then we find for the Euclidean action\footnote{As we are at zero temperature the Euclidean time is not a compact direction, and so all expressions for the Euclidean action contain a factor extensive in time that we are not explicitly writing out.} an expression of the form
\be
S_E \sim V(\phi_0) \int_{\infty}^{\Lam_{IR}^{-1}} d\zeta \sqrt{g} \sim \Lam_{IR}
\ee
which again scales as
$S_E \sim \Lam_{IR} \sim \exp \le[-{\pi \ov \sqrt{-u }} \ri]$. Note what has happened: even though the scalar is of $O(1)$ in the deep IR and so contributes to the potential in a large way, the infinite redshift deep in the AdS$_2$ horizon suppresses this contribution to the free energy, making it comparable to the UV part. A more careful calculation 
 also reveals that the free energy is indeed negative compared to the uncondensed state. We thus conclude that 
 \be 
 F \sim - \exp \le[-{\pi \ov \sqrt{-u }} \ri]
 \ee
 and  that the free energy is also continuous across the transition to an infinite number of derivatives, reminiscent of a transition of the Berezinskii-Kosterlitz-Thouless type.\footnote{The argument presented here are in agreement with the results of~\cite{Jensen:2010ga}.}

\subsection{
Thermal aspects} \label{sec:finTB}

We now look at the critical behavior near the bifurcating critical point at a finite temperature.  Our starting point is the expression for the finite-temperature susceptibility, which we reproduce below for convenience:
\be \label{roep22rpt}
\chi (\om, \vk, T) =\mu_*^{2\nu_U} {b_+ (k,\om,T) + b_- (k,\om,T) \sG_k^{(T)} (\om) \mu_*^{-2 \nu_k}  \ov a_+ (k,\om,T) + a_- (k,\om,T) \sG_k^{(T)} (\om) \mu_*^{-2 \nu_k}}, 
\ee
The finite temperature behavior mirrors the finite frequency behavior of last subsection. We simply repeat the analysis leading to~\eqref{finomsinh}, starting with \eqref{roep22rpt} rather than \eqref{roeprpt}; somewhat predictably, at $\om =0$ but finite $T$ we find
\be
\chi^{(T)}(\vk) = \chi_0 \frac{\sinh\le(\nu_k \log\le(\frac{T}{T_b}\ri)\ri)}{\sinh\le(\nu_k \log\le(\frac{T}{T_a}\ri)\ri)} , \label{finTsinh}
\ee
where $T_{a,b}$ differ from $\om_{a,b}$ by factors\footnote{For a charged scalar while $\om_{a,b}$ are complex , $T_{a,b}$ remain real.},  
\be 
{T_a  }=  {4 \mu_* \ov \pi} e^{\tilde \al \ov \al}  , \qquad 
{T_b  }=  {4 \mu_* \ov \pi} e^{\tilde \beta \ov \beta}
\ . \label{defTab}
\ee
Similarly to~\eqref{usmS}, the expression for $u < 0$ is obtained by analytically continuing~\eqref{finTsinh} to  obtain 
\be
\chi^{(T)} (\vk) =  \chi_0 \frac{\sin\le(\lam_k \log \le(\frac{T}{T_b}\ri)\ri)}{\sin\le(\lam_k \log \le(\frac{T}{T_a}\ri)\ri)} \ . \label{imagFinTsusc}
\ee
And again both~\eqref{finTsinh} and~\eqref{imagFinTsusc} are analytic at $u=0$ and reduce to 
the same function there
\be
\chi^{(T)} (\vk) =\chi_0  {\log {T \ov T_b} \ov \log {T \ov T_a}} + O(u, k^2)
\ .  \label{finiteTsusc}
\ee
Similar to~\eqref{zernu}, the pole in~\eqref{finiteTsusc} and~\eqref{finTsinh} at $T=T_a$ should not concern us as this expression is supposed to be valid only for $T \ll \mu_* \sim T_a$. For nonzero $\om$, \eqref{finTsinh} generalizes to 
\be
\chi^{(T)}(\om, k) = \chi_0 \frac{\sinh\le(\nu_k\le[ \log\le(\frac{2\pi T}{\om_b}\ri) + \psi\le(\ha - i\frac{\om}{2\pi T}\ri)\ri]\ri)}{\sinh\le(\nu_k \le[\log\le(\frac{2\pi T}{\om_a}\ri)  + \psi\le(\ha - i\frac{\om}{2\pi T}\ri)\ri]\ri)} \ ,\label{finTomsusc}
\ee 
where $\psi$ is the digamma function. It is easy to check using the identities $\psi(\ha) = -\ga_E - \log 4$ and $\psi(x \to \infty) \to \log x$ that this expression has the 
correct limiting behavior to interpolate between \eqref{finTsinh} and \eqref{finomsinh}.
Taking $\nu_k \to 0$ with $\om$ and $T$ fixed, we then find that
\be
\chi^{(T)}(\om,\vk) = \chi_0 \frac{\log\le(\frac{2\pi T}{\om_b}\ri) + \psi\le(\ha - i\frac{\om}{2\pi T}\ri)}{\log\le(\frac{2\pi T}{\om_a}\ri)  + \psi\le(\ha - i\frac{\om}{2\pi T}\ri)} \  . \label{finTexp0}
\ee

For $u > 0$, at a scale of
\be \label{uepp1}
T \sim \Lam_{\rm CO} \sim \mu_* e^{-{\# \ov \sqrt{u}}}
\ee
eq.~\eqref{finTsinh} crosses over to an expression almost identical to~\eqref{theoL} with $\om$ replaced by $T$. For $u < 0$, at such small temperature scales equation~\eqref{imagFinTsusc} has poles at (for $k=0$) 
\begin{align}
T_n & = T_a \exp\le(-\frac{n\pi}{\sqrt{-u}}\ri) \nonumber \\ &= {4 \mu_* \ov \pi}  \exp\le(-\frac{n\pi}{\sqrt{-u}}+ {\tilde \al \ov \al}\ri), 
\qquad n \in \mathbb{Z}^+ \ . \label{Tc} 
\end{align} 
Comparing to~\eqref{irsca2} and \eqref{doew1}, we see that $T_n\sim \Lam_n \sim 1/\zeta_n$. 
The first of these temperature should be interpreted as the critical temperature
\be 
T_c =  {4 \mu_* \ov \pi}  \exp\le(-\frac{\pi}{\sqrt{-u}}+ {\tilde \al \ov \al}\ri)
\ee
below which the scalar operator condenses. Including frequency dependence, one can check that 
$\chi^{(T)} (\om, \vk)$ has a pole at 
\be
\om_* = -\frac{2i}{\pi}(T-T_{n})
\ee
For $T > T_{n}$ this pole is in the lower half-plane, and it moves through to the upper half-plane if $T$ is decreased through $T_{n}$. Thus we see the interpretation of each of these $T_n$; as the temperature is decreased through each of them, one more pole moves through to the upper half-plane. There exist an infinite number of such temperatures with an accumulation point at $T = 0$; and indeed at strictly zero temperature there is an infinite number of poles in the upper half-plane, as seen earlier in~\eqref{polemm}.  Of course in practice once the first pole moves through to the upper half-plane at $T_c = T_1$, the uncondensed phase is unstable and we should study the system in its condensed phase.

 One can further study the critical behavior near the finite temperature critical point $T_c$.  Here one finds mean field behavior and we will only give results.  See Appendix~\ref{app:finiTC} for details. For example the uniform static susceptibility 
 has the form
 \be 
 \chi^{(T)} \approx \bca 
                  \frac{\chi_0}{2\nu_U \al\beta} \frac{T_c}{T -T_c} & T \to T_c^+ \cr
        \frac{\chi_0}{4\nu_U \al\beta} \frac{T_c}{T_c - T} & T \to T_c^-
       \eca
       \ee
The result that $\chi (T_c^-)$ has a prefactor twice as big as $\chi (T_c^+)$ is a general result of Landau theory. 

Similarly, the correlation length near $T_c$ is given by
\be
\xi^{-2} = \frac{6\mu_*^2(-u)^{\frac{3}{2}}}{ \pi T_c }(T-T_c) \  .
\ee
Note that the prefactor of $T-T_c$ diverges exponentially as $u \to 0$, and should be contrasted with the behavior~\eqref{zeroXi} at the quantum critical point. Finally we note that 
at the critical point $T = T_c$, we find a diffusion pole in $\chi^{(T)} (\om, \vk)$ given by  
\be \label{diffH}
\om_* = -i\frac{T_c}{3\mu_*^2(-u)^{\frac{3}{2}}}k^2
\ee
which is of the standard form for this class of dynamic critical phenomena (due to the absence of conservation laws for the order parameter, this is Model A in the classification of \cite{Hohenberg:1977ym}; see also \cite{Maeda:2009wv} for further discussion in the holographic context). Note that the diffusion constant goes to zero exponentially as the quantum critical point is approached. For a charged scalar, the factor multiplying $i k^2$ on the right hand side of~\eqref{diffH} becomes complex, reflecting the breaking of charge conjugation symmetry.  

\begin{figure}[!ht]
\begin{center}
\includegraphics[scale=0.4]{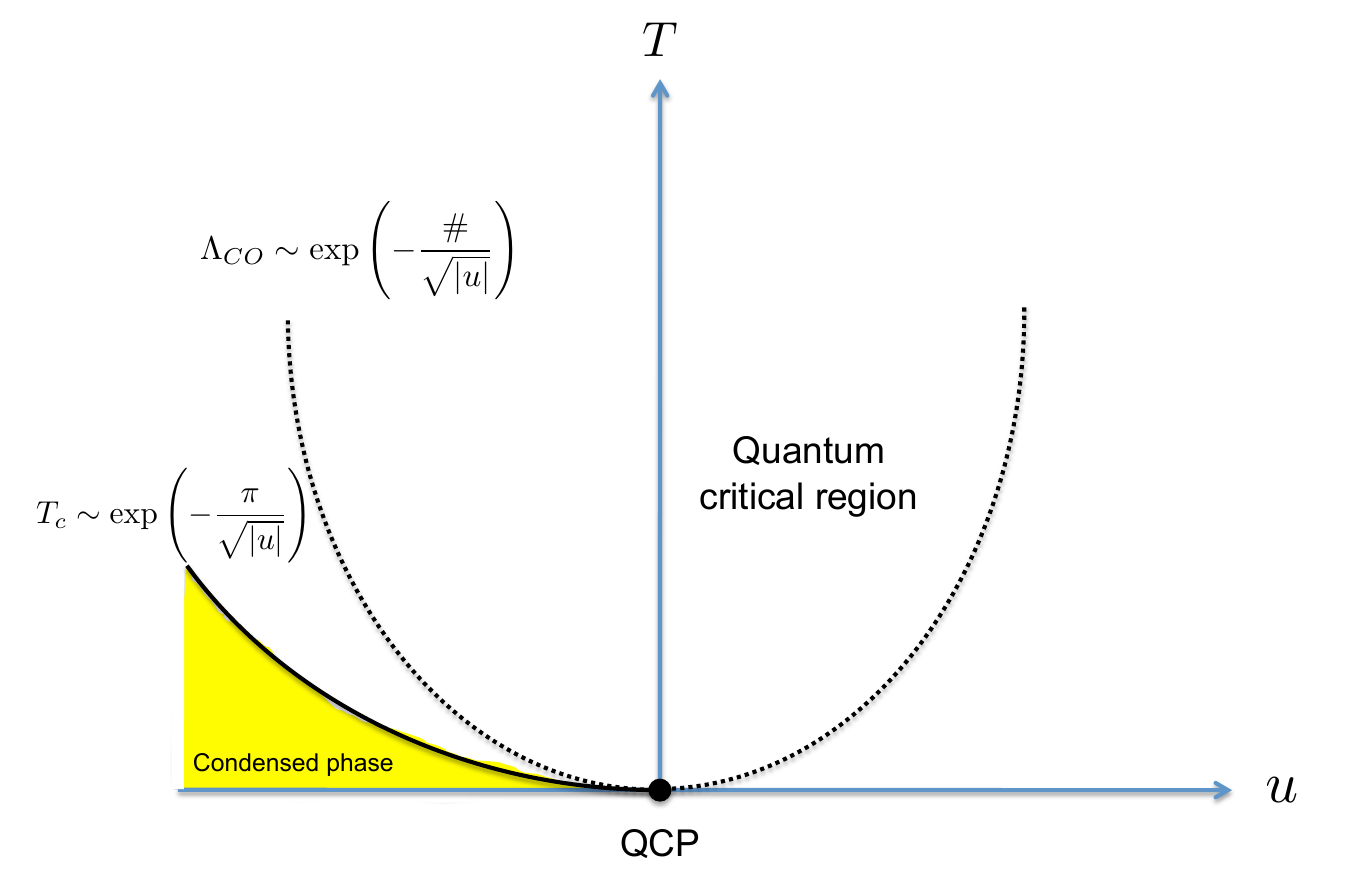}
\end{center}
\caption{Finite temperature phase diagram with the quantum critical region for bifurcating criticality as a function of $u$. The dotted line is given by $\Lam_{\rm CO}$ in~\eqref{uepp} and~\eqref{uepp1}. But note that the discussions there are not enough to determine the $O(1)$ factor in the exponent for $\Lam_{\rm CO}$. 
 The dynamical susceptibility in the bowl-shaped quantum critical region is given by~\eqref{finTexp0} with the zero temperature limit given by~\eqref{zernu}.}
\label{fig:funnel}
\end{figure}

In Fig.~\ref{fig:funnel} we summarize the finite temperature phase diagram.

\subsection{Summary and physical interpretation} \label{sec:bifsum}

In this section we studied the physics close to a ``bifurcating'' quantum critical point, i.e. the quantum critical point obtained by tuning the AdS$_2$ mass of the bulk scalar field through its Breitenlohner-Freedman bound. Here we briefly summarize the main results and discuss possible interpretations.

Much of the physics can be understood from the expression for the dynamic susceptibility at zero temperature \eqref{finomsinh}, 

\be
\chi (\om, \vk) = \chi_0 \frac{\sinh\le(\nu_k \log\le(\frac{-i\om}{\om_b}\ri)\ri)}{\sinh\le(\nu_k \log\le(\frac{-i\om}{\om_a}\ri)\ri)} + \dots  \label{rptfinom}
\ee
where $\nu_k = \sqrt{u + \frac{k^2}{6\mu_*^2}}$ and $u = 0$ is the location of the quantum critical point. 
$\om_{a,b}\sim \mu$ are some constants. This expression defines a crossover scale as in \eqref{uepp}
\be
\Lam_{CO} \sim \mu_* \exp\le(-\frac{\#}{\sqrt{|u|}}\ri), \label{rptco}
\ee 
with $\#$ some $O(1)$ number; for
$\om \gg \Lam_{CO}$,  one can expand the arguments of the hyperbolic sine to find
\be 
\chi(\om, \vk ) =\chi_0 \frac{\log\le(\frac{\om}{\om_b}\ri) - i\frac{\pi}{2}  }{\log\le(\frac{\om}{\om_a}\ri) - i\frac{\pi}{2} }  + O(u, k^2)\ . \label{quantcritexp}
\ee
with the spectral function given by 
\be \label{crispe}
\Im \chi (\om, k) = {\pi \chi_* \ov (\log \om)^2} + \dots \ . 
\ee

For $\om \ll \Lam_{CO}$, approaching the critical point from $u>0$ side, we find 
\be
\chi (\om \to 0, \vk) = \chi_0 - 2 \nu_k \chis - 4 \nu_k \chis \le({-i \om \ov 2 \mu_*}\ri)^{2 \nu_k} + \dots 
\ .
 \label{rptsusc}
\ee
Interestingly, the static susceptibility does {\it not} diverge approaching the critical point, but develops a branch point singularity at $u = 0$, as $\nu_{k=0} = \sqrt{u}$; it is trying to bifurcate into the complex plane as we cross $u = 0$.  Upon Fourier transformation to coordinate space, these singularities lead to a correlation length that diverges at the critical point,
\be \label{eoo}
\xi = \frac{1}{\sqrt{6}\mu_* \sqrt{u}} \ . 
\ee
While the exponent is the same as that of mean field, clearly the underlying physics is different. 
The coordinate-space expression is also different from that of the mean field, as shown in \eqref{bifcorr}.

 
For $\om \ll \Lam_{CO}$, approaching the critical point from $u<0$ side, 
in~\eqref{rptfinom},  the hyperbolic sine is replaced by a normal sine, and  
we find a 
geometric series of poles in the upper-half complex frequency-plane at 
\be \label{exppoles}
\om_n = i\om_a \exp \le(-\frac{n \pi}{\sqrt{-u}}\ri) \sim i \mu \le({\Lam_{IR} \ov \mu}\ri)^n,
\ee
with
\be  
\Lam_{IR} \equiv \mu \exp\le(-\frac{\pi}{\sqrt{-u}}\ri), 
\quad n=1,2, \dots
\ee
indicating that the disordered state is unstable and the scalar operator condenses in the true vacuum. 
Interestingly, one finds an infinite tower of  ``Efimov'' condensed states in one to one correspondence with the poles
in~\eqref{exppoles}
\be \label{vevex}
\vev{\sO}_n \sim \mu^\De  \exp \le(-\frac{n \pi}{2 \sqrt{-u}}\ri) 
\sim \mu^\De \le({\Lam_{IR} \ov \mu}\ri)^{n \ov 2}
\quad n=1,2, \dots \ .  
\ee
 Note that the factor $\ha$ in the exponent of~\eqref{vevex} compared with that of~\eqref{exppoles} is due to that $\sO$ has IR dimension $\ha$ at the critical point $u=0$.  $n=1$ state is the ground state with the lowest free energy which scales as  (with that of the disordered state being zero)
 \be
 F \sim - \Lam_{IR} \ .
\ee
A study of the full nonlinear response curve of the tower of ``Efimov states'' reveals a remarkable spiral structure, shown in Figure \ref{fig:spiral1}, which may be considered as a manifestation of a spontaneously broken discrete scaling symmetry in the time direction. 

At a finite temperature, in the quantum critical region $T \gg \Lam_{CO}$ (the bowl-shaped region in the right plot of Fig.~\ref{fig:funnel}), the zero temperature expression~\eqref{quantcritexp} generalizes to
\be
\chi^{(T)}(\om,\vk) = \chi_0 \frac{\log\le(\frac{2\pi T}{\om_b}\ri) + \psi\le(\ha - i\frac{\om}{2\pi T}\ri)}{\log\le(\frac{2\pi T}{\om_a}\ri)  + \psi\le(\ha - i\frac{\om}{2\pi T}\ri)} \  \label{finTexp}
\ee
which can now be applied all the way down to zero frequency.  Equation~\eqref{finTexp} reproduces~\eqref{quantcritexp} for $\om \gg T$. The $n=1$ pole in~\eqref{exppoles} provides the scale for the critical temperature 
\be
T_c \sim \mu  \exp \le(-\frac{\pi}{ \sqrt{-u}}\ri) \sim \Lam_{IR} \ . 
\ee

Now let us now try to interpret the above results. 
First we emphasize that nowhere on the uncondensed side do we see a coherent and gapless quasiparticle pole in the dynamical susceptibility, which usually appears close to a quantum phase transition and indicates the presence of soft order parameter fluctuations.  That at the critical point the susceptibility~\eqref{rptsusc} does not diverge and the spectral function~\eqref{crispe} is logarithmically suppressed at small frequencies are also  manifestations of the lack of soft order parameter fluctuations. 

We would like to argue that the quantum phase transition at a bifurcating critical point corresponds to a confinement/deconfinement transition~\cite{Iqbal:2011in}. At the critical point, two fixed points eCFT$_1^{\rm UV}$ and eCFT$_1^{\rm IR}$ of \Slql\ merge and annihilate, beyond which the scalar operator $\sO_{k=0}$  develops a complex dimension and conformality is lost. The  loss of conformality is realized through a dynamically generated ``confinement'' scale $\Lam_{IR}$ below which an infinite tower of geometrically separated bound states of operator $\sO$ form; 
 from this infinite tower the lowest energy bound state Bose condenses. There are infinitely many metastable vacua, where a higher Efimov state condenses~\eqref{vevex}. The physical picture here is similar to the BEC regime in a strongly interacting ultracold Fermi system where fermions form bound molecules and then Bose condense. 
This also explains why the susceptibility does not diverge at the critical point  and the 
spectral density is suppressed. The bifurcating QCP is characterized by the onset of forming bound states 
rather than by soft order parameter fluctuations. 

From this perspective, \Slql\ should be considered as a fractionalized state where degrees of freedom from which $\sO$ is formed become deconfined. Indeed this interpretation is consistent with
the power law behavior~\eqref{iiRc} of the spectral function of $\sO$ in the \Slql\ phase and finite entropy density of \Slql. 

Our story has some interesting differences with standard discussion of a confinement/decofinement transition (or crossover) which is driven by temperature:

\ben 

\item Here we use the term ``confinement'' in a somewhat loose sense, as in our context the ``confined'' state
still has gapless degrees of freedom left for both the condensation of a neutral and charged scalar. Thus in our story the  ``confinement'' only removes part of the deconfined spectrum. 

\item Here the transition is driven by an external parameter and thus is quantum mechanical in nature. 

\item The confinement (i.e. formation of bound states) and Bose condensation set in at the same point in parameter space. 


\een

To summarize, for a bifurcating QCP, while the phase transition can still be characterized by an 
order parameter, the order parameter remains gapped at the critical point and the phase transition is not driven by its fluctuations.  
Instead the phase transition appears to be driven by confinement coming from the merger of two different CFTs, 
in contrast with the Landau-Ginzburg-Wilson paradigm which is characterized by a single critical CFT with some relevant direction.

\section{Critical behavior of a hybridized QCP} \label{sec:hybrid}

In this section we examine the critical behavior around a hybridized QCP~\eqref{hybq}, reviewing and slightly generalizing an earlier discussion of~\cite{Faulkner:2010gj}. 

\subsection{Zero temperature: statics $(\mu_*=1 )$}

Let us first look at the the scaling of the expectation value and free energy on the condensed side.\footnote{In this subsection we will mainly use effective field theory arguments, hence we will set $\mu_*=1$ to alleviate the notation. In later subsections however we restore $\mu_*$.} This can be done by analyzing the condensed solution on the gravity side~\cite{Faulkner:2010gj}. Alternatively, one could use the low energy effective action~\eqref{thep}, which we copy here for convenience 
\be \label{thep1}
S_{eff} = \tilde S_{\rm eCFT_1} [\Phi] + \lam \int  \, \Phi \vp  + S_{LG} [\vp]+ \int \vp J \ .
\ee
We can generalize the Landau-Ginsburg action $S_{LG}$~\eqref{LGQ} in~\eqref{thep1} by including the next order nonlinear and time derivative terms 
\bea
S_{LG}  &= &  -  {1 \over 2} \int  \vp_{-\vk} (\ka_+ - \ka_c  + h_k  k^2)  \vp_{\vk}   \cr
&& -  u  \int \, \vp^4 +  h_t  \int \, (\p_t \vp)^2 + \dots
 \label{LGQ1}
\eea
with $u$ and $h_t$ some (positive) constants.\footnote{Their specific values can be worked out  from gravity. Here we are only interested in the scaling behavior and their values are not important.}  We will consider 
$k=0$ and denote $\nu_{k=0}$ simply as $\nu$.  The eCFT$_1$ operator 
 $\Phi$ has a scaling dimension $\ha + \nu$ and from the second term in~\eqref{thep1}, $\vp$ thus has dimension $\ha - \nu$.  Then from the last term in~\eqref{thep1} $J$ has  dimension $\ha + \nu$, the same as $\Phi$. Note that spatial coordinates or momenta  all have  zero IR dimension.
  Now let us imagine that $\vp$ develops some nonzero expectation value. From the relative scaling dimensions between $\Phi$ and $\vp$, we can then write $\Phi$ as 
\be 
\Phi \sim \vp^{\ha + \nu \ov \ha - \nu}
\ee
and the free energy density $F$ corresponding to~\eqref{thep1} can then be written as 
\be  \label{nifree}
F \sim C  \vp^{1 \ov \ha - \nu}  + \ha (\ka_+ - \ka_c) \vp^2 + u \vp^4 
\ee
where the first term comes from the $\Phi \vp$ term with $C$ some constant. Equation~\eqref{nifree} can also be derived from a detailed bulk analysis\footnote{See ~\cite{Faulkner:2010gj}. This expression was argued for in~\cite{Evans:2010np}.} which also gives that $C > 0$ for $\nu < \ha$.  Now notice that 
for $\vp$ small, the first term dominates over $\vp^4$ term if $\nu < {1 \ov 4}$, while the Landau-Ginsburg $\vp^4$ term dominates for $\nu > {1 \ov 4}$. In other words, since the first term is marginal by assignment, $\vp^4$ term becomes relevant when $\nu > {1 \ov 4}$.\footnote{Some readers might worry that higher powers like $\vp^6$ may also become relevant at some point (for example for $\nu > {1 \ov 3}$). But note that once the last two terms in~\eqref{nifree}
dominate we should reassign dimension of $\vp$ and the standard Landau-Ginsburg story applies.} 

For $\nu < {1 \ov 4}$ we can ignore the last term in~\eqref{nifree} and for $\ka_+< \ka_c$ find that 
\be \label{ovev}
\vev{\sO} \sim \vp \sim  (\ka_c - \ka_+)^{\ha - \nu \ov 2 \nu}
\ee
and as a result 
\be 
F \sim (\ka_c - \ka_+)^{1 \ov 2 \nu} \ . 
\ee
Including the source $J$, which has dimension $\ha + \nu$, the free energy should then be given by a scaling function 
\be \label{scalFr}
F = (\ka_c - \ka_+)^{1 \ov 2 \nu} f_1 \le(J  (\ka_c - \ka_+)^{-{\ha +\nu \ov 2 \nu}}\ri)
= \xi^{-{1 \ov \nu}} f_2 \le(J \xi^{\ha + \nu \ov \nu}\ri)
\ee
where in the second equality we have expressed the free energy in terms of the correlation length using~\eqref{colen}. From~\eqref{scalFr} we can also deduce that at the critical point we should have 
\be \label{onMexp}
\vev{\sO} \sim \vp \sim J^{\frac{\ha - \nu}{\ha + \nu}}  \ 
\ee 
which can again be confirmed by a bulk analysis.
From~\eqref{ovev},~\eqref{onMexp} and~\eqref{scalFr} we can collect the values of various scaling exponents (see Appendix~\ref{app:critexp} for a review of their definitions)
\be \label{varexp}
\al=  2-{1\ov 2\nu}, \quad  \beta= {\ha - \nu \ov 2\nu} , \quad 
\delta =  {\ha + \nu \ov \ha - \nu} \ .
\ee 

For $\nu > {1 \ov 4}$, we can ignore the first term in~\eqref{nifree} and the analysis becomes the standard Landau-Ginsburg one. As a result, the behavior near the critical point becomes that of the mean field, as pointed out earlier in~\cite{Faulkner:2010gj} from a detailed bulk gravity analysis. 
We thus find that for $\nu > {1 \ov 4}$, 
\be \langle\sO\rangle \sim \vp \sim  (\kappa_c - \kappa_+)^\ha, \; F \sim -(\kappa_c - \kappa_+)^{2}, \; \langle\sO\rangle_{\kappa _+= \kappa_c} \sim J^{\frac{1}{3}} \ 
\ee
and various exponents become
\be 
\al=  0, \quad  \beta= \ha , \quad 
\delta =  3 \ 
\ee 
which agree with the values of~\eqref{varexp} for $\nu = {1 \ov 4}$.

\subsection{Dynamical critical behavior}

Let us now examine the dynamical behavior near the critical point. 
 Expanding 
$\tilde a_+ (\om, k)$ around $\om =0$, $k=0$ and $\ka_+ = \ka_c$, we find that the full dynamical 
susceptibility~\eqref{roep21} can now be written as (for a neutral scalar)
\be 
\chi(\om, \vk) \approx \frac{\mu_*^{2\nu_U}}{\kappa_+ - \kappa_c + h_k \vk^2  - h_\om \om^2 +  h \sG_k(\om) } \label{hybridizedG1}
\ee
where $h_k$ was introduced earlier in~\eqref{hybridizedG0} and 
\be 
h_\om \equiv - {\tilde a_+^{(2)}  (k) \ov b_+^{(0)}  (k) } \biggr|_{k=0, \ka_+ = \ka_c}, \quad h \equiv { \mu_*^{-2 \nu_k} \tilde a_-^{(0)}  (k) \ov b_+^{(0)}  (k) } \biggr|_{k=0, \ka_+ = \ka_c} \ .
\ee
Recall that the \Slql\ retarded function $\sG_k (\om) \propto \om^{2 \nu_k}$. 
From explicit gravity calculation one finds various constants in~\eqref{hybridizedG1}  have the following behavior: $h_k> 0,\; h < 0$ and $h_\om > 0$~(for  $\nu_{k=0} > 1$).

The behavior of full dynamical susceptibility~\eqref{hybridizedG1} depends on the competition between 
the analytic contribution $h_\om \om^2$ and the 
non-analytic contribution $\sG_k (\om)$ from \ircft.  When $\nu \in (0,1)$, the non-analytic part 
dominates at low energies and the analytic contribution can be ignored, leading to 
\be
 \chi(\om, \vk) \approx \frac{\mu_*^{2\nu_U}}{(\kappa_+ - \kappa_c) + h_k k^2 + h C(\nu) (-i \om)^{2 \nu} } \label{univform}
\ee
with $C(\nu) < 0$. We will consider $k=0$ below.  At the critical point $\ka_+ = \ka_c$ we find that
\be 
\chi (\om, k=0) \sim (-i \om)^{-2 \nu} \ . \label{critom}
\ee
Away from the critical point, the relative magnitude of the two terms (with $k=0$) in the denominator of~\eqref{univform} defines a crossover energy scale 
\be 
\Lam_{\rm CO}^{(\om)} \sim |\kappa_c - \kappa_+|^{\frac{1}{2\nu}} \ . \label{Tcexp}
\ee 
For $\om \ll \Lam_{\rm CO}^{(\om)}$ we find that 
\be 
\chi (\om) \sim {\mu_*^{2\nu_U} \ov \ka - \ka_c} + O(\om^{2 \nu}) 
\ee
which is the typical behavior in the uncondensed phase (see e.g.~\eqref{gre}), 
while for $\om \gg \Lam_{\rm CO}^{(\om)}$ we recover the critical behavior~\eqref{critom}.  The crossover scale~\eqref{Tcexp}  defines the correlation time $\xi_\tau$ of the system
\be 
\xi_\tau \sim {1 \ov \Lam_{\rm CO}^{(\om)}} \sim  |\kappa_c - \kappa_+|^{-\frac{1}{2\nu}} \ .
\ee
Comparing the above expression with~\eqref{colen} we then find that $\xi_\tau  \sim \xi^z$ with
the dynamical exponent $z$ given by
\be \label{timeco}
z = {1 \ov \nu} \ . 
\ee
Of course this exponent can equivalently be seen by balancing the $k^2$ term and the $\om^{2 \nu}$ term in~\eqref{univform}.  Also note that when $\ka_+ < \ka_c$ equation~\eqref{univform} has a pole  in the 
upper half plane (since $h C(\nu) > 0$) at
\be 
\om_{\rm pole} \sim i \Lam_{\rm CO}^{(\om)} \ .
\ee

When $\nu > 1$, in~\eqref{hybridizedG1}, the non-analytic part $\sG_k(\om) \sim \om^{2\nu}$ from the \Slql\ becomes less important than the analytic corrections $\sim \om^2$ and one finds mean field like behavior.
Now the full dynamical susceptibility is given by 
\be 
 \chi(\om, \vk) \approx \frac{\mu_*^{2\nu_U}}{(\kappa_+ - \kappa_c) + h_k k^2 - h_\om \om^2 + h C(\nu) (-i \om)^{2 \nu} } \label{Nform}
\ee
which describes a long-lived (nearly gapless) relativistic particle with a small width $\Gamma \sim \om^{2\nu}$. The dynamical exponent is now $z=1$.

This crossover to mean field dynamical behavior at $\nu = 1$ can also  be readily seen from the effective action~\eqref{thep1}--\eqref{LGQ1}. For $\nu > 1$, the dimension for $\vp$ become smaller than $-\ha$, for which the kinetic term $(\p_t \vp)^2$   becomes relevant and  more important than the 
hybridization term $\Phi \vp$ (which is marginal by definition). Alternatively, we can now assign $-\ha$ as dimension of $\vp$  using  $(\p_t \vp)^2$, under which the 
hybridization term $\Phi \vp$ will have dimension $\nu$ which becomes irrelevant for $\nu > 1$. 

It is interesting to note that while the free energy already exhibits mean field behavior for $\nu > {1 \ov 4}$, the dynamical quantity exhibits mean field behavior only for $\nu > 1$.

\subsection{Finite temperature}

At a finite temperature $T \ll \mu$,  equation~\eqref{hybridizedG1} generalizes at leading order in $T/\mu$ to~(from expanding~\eqref{roep22}), 
\be 
\chi(\om, \vk; T) \approx \frac{\mu_*^{2\nu_U}}{\kappa_+ - \kappa_c + h_k \vk^2  - h_\om \om^2 + h_T T +  h \sG_k^{(T)} (\om)  } \label{hybridizedGT}
\ee
where $h_T T$ ($h_T$ a constant) comes from (analytic) finite temperature corrections to $a_+$ and $b_+$.  Finite temperature \Slql\ retarded function $\sG_k^{(T)} (\om)$ has the form  $\sG_k^{(T)} (\om)= T^{2 \nu_k} g({\om \ov T}, \nu_k)$ with $g$ a universal scaling function~(see~\eqref{finiteTgf21}).  Explicit gravity calculations give  $h_T > 0$~(for $\nu_{k=0} > \ha$). 
Let us first look at the static uniform susceptibility at finite $T$, which is 
\be \label{finrT}
\chi^{(T)} \sim {\mu_*^{2\nu_U} \ov \ka_+ - \ka_c + h_T T + h C(\nu) T^{2 \nu}} \ .
\ee 
It is interesting that the analytic contribution now dominates for $\nu > \ha$. 

For $\nu < \ha$ we find that there is a pole at 
\be 
T_c \sim (\ka_c - \ka_+)^{1 \ov 2 \nu} 
\ee
for $\ka_+ < \ka_c$. It should be interpreted at the critical temperature for a thermal phase transition, above which the instability disappears. From the uncondensed side, such a temperature scale gives the crossover scale 
\be
\Lam_{CO}  \sim  |\ka_c - \ka_+|^{1 \ov 2 \nu} 
\ee 
to the quantum critical behavior; for $T \gg \Lam_{CO}$
\be 
\chi(\om, \vk; T) \approx \frac{\mu_*^{2\nu_U}}{\kappa_+ - \kappa_c + h_k \vk^2 +  h  T^{2 \nu}   g({\om \ov T}, \nu) } 
\ee
which exhibits  $\om/T$ scaling. Note that 
 the finite temperature crossover scale tracks that of zero temperature equation~\eqref{Tcexp}.

For $\nu > \ha$, we find instead mean field behavior 
\be 
T_c \sim (\kappa_c - \kappa_+) \  \label{mfT_c}
\ee
and the finite temperature crossover scale becomes 
\be 
\Lam_{\rm CO}^{(T)} \sim  |\kappa_c - \kappa_+|
\ee
which no longer tracks that of zero temperature. In this regime, there is no $\om/T$ scaling and the non-analytic frequency dependence from the \Slql\ becomes irrelevant compared to leading temperature effects.

This crossover  at $\nu = \ha$ can again be readily seen from the effective action~\eqref{thep}. 
Finite temperature generates a term $\int \, T \vp^2$, which becomes relevant when the dimension of $\vp$ becomes smaller than zero, i.e. for $\nu > \ha$. Alternatively we can now use  $T  \vp^2$ term to assign dimension $0$ to $\vp$, under which the 
hybridization term $\Phi \vp$ then becomes irrelevant for $\nu > \ha$ as now the dimension for $\Phi$ becomes larger than $1$. 

\begin{figure*}[!ht]
        \begin{minipage}[r]{2.0\columnwidth}
        \begin{center}
\includegraphics[scale=0.35]{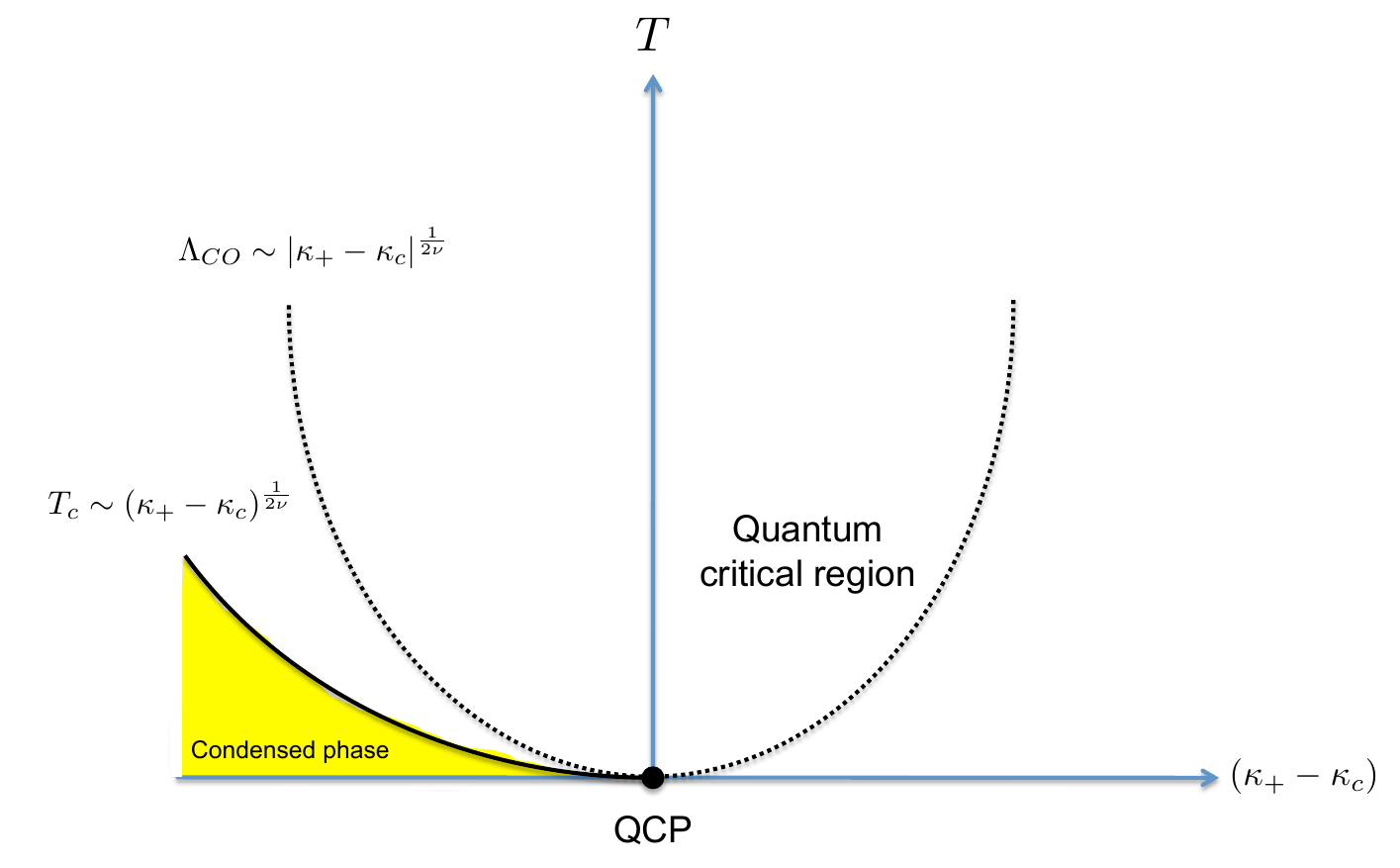}
\end{center}
\caption{Finite temperature phase diagram with the quantum critical region for a hybridized QCP for a fixed $0 < \nu < \ha$. In the quantum critical region the dynamical susceptibility is given by~\eqref{hybridizedGT} with~\eqref{univform} as the zero temperature limit.  } 
\label{fig:funnel2}
    \end{minipage}
\end{figure*}

\begin{figure*}[!ht]
        \begin{minipage}[r]{2.0\columnwidth}
        \begin{center}
\includegraphics[scale=0.33]{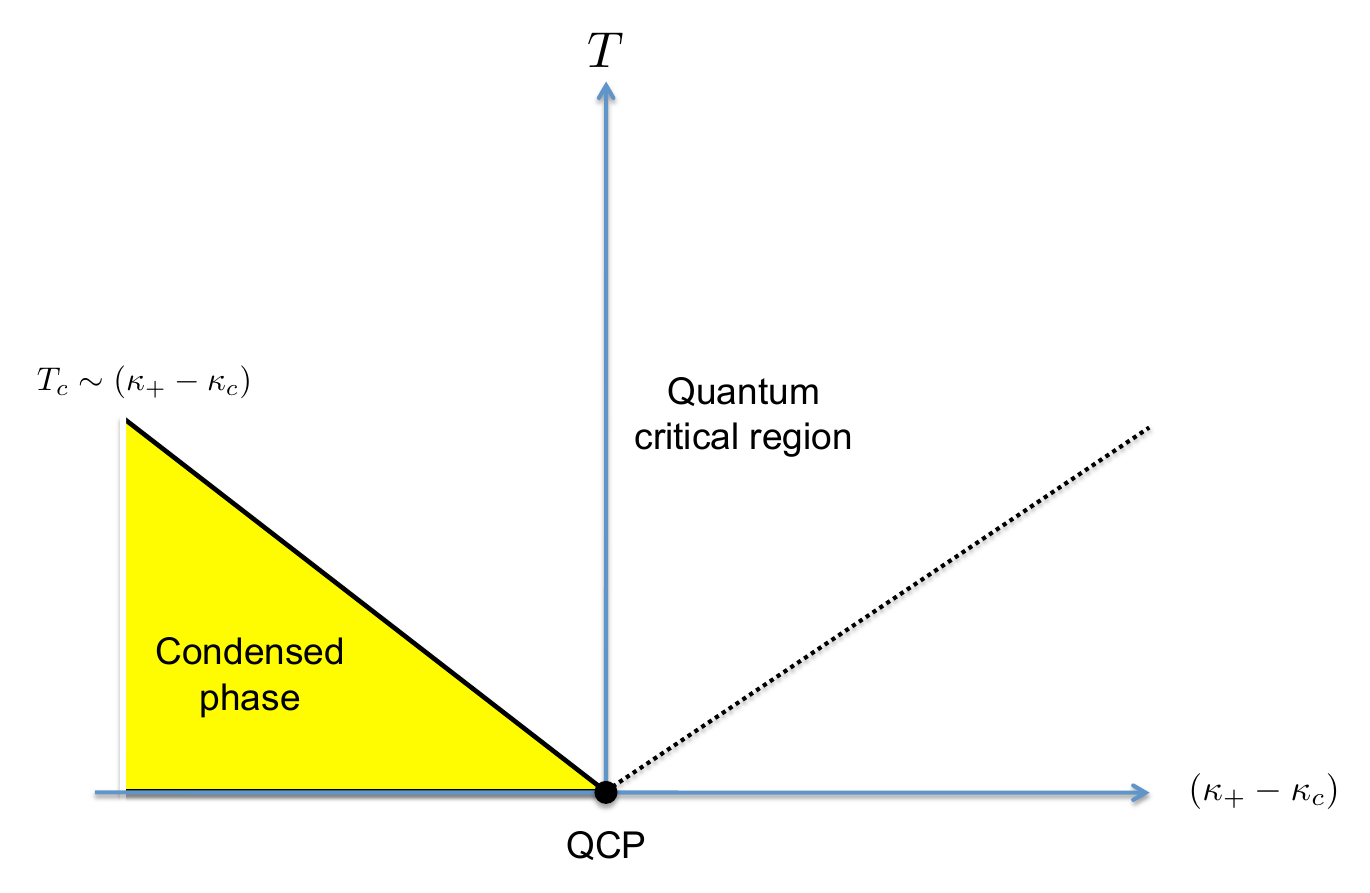} 
 \includegraphics[scale=0.33]{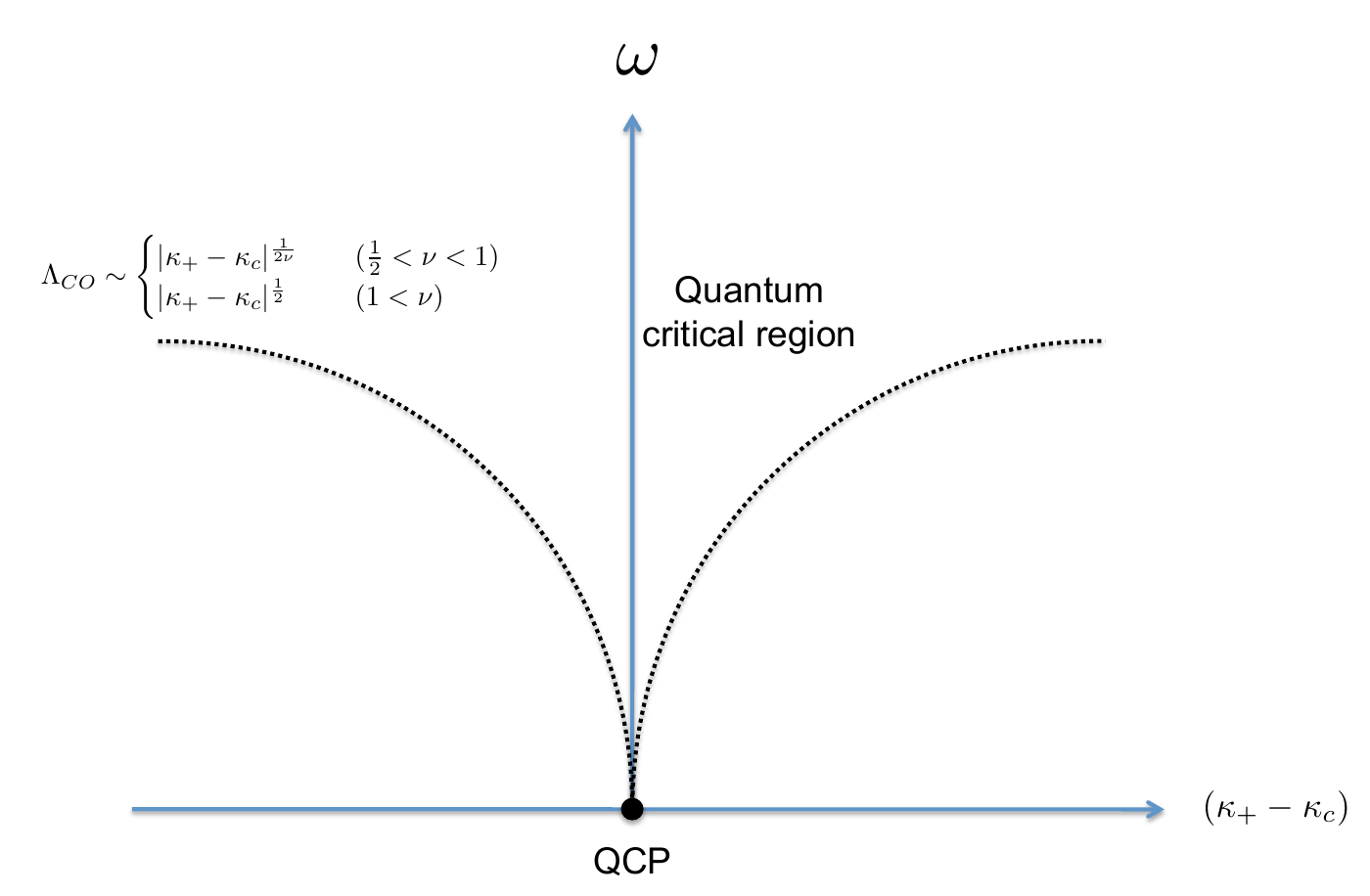}
\end{center}
\caption{The crossover diagrams for a hybridized QCP for $\nu > \ha$. Given the difference in the crossover scales between frequency (at zero temperature) and temperature, we plot them separately. For $\nu \in (\ha ,1)$, the zero temperature dynamical susceptibility exhibit nontrivial scaling (or for $\om^{2 \nu} \gg T$), but finite temperature behavior is given by that of mean field.  For $\nu \geq 1$, essentially everything is mean field at leading order. 
 } 
\label{fig:funnel3}
    \end{minipage}
\end{figure*}

We summarize various the finite $T$ phase diagram for various values of $\nu$ in Fig.~\ref{fig:funnel2} and~\ref{fig:funnel3}.


\subsection{Summary and discussion} \label{sec:hybsum}

We summarize the critical behavior near a hybridized QCP for various values of $\nu$ in the following table (see Appendix~\ref{app:critexp} for a review of definitions of various scaling exponents): 
\bwt
\begin{center} 
\begin{tabular}{c||c|c|c|c}
Quantity& $\nu\in\le(0,\frac14\ri)$ & $\nu\in\le(\frac14,\frac12\ri)$ &  $\nu\in\le(\frac12,1\ri)$  &  $\nu>1$ \\
\hline
$\om/T$ scaling & yes & yes & no & no \\ 
$\Lam^{(\om)}_{\rm CO}  $ & $ \le(\kappa_c-\kappa_+\ri)^{1 \ov 2\nu}$ & $ \le(\kappa_c-\kappa_+\ri)^{1 \ov 2\nu}$ & $ \le(\kappa_c-\kappa_+\ri)^{1 \ov 2\nu}$ & $ \le(\kappa_c-\kappa_+\ri)^{1 \ov 2}$ \\
$z$ & ${1 \ov \nu}$ & ${1 \ov \nu}$ & ${1 \ov \nu}$ & $1$ \\
$T_c $ & $ \le(\kappa_c-\kappa_+\ri)^{1 \ov 2\nu}$ & $ \le(\kappa_c-\kappa_+\ri)^{1 \ov 2\nu}$ & $\kappa_c-\kappa_+$ &  $\kappa_c-\kappa_+$ \\
$\vev{\sO}$ & $(\ka_c - \ka_+)^{\ha - \nu \ov 2 \nu}$ & $(\ka_c - \ka_+)^{\ha}$ & $(\ka_c - \ka_+)^{\ha}$ &$(\ka_c - \ka_+)^{\ha}$ \\
\multicolumn{5}{c}{$$} \\
\multicolumn{5}{c}{Critcal behavior for the neutral scalar.} 
\end{tabular} 
\end{center} 
\ewt
In the above the expressions which are independent of $\nu$ are all mean field behavior. 
For all values of $\nu$ the static susceptibility from 
the uncondensed side is always given by the mean field behavior
\be 
\chi_0 (\vk) \approx \frac{\mu_*^{2\nu_U}}{(\kappa_+ - \kappa_c) + h_k \vk^2  } 
\ee
with the spatial correlation length
\be 
\xi \sim |\ka_+ - \ka_c|^{-\ha} \ .
\ee
In the range  $\nu \in (0, {1 \ov 4})$ the free energy from the condensed side is given by the scaling form 
\be \label{freeN}
F = \xi^{-{1 \ov \nu}} f\le(J \xi^{\ha + \nu \ov \nu}, T \xi^{1 \ov \nu} \ri), 
\ee
with $f$ a universal scaling function, while for $\nu \in (0, \ha)$, the full dynamical susceptibility exhibits $\om/T$ scaling at a finite temperature 
\begin{align}
\chi(\om, \vk; T) & \approx  \frac{\mu_*^{2\nu_U}}{(\kappa_+ - \kappa_c) + h_k \vk^2   + \Sig (\om, k,T)  }, \nonumber \\ 
\Sig & =  h T^{2 \nu_k} g \le({\om \ov T}, \nu_k \ri) 
 \label{hybridizedGT3}
\end{align}
where $\Sig$ can be interpreted as self-energy and $g$ a universal scaling function given by~\eqref{finiteTgf21}. Note that since $\nu_k$ depends on $k$ through $k/\mu$, for $k \ll \mu$ we can approximate $\nu_k$ in the self energy $\Sig$~\eqref{hybridizedGT3} as $\nu_{k=0}$ and $\Sig$ becomes $k$ independent. 
As pointed out in~\cite{Faulkner:2010gj}, then equation~\eqref{hybridizedGT3} resembles 
 the dynamical susceptibility of $CeCu_{6-x} Au_x$ near the quantum critical point $x=0.1$~\cite{schroeder:00}.

From the definitions in Appendix~\ref{app:critexp} it is easy to check that various exponents here satisfy the relation 
\be
\gamma=(2-\eta)\nu_{crit} = \beta (\delta-1) \label{critrel1}
\ee
but the so-called hyperscaling relation 
\be
2\beta=(d-2+\eta)\nu_{crit} \  \label{hyper1}
\ee
is violated for the trivial reason that our results are independent of spacetime dimension $d$. 

Note that for a charged scalar, the discussion is very similar except that the perturbative corrections in 
$\om$ starts at linear order. As a result everything becomes mean field for $\nu > \ha$, i.e.  there are only three columns in the table  
below. Note that the mean field values of the dynamical quantities for the charged case are {\it different} from those of the neutral scalar, while the static exponents remain the same.
\begin{center} 
\begin{table}[!h]
\begin{tabular}{c||c|c|c}
Quantity& $\nu\in\le(0,\frac14\ri)$ & $\nu\in\le(\frac14,\frac12\ri)$ &    $\nu>\ha$ \\
\hline
$\om/T$ scaling & yes & yes & no  \\ 
$\Lam^{(\om)}_{\rm CO}  $ & $ \le(\kappa_c-\kappa_+\ri)^{1 \ov 2\nu}$ & $ \le(\kappa_c-\kappa_+\ri)^{1 \ov 2\nu}$ &  $ \kappa_c-\kappa_+$ \\
$z$ & ${1 \ov \nu}$ & ${1 \ov \nu}$  & $2$ \\
$T_c $ & $ \le(\kappa_c-\kappa_+\ri)^{1 \ov 2\nu}$ & $ \le(\kappa_c-\kappa_+\ri)^{1 \ov 2\nu}$ & $\kappa_c-\kappa_+$  \\
$\vev{\sO}$ & $(\ka_c - \ka_+)^{\ha - \nu \ov 2 \nu}$ & $(\ka_c - \ka_+)^{\ha}$ & $(\ka_c - \ka_+)^{\ha}$ \\
\multicolumn{4}{c}{$$} \\
\multicolumn{4}{c}{Critcal behavior for the charged scalar.} 
\end{tabular}
\end{table}
\end{center} 

For a hybridized QCP, the critical behavior is not described by the order parameter fluctuations alone as 
the order parameter is hybridized with degrees of freedom in \Slql.
The interplay between two sectors gives rise to a rich spectrum of critical behavior. In particular,  the semi-local nature of the \Slql\ leads to that the susceptibility of the order parameter has mean field behavior in the spatial sector, but exhibits nontrivial $\om/T$ scaling in some parameter range, reminiscent of local quantum critical behavior  observed in certain heavy fermion materials. Note that the mean field nature of the spatial sector and 
the independence of spacetime dimension of our results should be related to that gravity approximation corresponds to the large $N$ limit of the boundary theory. It would be interesting to understand better how the hybridized theory~\eqref{thep1} works when the $S_{LG}$ is below the upper critical dimension.

\section{Doubly fine-tuning to a marginal critical point} \label{sec:doucri}

Let us now consider the critical behavior around a marginal critical point, which can be obtained by tuning 
$u \to 0$ and $\ka_+ \to \ka_c^*$ at the same time, with $\kappa_c^*$ given by~\eqref{mcpv},
\be
\kappa_c^* = -\frac{\al}{\beta}  \ . \label{kappstar} 
\ee 
We have already studied the behavior as we vary $u$ through $0$ away from this point; for definiteness below we 
will fix $u=0$ and vary $\ka_+$. 

From earlier discussion in Sec.~\ref{sec:mcp}, at $u=0$ and general $\ka_+$, the system can be described by the low energy action~\eqref{effmarg}, which we copy here for convenience (keeping most essential terms), 
\be \label{effmarg1}
S_{eff} = S_{\rm eCFT_1}^{(\nu=0)} - {\tilde \xi_0^{(\ka)} \over 2} \int dt  \, \Phi^2 
  + \dots
\ee
where 
\be \label{xii}
\tilde \xi_0^{(\ka)} = {\al + \ka_+ \beta \ov \tilde \al + \ka_+ \tilde \beta}  = 2 \nu_U \beta^2 (\ka_+ - \ka_+^*) + \dots
\ .
\ee
In the second equality we have expanded $\tilde \xi_0^{(\ka)}$ around $\ka_+^*$ and used~\eqref{W1}.
The dynamical susceptibility for $\sO$ is given by the $u \to 0$ limit of~\eqref{roep23} 
\bwt
\bea  
 \chi ^{(u=0,\ka_+)}(\om,k) = \mu_*^{2\nu_U} {(\beta + O (k^2)) \sG_0 (\om) + \tilde \beta + O(k^2) \ov
 (\al + \ka_+ \beta + O(k^2)) \sG_0 (\om) + (\tilde \al + \ka_+ \tilde \beta) + O(k^2)}  
 =  \frac{\mu_*^{2\nu_U}  \log\le(\frac{- i \om}{\om_b}\ri) }{(\kappa_+ - \kappa_c^*)\log\le(\frac{-i \om}{\om_b}\ri) - \frac{1}{2\nu_U \beta^2}} + \dots
\label{sus14}
\eea 
\ewt
with $\sG_0$ given by~\eqref{eurp} and in the second line we have suppressed $k$ dependence which always comes with $k^2/\mu^2$ and is small when $k \ll \mu$. $\om_b$ was introduced earlier in~\eqref{defwab}.

The system develops an instability to the condensation of $\Phi$ (and thus $\sO$)\footnote{Recall that $\Phi$ is the operator in \Slql\ to which $\sO$ match in the IR.} when $\tilde \xi_0^{(\ka)}$ becomes negative (i.e. $\ka_+ < \ka_+^*$), where it becomes marginally relevant and generates 
an IR scale
\be 
\Lam_{IR} \sim \mu \exp \le({1 \ov \tilde \xi_0^{(\ka)}}\ri) = \mu \exp \le(\frac{1}{2\nu_U \beta^2} \frac{1}{\kappa_+ - \kappa_c^*} \ri) \ 
\ee
which can be seen from a pole of~\eqref{sus14} in the upper half $\om$-plane at\footnote{There is also a UV pole for $\ka_+ > \ka_+^*$ which is at an exponentially high energy scale. Our low-frequency formula breaks down far before the pole, and is thus of no concern to us.}
\be
\om_* = i\om_b \exp\le(\frac{1}{2\nu_U \beta^2} \frac{1}{\kappa_+ - \kappa_c^*}\ri) \sim i \Lam_{IR} \ .
\ee

Equation~\eqref{sus14} defines a crossover scale 
\be 
\Lam_{CO} \sim  \mu \exp \le(-\frac{1}{2\nu_U \beta^2} \frac{\#}{|\kappa_+ - \kappa_c^*|} \ri) \ 
\ee
where $\#$ is some $O(1)$ number. For $\om \gg \Lam_{CO}$ we can ignore the first term in the denominator 
and are thus in the quantum critical regime with 
\be
\chi^{(u=0, \ka_+)} (\om, k) = -2\nu_U\beta^2 \mu_*^{2\nu_U}  \le (\log\le(\frac{\om}{\om_b}\ri) - i \frac{\pi}{2}\ri) \label{zeroTatcrit}
\ee
The appearance of a pure logarithm in the {\it numerator} of this propagator at criticality is interesting. It leads to the spectral density 
\be \label{margc}
\Im \chi^{(u=0, \ka_+)} = \pi\nu_U\beta^2 \mu_*^{2\nu_U}  \mbox{sgn}(\om)
\ee
which is a pure step function.\footnote{The logarithm jumps by $i\pi$ as we cross through $\om = 0$, resulting in the step function; note that this was necessary in order to maintain the relation $\om \Im \chi(\om) > 0$, true for any bosonic spectral density.} This should be contrasted with the situation for a bifurcating critical point~\eqref{crispe} in which there is a logarithmic suppression at low frequencies.

When $\om \ll \Lam_{CO}$, the first term in the denominator of~\eqref{sus14} dominates and expanding in powers of the inverse logarithm, we find
\be
\chi^{(u=0, \ka_+)}(\om \to 0, k) = \frac{\mu_*^{2\nu_U} }{\kappa_+ - \kappa_c^*} + O\le(\frac{1}{\log{\om}}\ri). 
\ee
with a spectral density appears at order $\log^{-2}(\om)$,  
\begin{align}
\Im & \chi^{(u=0, \ka_+)} (\om \to 0, k) \nonumber \\ & = \frac{\pi \mu_*^{2\nu_U} }{4 (\kappa_+ - \kappa_c^*)^2 \nu_U \beta^2}\log^{-2}\le(\frac{\om}{\om_b}\ri) + O\le(\frac{1}{\log^{3}{\om}}\ri) \label{ImGnearcrit}
\end{align}
which is of course the generic behavior at a bifurcating critical point. Note that for $\ka_+ < \ka_+^*$, below $\Lam_{CO}$ the above equations no longer apply as the condensate sets in. Given that $\Phi$ has dimension $\ha$ we can easily deduce the expectation value for $\sO$ in the condensed side should scale as 
\be \label{mvev}
\langle \sO \rangle \sim \Lam^\ha_{IR}  \sim \mu_*^{\Delta}  \exp\le(\frac{1}{4\nu_U \beta^2(\kappa_+ - \kappa_c^*)}\ri) \ .
\ee
Similarly from~\eqref{effmarg1} the free energy should scale as 
\be \label{mfree}
F \sim \vev{\sO}^2 \sim \Lam_{IR} \sim \exp\le(\frac{1}{2\nu_U \beta^2(\kappa_+ - \kappa_c^*)}\ri) \ .
\ee
Both~\eqref{mvev} and~\eqref{mfree} can be confirmed by an explicit bulk analysis of the nonlinear solution for the condensed phase. 

From the $u\to 0$ limit of~\eqref{roep22}, Eq~\eqref{sus14} can be immediately generalized to a finite temperature 
\bwt
\be
\chi^{(u=0,\kappa_+)}(\om,k; T) =\mu_*^{2\nu_U}  \frac{\log\le(\frac{2\pi T}{\om_b}\ri) + \psi\le(\ha - i\frac{\om}{2\pi T}\ri)}{(\kappa_+ - \kappa_c^*)\le(\log\le(\frac{2\pi T}{\om_b}\ri) + \psi\le(\ha - i\frac{\om}{2\pi T}\ri)\ri) - \frac{1}{2\nu_U \beta^2}} \ . \label{margTf}
\ee  
From here we can see that for $\kappa_+ < \kappa_c^*$ the static susceptibility diverges at the critical temperature ($T_b$ was defined in~\eqref{defTab})
\be
T_c = T_b \exp\le(\frac{1}{2\nu_U \beta^2} \frac{1}{\kappa_+ - \kappa_c^*}\ri) \sim \Lam_{IR} \label{margTc}
\ee
above which the system is stable. The temperature is set by the same dynamically generated scale. 

Now for $T \gg \Lam_{CO}$  we find in the quantum critical region 
\be 
\chi^{(u=0,\kappa_+= \ka_+^*)}(\om,k; T) =- 2\nu_U \beta^2 \mu_*^{2\nu_U}  \le(\log\le(\frac{2\pi T}{\om_b}\ri) + \psi\le(\ha - i\frac{\om}
{2\pi T}\ri) \ri)
 \label{margTf1}
\ee 
Taking the imaginary part and using the identity\footnote{This can be proved using the reflection formula $\psi(1-x) - \psi(x) =\pi \cot(\pi x)$.} $\Im \psi\le(\ha + ix\ri) = \frac{\pi}{2}\tanh(\pi x)$ we find the expression
\be
\Im \chi^{(u=0,\kappa_+ =  \ka_+^*)}(\om, k; T) = \pi \nu_U \beta^2 \mu_*^{2\nu_U}  \tanh\le(\frac{\om}{2T}\ri) \label{finiteTatcrit}
\ee
\ewt
which is simply a smoothed-out version of the step function~\eqref{margc} that we find at zero temperature. 

Equation~\eqref{finiteTatcrit} implies that 
\be
\Im \chi^{(u=0,\kappa_+ = \kappa_+^*)}(\om, k;T) \sim \begin{cases} \frac{\om}{T} \quad &\om \ll T \\ \mbox{sgn}(\om) \quad &\om \gg T \end{cases}
\ee
which is precisely of the form for  spin and charge fluctuations in the phenomenological 
``Marginal Fermi liquid''~\cite{Varma89} description of High-$T_c$ cuprates in the strange metal 
region~(see also~\cite{SY,bgg}). Thus the marginal critical point can be viewed as a concrete realization of the bosonic fluctuation spectrum needed to support a Marginal Fermi liquid. In particular, this gives an alternative approach to construct holographic Marginal Fermi liquid.

\begin{figure}[!ht]
\begin{center}
\includegraphics[scale=0.36]{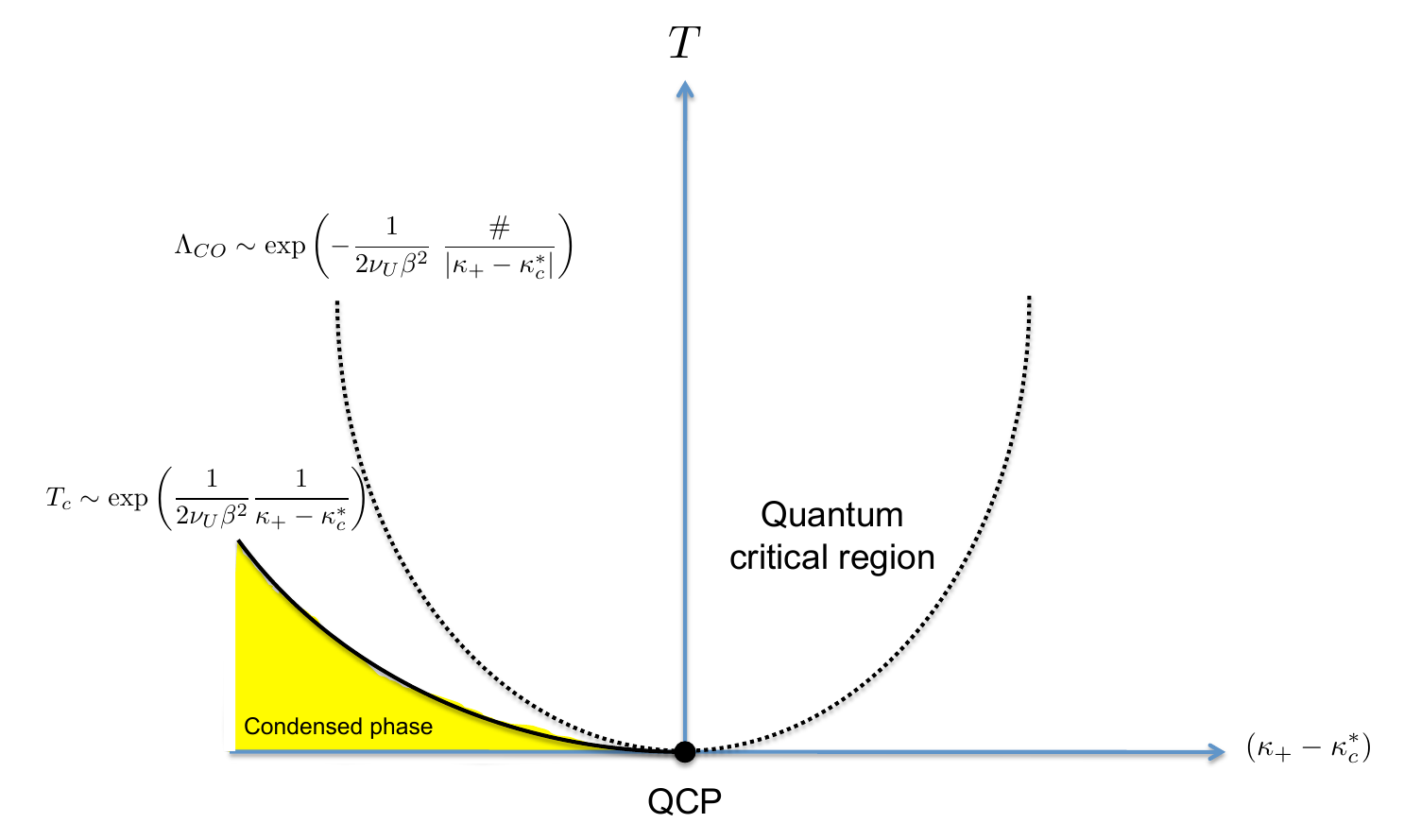}
\end{center}
\caption{Finite temperature phase diagram with the quantum critical region for marginal criticality at $u=0$ and changing $\le(\kappa_+-\kappa_+^*\ri)$. The susceptibility in the bowl-shaped quantum critical region is given by~\eqref{margTf1} with the $\om \gg T$ limit given by~\eqref{zeroTatcrit}}.  
\label{fig:funnel4}
\end{figure}

In Fig.~\ref{fig:funnel4} we summarize the phase diagram for a marginal critical point. 

\section{Discussion} \label{sec:disc}

In this paper we have discussed several types of quantum critical points from gauge-gravity duality which to different degrees lie  outside the Landau-Ginsburg-Wilson paradigm. Let us first  briefly summarize some key features:

\ben 

\item A hybridized QCP is described by an order parameter $\vp$ with a Landau-Ginsburg effective action $S_{LG}$  hybridized with degrees of freedom in \Slql, i.e. 
\be \label{hybid}
S_{eff} =   S_{\Slql} [\Phi] + \int \, \lam \Phi \vp  + S_{\rm LG} [\vp] \ .
\ee
The \Slql\ sector is strongly coupled (with no quasiparticle description). It has a scaling symmetry in the time direction only, and gapless excitations at generic finite momenta. 
Due to these features, the phase transition could exhibit a rich spectrum of critical behavior, including locally quantum critical behavior with nontrivial  $\om/T$ scaling, depending  on the scaling dimension of $\Phi$ in the \Slql. At the level of effective theory, this critical point lies mildly outside the standard Landau paradigm, as the phase transition is still driven by soft fluctuations of the order parameter and all the critical behavior is fully captured by~\eqref{hybid}, given (still mysterious) properties of the \Slql. 
 
On the gravity side the Landau-Ginsburg sector is associated with 
the appearance of certain scalar hair in the black hole geometry, which lies outside the AdS$_2$ region which describes the \Slql. 

\item A bifurcating QCP arises from instabilities of the \Slql\ itself to a confined state and appears not driven by soft order parameter fluctuations. On the condensed side, a scalar operator develops a complex scaling 
dimension in the \Slql, generating a tower of bound states, which then Bose-Einstein condense (at a geometric
series of exponentially generated scales).\footnote{ \Slql\ may be considered as a  ``deconfined'' state  in which the composite bound states deconfine and fractionalize into more fundamental degrees of freedom.} 
 In particular, one finds a finite critical susceptibility with a branch point singularity, and the response of condensed states is described by an infinite spiral. 
 
At a field theoretical level, underlying these features is the annihilation (and moving to the complex plane of a coupling constant) of two conformal fixed points (eCFT$_1^{\rm UV}$ and eCFT$_1^{\rm IR}$). We expect these critical phenomena generically occur in such a situation, where eCFT$_1$ can be replaced by some higher dimensional CFTs.  We emphasize that this is very different 
from the standard Landau-Ginsburg-Wilson paradigm of phase transitions, which can be characterized as a single critical CFT with some relevant directions. 

On the gravity side, for charged operators the instabilities of the \Slql\ manifest themselves as pair production of charged particles which then subsequently backreact on the geometry. For a neutral scalar operator, the instability is related to the violation of the BF bound in the AdS$_2$ region. 

\item A marginal QCP can be obtained by sitting at the critical point of a bifurcating QCP and then dialing  
the external parameter which drives a hybridized QCP. Here given the critical theory describing a bifurcating QCP (which comes from the merger of two fixed points), the phase transition can be described as the appearance of a marginally relevant operator.  
 Interestingly, the fluctuation spectrum that emerges is (when coupled to a Fermi surface) thought to underly the ``Marginal Fermi Liquid'' description of the optimally doped cuprates \cite{Varma89}, making this critical point of potential  importance. 

\een 

Note that while our results were found from gravity analysis, given the general field theoretical descriptions 
above, they likely correspond to generic phenomena, and it would be interesting to understand  
them better using  field theoretical methods.

\begin{figure}[!ht]
\begin{center}
\includegraphics[scale=0.6]{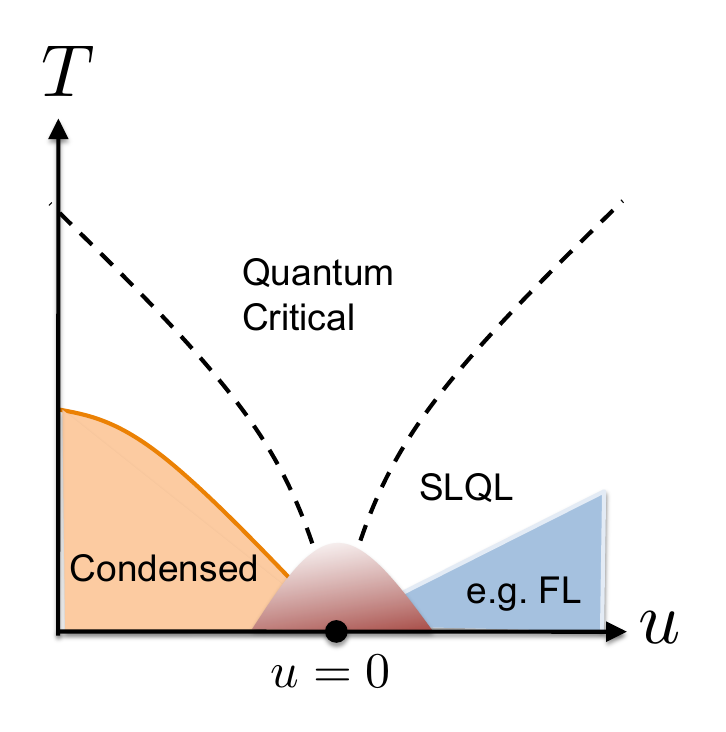}
\end{center}
\caption{How to interpret the results found in this paper. As discussed in~\cite{Iqbal:2011in},  \Slql\ should be interpreted as a universal intermediate phase which orders into some other phases, such as a Fermi liquid, at lower energies. Thus the results of the paper only describe the quantum critical behavior outside the 
dome-shaped region. In contrast to the results found in this paper, which do not depend on the details of a given system (both microscopically and macroscopically), what is inside the dome 
is model-dependent and likely requires understanding finite $N$ effects. 
} 
\label{fig:inter}
\end{figure}

Finally let us elaborate on an important point, which we have glossed over in our discussion so far. As emphasized recently in~\cite{Iqbal:2011in}, \Slql, which describes the disordered phase in our examples 
above, should be interpreted an intermediate-energy phase, rather than a genuine ground state. 
That is, we expect \Slql\ to order into some other phases at lower energies, which may not be visible 
at the large $N$ limit we are working with. An example discussed in~\cite{Iqbal:2011in} is that at some exponentially small scale in $N^2$, \Slql\ orders into a Fermi liquid phase.\footnote{See~\cite{Edalati:2011yv} for a recent discussion of nucleation of a neutral order parameter in a Fermi liquid-like phase.} Thus the quantum critical behavior found in this paper should be more correctly interpreted as describing the intermediate-region indicated in Fig.~\ref{fig:inter}. In the case of condensation of a neutral order parameter, as discussed at the end of Sec.~\ref{sec:cond}, even the condensed side may go to some other phase (e.g. a Fermi liquid phase
co-existence with AFM). Our discussion is nevertheless robust in region outside the dome-shaped region in
Fig.~\ref{fig:inter}.

\vspace{0.3in}   \centerline{\bf{Acknowledgements}} \vspace{0.2in} We thank T.~Faulkner, K.~Jensen, G.~Kotliar, P.~Lee, J.~Ren, D.~Park, M.~Roberts, S.~Sachdev, T.~Senthil, Q.~Si, D.~T.~Son and D.~Vegh for discussions. In particular, we would like to thank Ying Zhao for pointing out an error in our analysis of an earlier version. 
Work supported in part by funds provided by the U.S. Department of Energy
(D.O.E.) under cooperative research agreement DE-FG0205ER41360 and the OJI program.

\begin{appendix}

\section{Matching formulas and properties of $a_\pm, b_\pm$} \label{app:rev}

In this appendix we first briefly review some aspects of the derivation of the master formula~\eqref{roep21},

\be
G_R (\om, \vk) =  \mu_*^{2\nu_U} {b_+ (k,\om) + b_- (k,\om) \sG_k (\om) \mu_*^{-2 \nu_k}  \ov a_+ (k,\om) + a_- (k,\om) \sG_k (\om) \mu_*^{-2 \nu_k}},
\ee

This was first derived in \cite{Faulkner09}, and we refer readers to that work or the more recent review \cite{TASI} for a more in-depth discussion; here we simply recall some aspects of their treatment which require greater care in our analysis. We then discuss the properties of the functions $a_\pm, b_\pm$ appearing in the master formula.  

\subsection{Derivation of the master formula} \label{app:mat}

Recall that the equation of motion (in momentum space ) for $\phi$ in the charged black hole geometry~\eqref{RNmetric} can be written as
\be \label{eomM}
z^4 \p_z \le({f \ov z^2} \p_z \phi \ri) + z^2 \le({(\om+ qA_t)^2 \ov f} - k^2 \ri) \phi - m^2 R^2 \phi = 0 \ .
\ee 
As $z \to 0$ (i.e. to the AdS$_4$ boundary), $\phi$ has the standard asymptotic behavior 
\be \label{12}
\phi(z \to 0) \sim A z^{3-\Delta} + B z^{\Delta}
\ee
where $\De$ is the conformal dimension~\eqref{uvdim} for $\phi$ in the vacuum. The retarded function for $\sO$ in standard quantization can then be written as\footnote{Note we are using a nonstandard normalization for the Green's function, which differs from the standard one by a factor of $2\nu_U$. The same normalization difference applies to the Green's function in the AdS$_2$ region.  }  
 \be
 G_R(\om, \vk) =  \frac{B}{A}  \label{RetD}
\ee 
provided that $\phi$ is an in-falling wave at the horizon.

We are interested in the small-frequency expansion of \eqref{RetD}. The key point here, emphasized in the main text, is that near the horizon the geometry factors into an AdS$_2 \times \mathbb{R}^2$, where the AdS$_2$ involves the time and radial directions. This introduces nontrivial scaling in the time direction, meaning that great care must be taken in performing a small-frequency expansion. This problem was solved in \cite{Faulkner09}: we start with an exact solution in the AdS$_2   \times \mathbb{R}^2 $ region, in which the frequency dependence is treated exactly. We then evolve it outwards, eventually matching it to a solution in the UV region (away from the AdS$_2$) to determine the coefficients $A$ and $B$. In the UV region it is safe to treat frequency dependence perturbatively.

The scalar wave equation on  the AdS$_2   \times \mathbb{R}^2 $ region~\eqref{ads2M}  is
\be \label{ads2eq}
-\partial_\zeta^2 \phi_\vk + \frac{R_2^2 m_k^2}{\zeta^2}\phi_\vk = \le(\om + {q_* \ov \zeta}\ri)^2 \phi_\vk,
\ee
where $m_k^2 = {k^2 \ov \mu_*^2 R^2} + m^2$ and $q_* = {q g_F \ov \sqrt{12}}$.
Equation~\eqref{ads2eq} has solutions near the AdS$_2$ boundary   
\be  
\phi_\vk (\zeta) \sim \zeta^{\ha \pm \nu_k}, \quad  \zeta \to 0 
\ee
 with
 \be \label{nuej}
 \nu_k  =\sqrt{u  +  {k^2 \ov 6 \mu_*^2} }, \qquad u \equiv {m^2 R^2_2}- q_*^2 + {1 \ov 4}
 \ee
which implies that  the corresponding CFT$_1$ operator $\oir_{\vk} (t)$ dual to $\phi_\vk (t, \zeta)$  has a conformal dimension\footnote{This is the dimension in the AdS$_2$ standard quantization; the full UV answer is of course insensitive to this choice.} $\de_k = \ha + \nu_k$. 

Expanding the in-falling solution, which behaves as $e^{i \om \zeta}$ for $\zeta \to \infty$,  near the AdS$_2$ boundary for small $\zeta$, we find (up to an overall normalizing constant)
\be \label{irsol}
\phi(\zeta) = \zeta^{\frac{1}{2} - \nu_k} + \sG_k(\om) \zeta^{\frac{1}{2}+\nu_k}, 
\ee
where by definition $\sG_k(\om)$ is the retarded Green's function for $\oir_\vk (t)$ in the \ircft; it can be 
found by directly solving~\eqref{ads2eq} and has been given earlier in~\eqref{iiRc}.

We now match the IR solution~\eqref{irsol} to a solution in the UV region. To leading order we can set $\om =0$ 
in~\eqref{eomM} in the UV region. The resulting equation has two independent solutions $\eta_\pm^{(0)}$ which can be specified by their behavior near $z \to z_*$ as
\be
\eta_{\pm}^{(0)} (z) \to  \le(\frac{6 (z_* -z) }{z_*}\ri)^{-\frac{1}{2}\pm \nu_k} = \le({\zeta \ov z_*}\ri)^{\ha \mp \nu_k} , \quad z \to z_*  \ 
\label{IRdefUVsoln}
\ee
with the corresponding asymptotic behavior as $z \to 0$ as 
\be \label{12as}
\eta_{\pm}^{(0)} (z ) \approx  a_{\pm}^{(0)} (k) \le({z\ov z_*}\ri)^{3-\Delta }  + b_{\pm}^{(0)} (k) \le({z\ov z_*}\ri)^{\Delta}  \ .
\ee
$a_{\pm}^{(0)} (k)$ and $b_\pm^{(0)} (k)$ thus defined are (dimensionless) functions of $k$ which can be computed numerically. 

At small $\om$, there is an overlapping region in which both both~\eqref{irsol} and~\eqref{IRdefUVsoln} should apply, which determines the full UV solution to be 
\be \label{match}
\phi(z) = \eta_{+}^{(0)} (z) + \sG_k(\om) z_*^{2 \nu_k}  \eta_{-}^{(0)} (z) \ . 
\ee
Equation~\eqref{match} can be generalized  to higher orders in $\om$\be \label{match1}
\phi(z) = \eta_{+} (z) + \sG_k(\om) z_*^{2 \nu_k}  \eta_{-} (z) \  
\ee
where 
\be  \label{omex1}
\eta_\pm = \eta^{(0)}_\pm + \om  \eta^{(1)}_\pm + O(\om^2) 
\ee
are the two linearly independent perturbative solutions to the full UV region equation. We have glossed over several details here and again refer the interested reader to \cite{Faulkner09}. The key point here is that the $\eta_{\pm}$ do depend on frequency, but {\it analytically} with smooth Taylor expansions near $\om = 0$; the non-trivial scaling behavior all arises from the AdS$_2$ region. If $\nu_k$ is real than the $\eta_{\pm}$ are also real, as they obey a real equation with real boundary conditions. Nevertheless, the case where $\nu_k$ is imaginary is important in our analysis and is discussed below. 

Near $z =0$, the $\eta_\pm$ have the expansion of the form~\eqref{12as} with various coefficients $a_\pm^{(0)}, b_\pm^{(0)}$ replaced by $a_\pm, b_\pm$ which also have an analytic $\om$-expansion such as 
\be \label{omex2}
a_+ (k, \om) = a_+^{(0)} (k) + \om a_+^{(1)} (k) + \dots \  .
\ee  
Note that since both the boundary conditions and the equation~\eqref{eomM} are real, $a_\pm, b_\pm$ are real. From~\eqref{match1} and the expansion of $\eta_\pm$ near $z=0$ we thus find the boundary theory Green's function 
to be 
\be \label{roep1}
G_R (\om, \vk) =  \mu_*^{2\nu_U}  {b_+(\om, k)+ b_- (\om, k) \sG_k (\om) \mu_*^{-2 \nu_k} \ov
 a_+ (\om, k) + a_- (\om, k) \sG_k (\om) \mu_*^{-2 \nu_k}} \ .
\ee

We conclude this discussion with some remarks:

\ben

\item Note that for a neutral scalar with $q=0$, equation~\eqref{eomM} only depends on $\om^2$ and the expansion parameter in~\eqref{omex1} and~\eqref{omex2} should be $\om^2$, i.e.
 \be
a_+ (k, \om) = a_+^{(0)} (k) + \om^2 a_+^{(2)} (k) + O(\om^4) \ 
\ee 
and so on. 

\item  For $\nu_k =0$, 
which happens for $k=0$ and $u=0$ (see~\eqref{nuej}), some new elements arise. 
 At $\nu_k=0$, the basis of functions in~\eqref{IRdefUVsoln} should be replaced by
\be
\eta^{(0)} (z) = \le({\zeta \ov z_*}\ri)^\ha  \qquad \tilde \eta^{(0)} (z) = -\le({\zeta \ov z_*}\ri)^\ha  \log {\zeta \ov z_*} \label{nu0etas}
\ee
where the asymptotic behavior for them at $z \to 0$ is
\begin{align} \label{11as}
\eta^{(0)} (z) &\approx  \al  \le({z\ov z_*}\ri)^{3-\Delta }  + \beta  \le({z\ov z_*}\ri)^{\Delta }, \nonumber \\  
\tilde \eta^{(0)} (z ) &\approx  \tilde \al  \le({z\ov z_*}\ri)^{3-\Delta } + \tilde \beta   \le({z\ov z_*}\ri)^{\Delta }
\end{align}
$\al, \beta, \tilde \al, \tilde \beta$ are now dimensionless real numbers which can again be found numerically. They play a key role in understanding the analytic properties of $a_{\pm}$, $b_{\pm}$ as is discussed in some detail below.

\item For $u < 0$, $\nu_k =- i \lam_k$ is pure imaginary for small enough $k$, and the basis of solutions~\eqref{IRdefUVsoln} now has the form 
\be
\eta_{\pm}^{(0)} (z) \to  \le(\frac{6 (z_* -z) }{z_*}\ri)^{-\frac{1}{2} \mp i   \lam_k} = \le({\zeta \ov z_*}\ri)^{\ha \pm i \lam_k} , \quad z \to z_*  \ 
\label{UVsolnS}
\ee
These boundary conditions are now complex, and thus so are the $\eta_{\pm}$. As the $\eta_{\pm}$ actually obey a real wave equation, the full analytic structure is determined by the boundary conditions in the infrared; thus we find that now $\eta_{+} = \eta_{-}^*$. This also implies that $a_\pm, b_\pm$ are complex and 
\be 
a_+^* = a_-, \qquad b_+^* = b_- \ .
\ee

\een
We now discuss some further properties of the UV expansion coefficients $a_{\pm}, b_{\pm}$. 

 \subsection{Analytic properties of $a_\pm,\  b_\pm$} \label{app:abpro}
 
 The functions $a_\pm (\om, k), \ b_\pm (\om,k)$ are obtained by solving equation~\eqref{eomM}
perturbatively in $\om$ in the UV region. Their $k$-dependence comes from two sources, from dependence on $\nu_k$ via the boundary condition~\eqref{IRdefUVsoln} and from $k^2$ dependence in the equation~\eqref{eomM} itself. Since the geometry is smooth throughout the UV region we expect the dependence on both $\nu_k$ and $k^2$ to be analytic.
In fact we can think of $b_{\pm}$ and $a_{\pm}$ as {\it functions} of $\nu_k$; i.e. there exists a function $b(\nu_k,k^2,\om)$, {\it analytic in all its arguments}, such that $b_{\pm} = b(\pm\nu_k,k^2,\om)$. This is clear from the boundary condition~\eqref{IRdefUVsoln} (and its generalization for higher orders in $\om$) and from the fact that there is no other dependence on $\nu_k$ from the equation of motion itself.  

Let us now look at the behavior of $a_\pm^{(0)}, b_\pm^{(0)}$ in the limit of $\nu_k \to 0$ in some detail. Note that this limit should be considered as a double limit 
$k^2 \to 0$ and $u \to 0$.  First, we note that in the limit $\nu_k \to 0$, the basis of functions introduced in~\eqref{IRdefUVsoln} can be expanded as
\be \label{expsi}
\eta_\pm^{(0)} = \eta^{(0)} (z) \pm \nu_k \tilde \eta^{(0)} (z) + O(\nu_k^2) \ ,
\ee
where $\eta^{(0)}$ and $\tilde \eta^{(0)}$ were introduced in~\eqref{nu0etas}. This leads to 
\begin{align} \label{req}
b_\pm^{(0)} & = \beta \pm {\nu_k} \tilde \beta + (c_1 k^2 + d_1 u) + \dots \ , \nonumber \\    
a_\pm^{(0)} & = \al \pm {\nu_k} \tilde \al + (c_2 k^2 + d_2 u) + \dots \ .
\end{align}
In the above equations the linear order terms directly come from the linear order term in~\eqref{expsi}, while the quadratic order terms also receive contributions from equation of motion itself (not just the boundary conditions) and cannot be expressed in terms of $\nu_k^2$ alone. The important point is that the quadratic order terms are independent of the signs before $\nu_k$ and thus are the same for $a_\pm$ and $b_\pm$ repsectively. Similarly approaching $\nu_k=0$ from the imaginary $\nu_k = - i \lam_k$ side, we have
for small $\lam_k$,  
\be
\label{req1}
b_\pm^{(0)} = \beta \mp i {\lam_k} \tilde \beta + \dots, \qquad   a_\pm^{(0)} = \al \mp i {\lam_k} \tilde \al + \dots \ .
\ee
Again the quadratic order terms should be the same for $a_\pm$ and $b_\pm$. 

Note also that $\nu_k$ itself becomes {\it non-analytic} in $k^2$ at $u=0$ (see~\eqref{nuej}) and as a result through formula such as \eqref{req} $a_\pm, b_\pm$ will also develop non-analytic behavior in $k^2$ at $u=0$. 
This fact is important for understanding the critical behavior around the critical point $u=0$ discussed in the main text. 

We conclude this section by noting that coefficients $a_\pm, b_\pm$ are not independent. For example evaluating the Wronskian of~\eqref{eomM} (for $\om =0$)\footnote{The Wronskian of equation~\eqref{eomM} is given by 
\be
W [\phi_1, \phi_2]=  {f \ov z^2}  (\phi_1 \p_z \phi_2 - \phi_2 \p_z \phi_1)
\ee
which is independent of $z$.} for $\eta_\pm^{(0)}$ and demanding that it be equal at infinity and at the horizon, we find the elegant relation:
\be
a_+^{(0)} (k) b_-^{(0)} (k) - a_-^{(0)} (k)  b_+^{(0)} (k) = {\nu_k \ov \nu_U}  \ .
\ee
A similar analysis on $\eta, \tilde\eta$ results in
\be \label{W1}
\al \tilde \beta - \beta \tilde \al = -{1 \ov 2\nu_U}  \ .
\ee
Interestingly, this particular combination of coefficients appears many times throughout this paper. 

We conclude this section by specifying the explicit values for these constants for a neutral scalar moving on the pure Reissner-Nordstrom background; in this model by requiring $\nu_{k=0} = 0$ we fix the value of the mass to be $m^2 R^2 = -\frac{3}{2}$, and thus we can (numerically) compute the coefficients once and for all to be:
\begin{align}
\label{pp1}
\al&=0.528  &\tilde \al&=0.965\\
\beta&=0.251  &\tilde \beta&=-0.640 \ .
\label{pp2}
\end{align}
One can check that within numerical error these values satisfy \eqref{W1} with $\nu_U = \frac{\sqrt{3}}{2}$.

\section{AdS correlators and instabilities} \label{app:dt}

In this Appendix we give a more detailed discussion of various scalar instabilities of a $(d+1)$-dimensional AdS spacetime mentioned in Sec.~\ref{sec:insta}. We will first consider general $d$ and then specify to $d=1$, i.e. AdS$_2$, for which more can be said and which plays an important role in this paper. Our discussion for $d > 1$ applies to any scalar operator, but for AdS$_2$ (which often contains a background electric field) the results for charged and neutral scalars are different and we will comment on this. We treat only linear response and so will describe the nature and onset of the instability, not the endpoint of the condensate (which depends on the specific model.) 

\subsection{General $d$}

As in  Sec.~\ref{sec:insta}, consider a  scalar field $\phi$ in AdS$_{d+1}$, which is dual to an operator~$\Phi$ in some boundary CFT$_d$. The possible conformal dimensions of $\Phi$  are given by~\eqref{dim1} which we reproduce here for convenience, 
\be \label{dim2}
\De_\pm = {d \ov 2} \pm \nu  , \qquad \nu = \sqrt{M^2 R^2 + {d^2 \ov 4}}
\ee
where $M^2$ is the mass square for $\phi$. As discussed below~\eqref{dim1},
for  $\nu \in (0,1)$, there are two ways to quantize $\phi$ in AdS, giving rise to the two theories CFT$_d^{\rm IR}$
and CFT$_d^{\rm UV}$, in which the corresponding operators $\Phi_\pm$ have dimension $\De_+ = {d \ov 2} + \nu$ and $\De_- = {d \ov 2} - \nu$ respectively. These are often called the ``standard'' and ``alternative'' quantizations; for $\nu > 1$ only the standard quantization is allowed. This can be seen via a simple normalizability argument and as we will see below is modified for a charged scalar in AdS$_2$. 

The two-point retarded Green's function for $\Phi_+$ in the CFT$_d^{\rm IR}$ (``standard quantization'') 
is given by 
\be \label{adsco}
G_+ (\om, \vec k) =  C(\nu)  \le({ \vec k^2 - (\om + i \ep)^2 \ov 4} \ri)^{\nu}, \quad
C(\nu) \equiv \frac{\Ga(-\nu)}{\Ga(\nu)} 
\ee 
and at zero spatial momentum 
\be \label{adsco1}
G_+ (\om, \vec k=0) =  \le(-\frac{i \om}{2}\ri)^{2\nu} 
\frac{\Ga(-\nu)}{\Ga(\nu)} \  
\ee 
That for $\Phi_-$ in the ``alternative quantization'' CFT$_d^{\rm UV}$ (which we denote by $G_- (k_\mu)$) is given by 
\be \label{alco}
G_-(k_{\mu}) = -G_+(k_{\mu})^{-1} \ .
\ee
Also note that for a theory deformed by a double-trace operator\footnote{Note that in this Appendix we use a dimensionful $\kappa$ as opposed to the main text.}
\be \label{defF}
\de S =  {\ka \ov 2} \int d^d x \, \sO^2
\ee
in the large $N$ limit, the retarded correlation function for $\sO$ becomes 
\be \label{douGF}
G_R^{(\ka)} (\om,\vk) = {1 \ov G_R^{-1} (\om,\vk) + \ka }
 \ 
\ee
where $G_R (\om,\vk)$ is the retarded function in the absence of deformation.

\subsubsection{Instabilities from double trace deformations} \label{app:douins}

We now consider instabilities induced by a double trace deformation of the system. We begin by considering $\nu \in (0,1)$ with the alternative quantization, CFT$_d^{\rm UV}$ and turn on the following double trace deformation:
\be \label{douir1}
{\kappa_- \ov 2}  \int d^d x \, \Phi_-^2  \ . 
\ee
From~\eqref{douGF} the retarded Green's function of the operator $\Phi_-$ now becomes
\be \label{gmdt}
G^{(\kappa_-)}_-(\om,\vk) = \frac{1}{G_-^{-1}(\om,\vk) + \kappa_-}\ ,
\ee
where $G_-(\om, \vk)$ is the undeformed correlator in alternative quantization. The double-trace deformation \eqref{douir1} has dimension $d-2\nu$ and so is relevant. For $\kappa_- > 0$ it triggers a flow which leads to the standard quantization, CFT$_d^{\rm IR}$ (and hence the respective names of the two theories). This can be seen by expanding \eqref{gmdt} at small momentum; for any nonzero $\kappa_-$ the resulting answer will coincide (up to a contact term) with the correlator $G_+(\om, \vk)$.

For $\kappa_- < 0$, the theory develops an IR instability~\cite{Vecchi:2010jz,Vecchi:2010dd},
as the static susceptibility becomes negative at $k=0$, signaling a tachyonic mode. The scale at which the instability sets in can be found either by examining the beta function for the running double-trace coupling which develops an IR Landau pole at certain scale or from that equation~\eqref{gmdt}  develops a pole in the upper half complex plane. 
We now elaborate on the latter a bit more. With $\vec{k} = 0$,  from \eqref{adsco1}
\be
G_-(\om) = -\frac{1}{G_+(\om)} = -\frac{1}{C(\nu)} \le(-\frac{i \om}{2}\ri)^{-2\nu} \ .
\ee
Plugging this into \eqref{gmdt} we find that $G^{(\kappa_-)}_-(\om,k)$ has a pole at 
\be \label{polalt}
\om = i\om_\ka \qquad \om_\ka = 2\le(\frac{\kappa_-}{C(\nu)}\ri)^{\frac{1}{2\nu}}
\ee
As $C(\nu)$ is negative for  $\nu \in (0,1)$, this pole is in the upper-half plane when $\ka_- < 0$, indicating a {\it dynamical} instability, meaning the bulk scalar field will condense.\footnote{Turning on a finite temperature can stabilize the system, and the critical temperature scales as $T_c \sim (-\kappa_-)^\frac{1}{2\nu}$~\cite{Faulkner:2010gj}.} Note that as we increase $\ka_-$ from zero, the pole appears from the origin; thus it is an IR instability, consistent with our understanding of the double-trace deformation \eqref{douir} as a relevant operator. 

To summarize, $\kappa_- = 0$  can be considered as a quantum critical point
separating two different phases; for $\kappa_- > 0$ we flow to a conformal phase which in the IR is simply the standard quantization theory CFT$_d^{\rm IR}$, whereas for $\kappa_- < 0$ we have a different phase in which the bulk scalar field condenses; the final endpoint of the instability cannot be answered without knowing more details about the nonlinear structure of the theory. 

The story for the standard quantization with a double trace deformation $\frac{\ka_+}{2} \int d^dx\;\Phi_+^2$ can be similarly worked out. For $\nu \in (0,1)$, the system is stable for $\ka_+ < 0$ and the alternative quantization is obtained in the $\ka_+ \to -\infty$ limit. 
However for positive $\ka_+$ we find a pole in the upper half plane at\footnote{It is intriguing that $C(\nu)$ oscillates; this implies for example that there are poles in the upper-half plane for negative $\ka_+$ and $\nu \in (1,2)$. It would be good to understand this better.}
\be \label{polk1}
\om = i \om_\ka, \qquad \om_\ka = 2 (-\ka_+ C(\nu))^{-{1 \ov 2 \nu}} \ . 
\ee
Note that as we increase $\ka_+$ from zero, the pole~\eqref{polk1} moves in from $+ i \infty$, which means that it is a {\it UV instability}. This is consistent with our RG understanding, as $\Phi_+^2$ has dimension $d + 2\nu$; it is irrelevant and so is expected to be important only at large energies. This UV instability can also be seen from a Landau pole in the running coupling of the double-trace operator $\Phi_+^2$~\cite{Vecchi:2010jz,Vecchi:2010dd}.  See Fig.~\ref{fig:phs} which summarizes the above conclusions. 

For $\nu > 1$, there is only standard quantization. For $\nu \in (1,2)$, $C(\nu)$ in~\eqref{adsco} becomes positive. Thus in contrast with $\nu \in (0,1)$, now the system develops a UV instability for $\ka_+ < 0$. 
While there is no apparent instability for $\ka_+ > 0$, the UV completion is not currently known. 
The sign of $C(\nu)$ oscillates with integer $n$ for $\nu \in (n, n+1)$ and the instability region oscillates between $\ka_+ > 0$ and $\ka_+ < 0$ depending on whether $n$ is even or odd.

\subsubsection{Instabilities from the Breitenlohner-Freedman bound}\label{app:A}

We now study a different mechanism for an instability. As $\nu \to 0$, i.e. $M^2 \to -{d^2 \ov 4} \equiv M_c^2$, the two possible quantizations -- two different CFT$_d$'s -- merge into one at $\nu =0$.  If $M^2$ further drops {\it below} $M_c^2$, the so-called Breitenlohner-Freedman bound, $\nu$ becomes complex and 
 $\Phi$ develops exponentially growing modes~\cite{bf}. The system becomes unstable to the 
 condensation of $\Phi$ modes. Introducing a UV cutoff  $\Lam$, then there is an exponentially generated  IR energy scale $\Lam_{IR}$~\cite{Kaplan:2009kr}
   \be \label{irsca1}
\Lambda_{IR} \sim \Lam \exp\le(-\frac{\pi}{\sqrt{M^2_c - M^2}}\ri)  \ 
\ee
below which the physics of condensate sets in. 
 
Now instead of going below the BF bound, consider sitting precisely at $\nu =0$, we find that the double-trace deformation $\ka \Phi^2$ is marginal. Whether it is marginally relevant or irrelevant depends on its sign~\cite{Witten:2001ua}.
To see this, note that at $\nu=0$, equation~\eqref{adsco1} should be replaced by\footnote{Note that~\eqref{zeronu} is obtained by directly solving the bulk equation of motion at $\nu=0$. Taking $\nu \to 0$ limit in~\eqref{adsco1} one instead finds 
\be \label{zeronuL}
G_{+}  (\om) \to - 1 + 2\nu G_0 (\om) + \dots, \quad \nu \to 0 \ .
\ee}
\be \label{zeronu}
G_{0} (\om) = - \log \le(- {i \om \ov \Lam} \ri)  
\ee
where $\Lam$ is a UV regulator. Under a double trace deformation with coupling $\ka$ we find 
\be \label{gmdt1}
G^{(\kappa)}_0 (\om) = \frac{1}{G_0^{-1}(\om) + \kappa}\ .
\ee
The above equation has a pole in the upper half frequency plane at
\be 
\om_\ka = i \Lam \exp \le({1 \ov \ka} \ri) \equiv i \Lam_\ka  \ .
\ee
Considering increasing $\ka$ from zero to some positive value, then the above pole emerges from $+i \infty$, implying a UV instability. We thus conclude that $\ka > 0$ is marginally irrelevant. Similarly a negative $\ka$ is marginally relevant and leads to an IR instability. In both cases we can identify $\Lam_\ka$ as a dynamically generated scale. 

\begin{figure}[!ht]
\begin{center}
\includegraphics[scale=0.5]{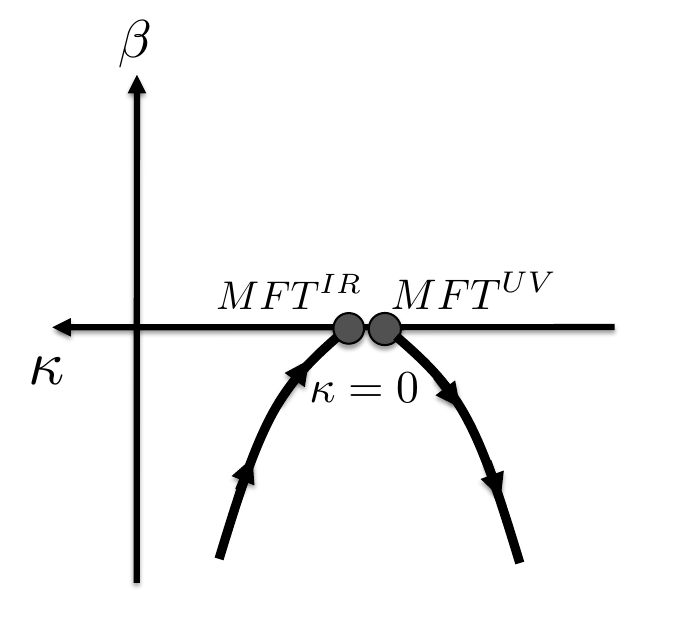}
\end{center}
\caption{The RG flow diagram at $\nu=0$ for double trace coupling $\ka$. Note the positive $\ka$-axis is pointed to the left and the arrow denotes flowing to IR. MFT$^{UV}$ denotes the fixed point in which the retarded function for $\Phi$ is given by~\eqref{zeronu} and MFT$^{IR}$ denotes the theory in which the retarded function for $\Phi$ is proportional 
to $ {1 \ov \log (- i \om)}$. 
 } 
\label{fig:MFT}
\end{figure}

Here we note a curious fact. Suppose we deform the system with $\ka > 0$
at some scale far below $\Lam_\ka$ where the UV instability sets in. Naively we would expect that the system should flow back to $\ka=0$ in the IR. However, it follows from~\eqref{gmdt1} that in the $\om \to 0$ limit, the retarded function is given by $G^\ka_0 \sim {1 \ov \log(-i \om)}$ instead of $G_0$. The situation is summarized 
in Fig.~\ref{fig:MFT}.  Starting from MFT$^{UV}$, under a double trace deformation $\ka$, the theory flows to MFT$^{IR}$ for positive $\ka$ in the IR, while for $\ka < 0$, the theory develops an IR instability and flows to a condensed phase. Thus MFT$^{UV}$ is a multi-critical point (since we need to tune to $\nu=0$ and $\ka =0$
at the same time).

\subsection{Specializing to AdS$_2$} \label{app:ad2i}

We now specialize to $d = 1$, i.e. an AdS$_2$ bulk geometry, for which there are some new elements. We work with an AdS$_2$ metric with the form
\be
ds^2 = -\frac{R_2^2}{\zeta^2}(-dt^2 + d\zeta^2) \qquad A = \frac{e_d}{\zeta} dt \label{appads2}
\ee
Note that if are finding this as the near-horizon limit of the Reissner-Nordstrom black hole in an asymptically AdS$_4$ spacetime the value of $e_d$ is fixed to be $\frac{g_F}{\sqrt{12}}$, as in \eqref{ads2M}. Note that the gauge field actually blows up as we go to the boundary $\zeta \to 0$.
This influences various properties of operators dual to bulk charged fields; for example, the conformal dimension of the operator can now depend on the charge $q$. For a scalar with mass $m$ we find using the metric \eqref{appads2}
\be
\Delta_{\pm} = \ha \pm \nu \qquad \nu = \sqrt{m^2R_2^2 - q_*^2 + \frac{1}{4}}. \label{newexp}
\ee
where $q_* \equiv q e_d$.

The allowable $\Delta_-$ range that can be reached using alternative quantization is also different; for a neutral scalar in any AdS$_{d+1}$ we find that alternative quantization is allowed if $\nu \in (0, 1)$, but for a charged scalar in AdS$_2$ normalizability of the wave function requires $\nu \in (0, \ha)$.  Note this imposes a new ``unitarity bound'' on the lowest possible dimension of a charged operator in AdS$_2$/CFT$_1$: $\Delta_- > 0$; this is {\it stronger} than the usual unitarity bound, which is $\Delta_- > \frac{d-2}{2} \to -\ha$ for $d = 1$.  
There is a heuristic way to understand this new bound; usually the field theoretical unitarity bound coincides with the dimension of a free massless (field theoretical) scalar in $d$-dimensions. In one-dimensional quantum mechanics if we turn on a chemical potential for a charged scalar $X$ its scaling is determined not by the $\int dt \dot{X}^{\dagger} \dot{X}$ term but rather by the term $\int dt A_t X^{\dagger} \dot{X}$, which results in a free charged scalar of dimension $[X] = 0$, coinciding with the bound derived above. 

\section{Details of the phase diagram}\label{app:phase}

\begin{figure}[!h]
\begin{center}
\includegraphics[scale=0.69]{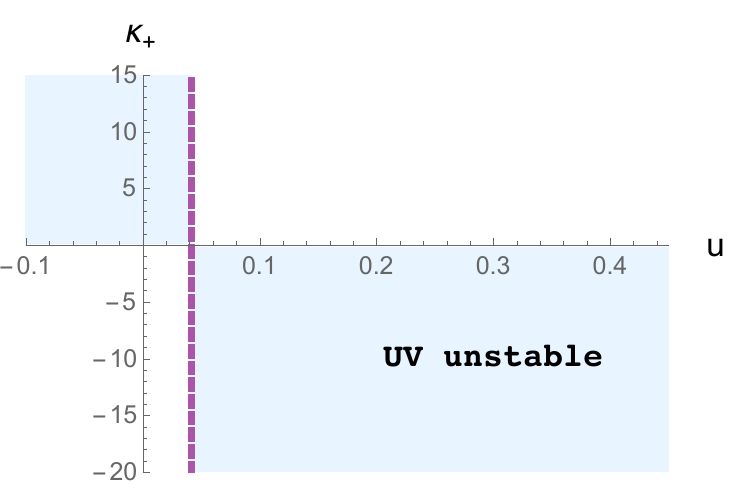} 
\end{center}
\caption{ Phase diagram of the neutral scalar system in the standard quantization at zero density.}
\label{fig:UVphase}
\end{figure}

In this Appendix we go through the steps of constructing the standard quantization phase diagram Fig.~\ref{fig:fullphase}. We begin by drawing the phase diagram of the system at zero density. As discussed in Appendix~\ref{app:dt}, for $\nu_U \in (n, n+1)$ the instability region oscillates between $\ka_+ > 0$ and $\ka_+ < 0$ depending on whether $n$ is even or odd. Translating between $\nu_U$ and $u$ gives us the vacuum phase diagram for a neutral scalar, Fig.~\ref{fig:UVphase}. By introducing finite density, on top of the UV instabilities determined by the vacuum structure the system can develop other IR instabilities. We devote the rest of the Appendix to their study.

\begin{figure*}[!ht]
        \begin{minipage}[r]{2.0\columnwidth}
        \begin{center}
{\includegraphics[scale=0.75]{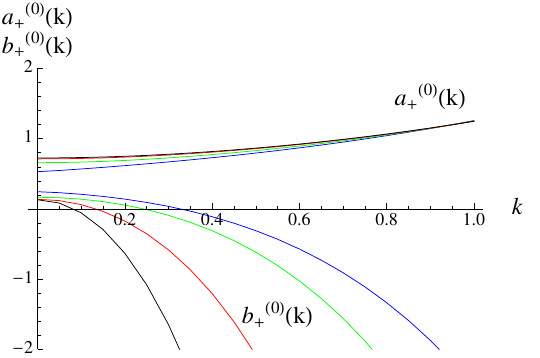}}
\includegraphics[scale=0.91]{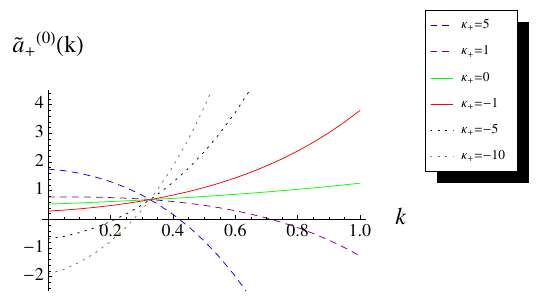} 
\end{center}
\caption{ {\it Top plot:} $a_+^{(0)} (k)$ and $b_+^{(0)} (k)$ plotted for different values of $m^2R^2 < -{5 \ov 4}$ with $q=0$; 
blue is $m^2 R^2=-1.4999$, green $m^2 R^2=-1.4$, red $m^2 R^2 =-1.3$ and black $m^2 R^2 = -1.27$. 
 $a_+^{(0)} (k)$ is positive and monotonically increases with $k$, while 
 $b_+^{(0)} (k)$ monotonically decreases with $k$. Thus for $\ka_+ < 0$, $\tilde a_+^{(0)} (k)
 = a_+^{(0)} (k) + \ka_+ b_+^{(0)} (k) $ is a monotonically increasing function of $k$.
 \\
  {\it Bottom plot:} $\ta_+^{(0)} (k)$ for $ m^2 R^2 =-1.4999 $ and $q=0$. $\ta_+^{(0)} (k)$ 
has a zero for some $k$ when $\ka_+ < \ka_c = -2.13$, which implies an IR instability. 
 For $0 > \ka_+ > \ka_c$ there is no instability (see the $\ka_+ =-1$ curve). 
When $\ka_+ > 0$, $\ta_+^{(0)} (k)$ can again develop a zero for some $k_F$ (with the value of $k_F$ approaching infinity for $\ka_+ \to 0_+$); this is the {\it UV} instability discussed in Sec.~\ref{sec:insta} which is already present in the vacuum.   To lead the eye we use solid lines for stable values of $\ka$ (i.e. $\ta_+^{(0)} (k)$ does not have any zero), dotted lines for those with an IR instability, and dashed lines for those with a UV instability. 
\label{fig:neu}}
 \begin{center}
\includegraphics[scale=0.75]{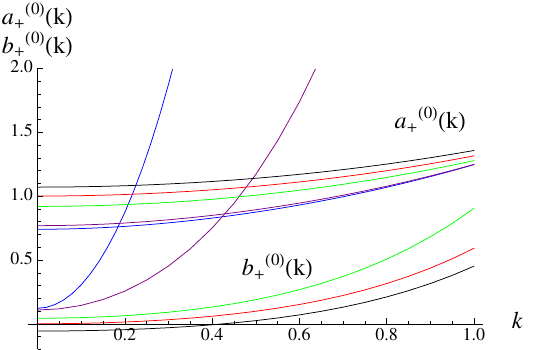}\\
\includegraphics[scale=0.91]{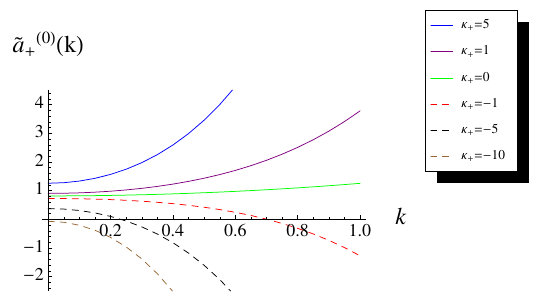}
\includegraphics[scale=0.91]{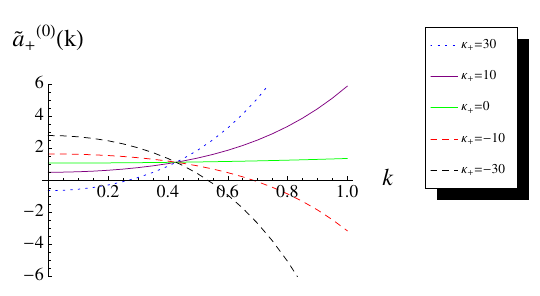}
\end{center}
\caption{
  {\it Top plot:} $a_+^{(0)} (k)$ and $b_+^{(0)} (k)$ plotted for different values of $-{5 \ov 4} < m^2R^2 < {7 \ov 4}$ with $q=0$; 
blue is $m^2 R^2=-1.23$, purple $m^2 R^2 = -1.15$, green $m^2 R^2=-0.5$, red $m^2 R^2 =0$ and black $m^2 R^2 = 0.5$. While as in Fig.~\ref{fig:neu}, $a_+^{(0)} (k)$ is positive and monotonically increases with $k$, for this mass range $b_+^{(0)} (k)$ becomes monotonically {\it increasing} with $k$. 
Note that $b_+^{(0)} (k=0) =0$ for $m^2 =0$ and $b_+^{(0)} (k=0) < 0$ for $m^2 > 0$.  
\\
 {\it Bottom left plot:} $\ta_+^{(0)} (k)$ for $m^2 R^2 =-1$ and $q=0$. For $0 \geq m^2 R^2 > -{5 \ov 4}$ the system is stable for any $\ka_+ > 0$, but develops a UV instability for $0 > \ka_+ > \ka_c$,  $\ta_+^{(0)} (k)$ 
 has a zero which approaches infinity for $\ka_+ \to 0_-$. For $\ka < \ka_c$, $\ta_+^{(0)} (k)$ becomes negative definite. This is the only instance, when we cannot resort to the logic of the second point in Sec.~\ref{sec:insta}. However, because the UV and IR effects are independent, we can conclude that phase has {\it both a UV and an IR instability}.\\
  {\it  Bottom right plot:}  $\ta_+^{(0)} (k)$ for $m^2 R^2 =0.5$ and $q=0$. For $m^2 >0$, the system now develops an IR instability for 
 $\ka_+ > \ka_c > 0$. For $\ka_+ < 0$, there is a UV instability.
\label{fig:neu2}}
\end{minipage}
\end{figure*}

To determine when finite density instabilities occur, we have to solve for $a_+^{(0)} (k)$ and $b_+^{(0)} (k)$ numerically. As explained in Sec.~\ref{sec:finin}, in the double trace deformed theory we look for the zeros of $\tilde a_+^{(0)} (k) = a_+^{(0)} (k) + \ka_+ b_+^{(0)} (k) $ as a function of $k$. At these special values of $k$ there is a pole crossing over to the upper half $\om$-plane resulting in a dynamical instability. The phase boundaries are then easily mapped out. In Fig.~\ref{fig:neu},~\ref{fig:neu2} we plot $a_+^{(0)} (k)$, $b_+^{(0)} (k)$, and $\tilde a_+^{(0)} (k)$ for various values of $u$ (equivalently $m^2 R^2$) and $\ka_+$. 
 
\begin{figure}[!h]
\begin{center}
\includegraphics[scale=0.51]{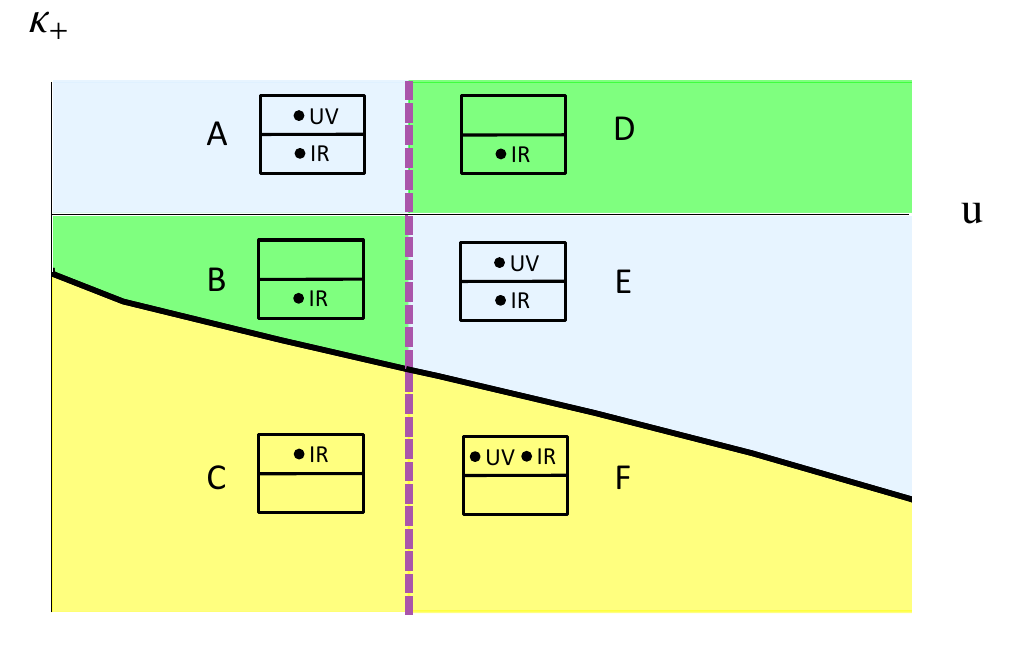} 
\end{center}
\caption{Cartoon illustration of movement of UV and IR poles in the complex $\om$-plane. Each box with two halves represents the upper and lower complex plane, with dots indicating where the UV and IR poles sit. The color coding is the same as on the complete phase diagram Fig.~\ref{fig:fullphase}. Note that when moving across certain lines on the phase diagram it is evident from Fig.~\ref{fig:neu},~\ref{fig:neu2} that the asymptotic large k structure of $\ta_+^{(0)} (k)$ changes completely, allowing the UV pole to move (or return) from infinity: e.g. from A to B it moves to infinity, and from B to E it returns from infinity. On the other hand the IR pole crosses from the upper to the lower half plane (or vice-versa) through $\om = 0$ whenever the susceptibility changes sign (e.g. from B to C, or from F to E). Consideration of the functions $\ta_+^{(0)} (k)$ results eventually in this assignment of poles. 
}
\label{fig:polephase}
\end{figure}

From these plots we can read off the movement of the poles on the $\om$-plane. There can be two sources of instabilities and correspondingly two types of poles can make appearance in the upper half plane. The position of the UV poles is determined by vacuum physics, and gives the vacuum phase diagram Fig.~\ref{fig:UVphase}. The IR poles are a result of finite temperature physics. When such poles cross the real line the static susceptibility, $\chi$ diverges. We show where these poles are in the various phases in Fig.~\ref{fig:polephase}. We emphasize that the physics of UV and IR instabilities is different, and correspondingly the movement of IR and UV poles is independent.

Finally, we complete the phase diagram with the bifurcating critical line at $u_c=0$, see Fig.~\ref{fig:fullphase}.

$\\ $

\section{Nonlinear solution near bifurcating critical point: Efimov spiral} \label{app:COND}

In this Appendix we give the gravity analysis of the critical behavior near the bifurcating quantum critical point approaching from the condensed side,  i.e. $u< 0$. As discussed in~\cite{Iqbal:2010eh,Jensen:2010ga}, for the lowest $n=1$ state there is a new  exponentially generated scale 
\be \label{irsca}
\Lambda_{IR} \sim \mu \exp\le(-\frac{\pi}{\sqrt{-u}}\ri)
\ee
and when the AdS$_2$ radial coordinate $\zeta$ satisfies $\Lam_{IR}\zeta \gg 1$ (i.e. deep in the  
AdS$_2$ region), $\phi$ becomes of order $O(1)$. Thus at zero temperature, no matter how close one is to the critical point and even though the vacuum expectation value of the condensed operator is very small near the critical point, the nonlinear dynamics of $\phi$ and the backreaction to the bulk geometry  will be needed deep in the AdS$_2$ region.
Nevertheless, we will find that a great deal of information can be obtained even without detailed analysis of the nonlinear equations and  backreaction. For illustration purpose, as in~\cite{Iqbal:2010eh} we will consider an action for $\phi$ of the form 
 \be \label{newp}
\sL_\phi = {1 \ov 2 \kappa^2 g } \le[- {1 \ov 2} (\p \phi)^2 - V(\phi) \ri]
 \ee
where $g$ is a coupling constant. The precise form of the potential $V(\phi)$ is not important for our discussion below except that $V(0) =0$, $V'' (0) = m^2$ and it has a minimum at some $\phi_0 \neq 0$.  To be close to $u=0$ critical point on the condensed side, we will thus take $m^2$ to be slightly smaller than the value in~\eqref{crimas} and $u = (m^2 - m_c^2)R_2^2 < 0$. 

We will now proceed to compute the response of the system to a static, uniform external source.
Thus we consider equation of motion following from~\eqref{newp} with $\phi$ depending on radial coordinate only. The analysis is a slight generalization of that in~\cite{Kaplan:2009kr,Jensen:2010ga,Iqbal:2010eh}. To describe the behavior of the bulk solution describing a condensed phase, we separate the spacetime into three regions: 

\ben

\item  IR region I: $\zeta > \Lam_{IR}^{-1}$. 
Here the nonlinear effect of the scalar potential is important and  the value of $\phi$ is of O(1). We note that the boundary condition at the horizon is given by 
\be
\phi (\zeta \to \infty) = \phi_0 
 \label{irbc}
\ee
where 
$\phi_0$ is the minimum of the potential $V(\phi)$. Thus as $\zeta \to \infty$, the spacetime metric approaches $\widetilde{AdS}_2 \times \RR^2$ where $\widetilde{AdS}_2$ has a different curvature radius from the  near-horizon AdS$_2$ region of the condensed phase, 
\be \label{newcos}
{1 \ov \tilde R^2_2} = {1 \ov R_2^2} - {V(\phi_0) \ov g} \ .
\ee

\item IR region II:  AdS$_2$ region with $\zeta < \Lam_{IR}^{-1}$ (but still $\mu\zeta \gg 1$ so that the AdS$_2$ scaling is appropriate). 
In this region the value of $\phi$ is small, and we can treat it linearly and neglect its backreaction on the geometry.  Here $\phi$ has a well-defined but complex conformal dimension in \ircft\ dual to the original AdS$_2$ (with $k=0$) :
\be \label{compd}
\de_\pm = \ha \pm i \sqrt{-u}, \qquad - u \ll 1 
\ee  
and general solution to linearized equation can be written as $a \zeta^{\de_+} + b \zeta^{\de_-}$.

\item UV region: the rest of the black hole spacetime. Again in this region linear analysis suffices.  
 
\een
Note that the IR region II is not guaranteed to exist {\it a priori}, but will be justified by the results, i.e.~\eqref{irsca} (right now $\Lam_{IR}$ should be considered just as a parameter we introduce to distinguish IR region I and II).

We will solve the nonlinear equation following from~\eqref{newp} starting in IR region I and moving towards the boundary of the spacetime.  Note that the horizon boundary condition~\eqref{irbc} fixes one of the integration constants in the second-order equation for $\phi$.  As we move outwards, the scalar becomes smaller and smaller until around $\zeta \sim \Lam_{IR}^{-1}$,  where we can neglect its backreaction on the geometry and treat it linearly.  Note that the solution to the nonlinear equation in IR region I should be insensitive to the precise value of $m^2$ (which is the mass square near $\phi=0$) and thus when $|u|\ll 1$, we could set $u$ to $0$ in solving it. 
This implies that, near $ \Lam_{IR}^{-1}$, it should be a good approximation to solve the linearized equation (around $\phi =0$) with $u=0$, and a general solution can be written as\footnote{Note at $u=0$, the two exponents in~\eqref{compd} become degenerate and the independent solutions to the linear equation become $\zeta^\ha$ and $\zeta^\ha \log \zeta$ respectively.} 
\be  \label{nonla}
\phi (\zeta) =\ga \sqrt{\zeta \ov \zeta_*}  \log {\zeta \ov \zeta_*} + O (\sqrt{-u}) 
\ee
where  $\ga \sim O(1)$ and $\zeta_* \sim \Lam_{IR}^{-1} $ are integration constants.

In the limit of no backreaction (e.g. $g \to \infty$ in~\eqref{newcos}), it can be readily checked that the full nonlinear problem in AdS$_2$ region has an AdS$_2$ scaling symmetry under which both the IR boundary condition \eqref{irbc} and the equations of motion are invariant. Recall that the horizon boundary condition~\eqref{irbc} fixed only one of the two integration constants, leaving a one-parameter family of acceptable solutions. We conclude that in this case because of the scaling symmetry this family is parametrized by $\zeta_*$, and the number $\ga$ must be fixed by~\eqref{irbc} to be an $O(1)$ constant (as there are no small parameters in the nonlinear analysis).
If we allow backreaction then these statements are no longer strictly true, in that as we traverse the remaining one-parameter family of solutions we will likely move through a nontrivial trajectory in the $(\ga, \zeta_*)$ space. This should be kept in mind; however in the remainder of the analysis for simplicity we will assume that backreaction is small and so we can assume that $\ga$ is fixed by~\eqref{irbc} and the remaining solutions are parametrized by $\zeta_*$.

Now from~\eqref{compd} the most general solution to the linearized equation in IR region II can be written as 
 \be \label{linsol}
\phi (\zeta) = d_1 \sqrt{\zeta \ov \zeta_*} \cos \le[ \sqrt{-u} \log {\zeta \ov \zeta_*}  + d_2 \ri] \ 
\ee
where we have chosen $\zeta_*$ as a reference point and $d_1, d_2$ are numerical integration constants. For  $\zeta \sim \zeta_*$, expanding~\eqref{linsol} in $\sqrt{-u}$ and comparing with~\eqref{nonla} we  conclude that $d_1 \sim {1 \ov \sqrt{-u}}$ and $d_2= {\pi \ov 2} + O(-u) $ and~\eqref{linsol} can be written as 
\be \label{finLsol}
\phi(\zeta) = \frac{\ga}{\sqrt{-u}}\sqrt{\frac{\zeta}{\zeta_*}}\sin\le(\sqrt{-u} \log\frac{\zeta}{\zeta_*}\ri) \ .
\ee
It is important to emphasize that  the $\sqrt{-u} \log\frac{\zeta}{\zeta_*}$ term may not be small, as $\zeta$ may vary over exponentially large distance in $1/\sqrt{-u}$. 

Finally we now consider matching~\eqref{finLsol} to the solution in the UV region near $\mu \zeta \sim O(1)$ with identification $\zeta = {z_*^2 \ov 6(z_*-z)}$. This is exactly the same as the linear matching problems discussed in Appendix~\ref{app:mat} and so we will be brief. In terms of 
the basis of solutions introduced in~\eqref{UVsolnS} we can write $\phi$ as 
\be
\phi (z) = {\ga \ov \sqrt{-u}}   \sqrt{z_* \ov \zeta_{*}} {1 \ov 2i} 
 \le(e^{-i \sqrt{-u} \log {\zeta_{*} \ov z_*} } \eta_+^{(0)}  - e^{i\sqrt{-u} \log {\zeta_{*} \ov z_*} } \eta_-^{(0)} \ri)  \ .
\ee
Using the expansion~\eqref{12as} and the following definitions and properties of $a_\pm,b_{\pm}$: 
\be
a_+ = |a_+| e^{i \th_a}, \quad b_+ = |b_+| e^{i \th_b}, \quad a_- = a_+^*, \quad b_- = b_+^*  \label{imagABdef}
\ee
we then conclude that the coefficients $A$ and $B$ in~\eqref{12} are given by
\begin{align} \label{ABexpr}
A  &= -z_*^{3-\De} {\ga \ov \sqrt{-u}} |a_+|  \sqrt{z_{*} \ov \zeta_{*}}  \sin\le(\sqrt{-u} \log {\zeta_{*} \ov z_*}  -\th_a\ri),\cr
B &=  -z_*^{-\De} {\ga \ov \sqrt{-u}} |b_+|  \sqrt{z_{*} \ov \zeta_{*}}  \sin\le(\sqrt{-u} \log {\zeta_{*} \ov z_*}  - \th_b\ri) \ .
\end{align}
Recall that $\zeta_{*}$ parametrizes movement through the solution space; as we vary $\zeta_*$, we see that we trace out a {\it spiral} in the $(A,B)$ plane. See fig.~\ref{fig:spiral1}. If we are studying a normalizable solution (in standard quantization), then we require $A = 0$: the spiral will cross this axis an infinite number of times as we take $\zeta_* \to \infty$, giving as an infinite tower of states.  These are the ``Efimov'' states described in the main text. Note that no matter how small we consider $A$ or $B$ to be, the curve continues to spiral and nonlinear dynamics remains important--this is because the scalar in the deep interior is always of $O(1)$. 
Comparing with~\eqref{req1} we find that as $\sqrt{-u} \to 0$,
\begin{align}
|a_+| &= \al, & \th_a  &= - \sqrt{-u} {\tilde \al \ov \al},  \nonumber \\
|b_+| &= \beta, & \th_b  &= - \sqrt{-u} {\tilde \beta \ov \beta} \   \label{imagSmallLam}
\end{align}
giving~\eqref{ABexpr12} quoted in the main text.


For the case of the double-well potential: 
\be
 V(\phi) = {1 \ov 4R^2}  \le(\phi^2 + {m^2 R^2} \ri)^2
- {m^4 R^2 \ov 4 } \ .
\ee
there is a $\phi \to - \phi$ symmetry which results in the symmetry $A, \ B \to -A, \ -B$ of fig.~\ref{fig:spiral1}. For this potential in the limit of no backreaction we find $\gamma \approx 2$.


\pagebreak
\section{Finite-temperature line near bifurcating critical point} \label{app:finiTC}

The bifurcating quantum phase transition is the endpoint of a line of finite-temperature phase transitions. In this Appendix we present some calculations near this line. As argued earlier this is a rather standard mean field second order transition, so we do not present much detail. One novel feature is that close to the quantum critical point then we are at exponentially small temperatures and so we have a great deal of analytic control over the calculations, allowing us to verify explicitly many of the features expected of such a transition.

\subsection{Dynamic critical phenomena near finite-$T$ transition} 

We first turn on a finite $\om$ and $k^2$ and study the critical behavior close to the finite-temperature critical line. The leading $\om$ behavior comes from the dependence of the IR Green's function in \eqref{roep22rpt} on $\om/T$. At finite $\om$ this IR Green's function is no longer a pure phase, and to lowest order we find
\bwt
\be
 \sG_k(\om;T) = \le(\pi T\ri)^{-2i\lam_k}\frac{\Ga(i\lam_k)}{\Ga(-i\lam_k)}\frac{\Ga\le(\frac{1}{2} - i\lam_k\ri)}{\Ga\le(\frac{1}{2} + i\lam_k\ri)}\le(1 - \frac{\pi\om}{2 T}(\lam_k + O(\lam_k^2))\ri)
\ee
\ewt 
Recall that $u$ measures the distance from the critical point. The leading $k^2$ dependence comes from expanding $\lam_k$ in powers of $k^2$ close to the critical point:
\be
\lam_k = \sqrt{-u} - \frac{k^2}{6\mu_*^2\sqrt{-u}} + O(k^4)  
\ee
Note that a sufficiently large $k$ will take us out of the imaginary $\nu$ phase and invalidate this expansion; while this could presumably be dealt with, it would complicate the analysis, and thus throughout we will simply assume that $k^2$ is parametrically small: $k^2 \ll (g_c -g)$. In this regime the UV contributions to the $k^2$ dependence can be ignored, as they will be higher order in $u$. 

We now insert these expansions into \eqref{roep22rpt}. The denominator of the Green's function then takes the form
\bwt
\be
G_R(\om, k;T)^{-1} \sim \sin\le(\log\le(\frac{T}{T_a}\ri)\le(\lam_0 + \frac{d\lam}{dk^2} k^2\ri)\ri) - i\frac{\om\lam_0 \pi}{2 T}e^{-i\lam_0\log\le(\frac{T}{T_a}\ri)} \ ,
\ee
\ewt
where $\lam_0=\lam_{k=0}=\sqrt{-u}$. We now further expand the \emph{temperature} in the vicinity of the $n$-th ``Efimov temperature'' $T_n$, defined in \eqref{Tc}. We now find
\be
(-1)^n G_R(\om, k; T)^{-1} \sim   \frac{\lam_0(T - T_c^{(n)})}{T_c^{(n)}} + \frac{n\pi}{6\mu_*^2\lam_0^2}k^2 - i\frac{\om\pi\lam_0}{2 T_c^{(n)}} 
\ee
Let us now study this expression, first setting $k \to 0$; we find then that the Green's function has a pole at 
\be
\om_* = -\frac{2i}{\pi}(T-T_n)
\ee
For $T > T_n$ this pole is in the lower half-plane, and it moves through to the upper half-plane if $T$ is decreased through $T_n$.

Of course in practice once the first pole moves through to the upper half-plane, the uncondensed phase is unstable and we should study the system in its condensed phase; thus we see that the true critical temperature is precisely at the first Efimov temperature, $T_c = T_1 = T_a \exp\le(-\frac{\pi}{\sqrt{-u}}\ri).$

We can also set $\om \to 0$ and study the static correlation length; we see that near each Efimov temperature (including the critical temperature) we have a standard finite correlation length $\zeta$ with a mean field scaling in $(T - T_c)$:
\be
\zeta^{-2} = \frac{6\mu_*^2(-u)^{\frac{3}{2}}}{T_n n\pi}(T-T_n)
\ee
This correlation length exhibits an intriguing scaling in $-u$. 

Finally, we can keep both $\om$ and $k^2$ nonzero and sit at the critical point $T = T_n$; we then find a diffusion mode
\be
\om_* = -i\frac{n}{3\mu_*^2(-u)^{\frac{3}{2}}}k^2
\ee
which is of the standard form for this class of dynamic critical phenomena (due to the absence of conservation laws for the order parameter, this is Model A in the classification of \cite{Hohenberg:1977ym}; see also \cite{Maeda:2009wv} for further discussion in the holographic context).

\subsection{Susceptibility across the critical point}

We now compute the linear susceptibility near the critical point as we approach from the uncondensed side, i.e. $T > T_c$. We already have all of the ingredients; from \eqref{imagFinTsusc} we have

\be
\chi^{(T)} =  \chi_0\frac{\sin\le(\lam_0 \log \le(\frac{T}{T_b}\ri)\ri)}{\sin\le(\lam_0 \log \le(\frac{T}{T_a}\ri)\ri)} 
\ee
Now expanding near $T = T_c = T_a \exp\le(-\frac{\pi}{\lam_0}\ri)$ we find
\be
\chi^{(T)} \approx \chi_0\frac{T_c\log \le(\frac{T_a}{T_b}\ri)}{T -T_c}=\frac{\chi_0}{2\nu_U \al\beta} \frac{T_c}{T -T_c} , \label{suscTc}
\ee
where as usual we have used \eqref{W1}. 

We will now perform the analogous calculation from the {\it condensed} side. This will require some understanding of the nonlinear solution close to the critical point. We will use analyticity properties of nonlinear classical field configurations on black hole backgrounds; these are precisely analogous to the analyticity arguments in the Landau theory of phase transitions. Similar arguments led us in~\cite{Iqbal:2010eh} to conclude that for finite temperature phase transitions we find mean field critical exponents. 

First we express $A,\ B$ as functions of the horizon value of the scalar field, $\phi_h$. We have
\be
\frac{B}{\mu_*^{-\Delta}} = b_+(T) \phi_h + b_3(T) \phi_h^3 + \dots \label{Bexp}
\ee
and the corresponding expression for $A$:
\be
\frac{A}{\mu_*^{\Delta - 3}} = a_+(T) \phi_h + a_3(T) \phi_h^3+\dots \label{Aexp}
\ee
For small values of the scalar linear response must apply, and thus the $a_+$ and $b_+$ appearing above are the same as those used throughout this paper in calculating linear response functions. Now from the calculation above we know that close to the critical temperature we have $a_+(T) \sim \tilde{a}(T - T_c)$; matching to \eqref{suscTc} above we see that
\be
\frac{b_+(T_c)}{\tilde{a}} = \frac{T_c}{2\nu_U \al^2}
\ee
Now we see that for $T < T_c$ we have a nontrivial zero in $A$ (and thus a normalizable bulk solution) at 
\be
\phi_h = \le(\frac{\tilde{a}(T - T_c)}{a_3}\ri)^{\ha} \ \equiv \phi_{norm}
\ee
The definition of the nonlinear susceptibility $\chi_L$ is the derivative of the vacuum expectation value (i.e. $B$) with the source as we approach the normalizable solution on the condensed side, i.e.
\be
\chi_L = \mu_*^{2\nu_U} \frac{d B}{d A}\bigg|_{A \to 0} = \mu_*^{2\nu_U}  \frac{d B}{d A}\bigg|_{\phi_h = \phi_{norm}} \ .
\ee
Evaluating the derivatives this works out to be
\be
\chi_L = \mu_*^{2\nu_U} \frac{dB}{d\phi_h}\frac{d\phi_h}{d A}\bigg|_{\phi_h = \phi_{norm}} = \frac{\chi_0}{4\nu_U \al\beta} \frac{T_c}{T_c -T} 
\ee
Compare this to the linear susceptibility $\chi$ calculated in \eqref{suscTc}; we see that the leading divergence in $\chi_L$ has a a prefactor that is half that of $\chi$. This fact is a general result of Landau theory and follows from the symmetry and analyticity arguments that allowed us to write down \eqref{Bexp} and \eqref{Aexp}.

\section{Review of critical exponents} \label{app:critexp}

In the vicinity of a critical point we observe scaling behavior of various observable quantities, which is characterized by a set of critical exponents.  We list some of the most commonly used exponents in the following. We will denote the external tuning parameter $g$ with which we tune the system to the critical point $g=g_c$. Near the critical point the spatial correlation length diverges as
\be
\xi \sim \le| g-g_c  \ri|^{-\nu_{crit}} \ .
\ee
The energy gap for elementary excitations scales as
\be
E_{gap} \sim \xi^{-z}\sim \le| g-g_c  \ri|^{-z \nu_{crit}} \ ,
\ee
where $z$ is called the dynamic critical exponent and determines the characteristic time scale of the approach to equilibrium via $\tau_{eq} \sim 1/E_{gap} $. On the condensed side the order parameter $\sO$ also exhibits scaling near the critical point; the corresponding exponent is: 
\be
\le\< \sO \ri\>\sim \le| g-g_c  \ri|^{\beta} \ ,
\ee
and exactly at the critical point it will depend on the source as
\be
\le\< \sO \ri\>\sim J^{1/\delta} \ ,
\ee
where the coupling to the external source is $J \sO$. The correlation function $\chi=\le\< \sO \sO \ri\>$ can also be used to probe the physical properties of the critical point. The next critical exponent we introduce is for $\chi$ at zero momentum:
\be
\chi(k=0,\om=0) \sim \le| g-g_c  \ri|^{-\gamma} \ .
\ee
The decay of $G_R$ at the critical point in the free theory would be $1/x^{d-2}$, the deviation from this is characterized by $\eta$:
\be
\chi (x,\om=0)\vert_{g=g_c} \sim {1 \ov x^{d-2+\eta}} \ .
\ee
To study the scaling of thermodynamic functions we introduce $\al$ as:
\be
f\sim \le| g-g_c  \ri|^{2-\al} \ ,
\ee
where $f$ is the free energy density.

The scaling exponents obey scaling relations which can be derived from scaling arguments. 
\be
\gamma=(2-\eta)\nu_{crit} = \beta (\delta-1) \label{critrel}
\ee
With some additional input one can derive the hyperscaling relation which is obeyed by critical theories in the Landau-Ginsburg-Wilson paradigm in the absence of dangerously irrelevant operators:
\be
2\beta=(d-2+\eta)\nu_{crit} \ . \label{hyper}
\ee
We will see in the bulk of the paper that hybridized criticality violates the hyperscaling relation, hence we only accept \eqref{critrel} as valid equations. Choosing the independent exponents to be $\beta, \ \gamma , \ \nu_{crit}$ we can express all other exponents with them:
\bea
\al&=&2- 2\beta - \gamma\\
\delta&=&{\beta+\gamma \ov \beta} \\
\eta&=&2-{\gamma \ov \nu_{crit}} \label{etarel} \ .
\eea
Hyperscaling would give us an additional relation between $\beta, \ \gamma , \ \nu_{crit}$.

\section{Index of Symbols} \label{app:symindex}
For convenience here we compile (in rough alphabetical order) the important symbols used in this paper, with brief definitions and a reference to the equation number where they are defined.

\ben
\item  $A$, $B$: UV expansion coefficients of a general solution to the bulk wave equation. Defined in \eqref{12}.
\item $a_{\pm}$, $b_{\pm}$: UV expansion coefficients of a particular basis of solutions to the bulk wave equation, chosen to have definite scaling behavior in the AdS$_2$ geometry. Defined in \eqref{12as}.
\item $\al, \tilde{\al}, \beta, \tilde{\beta}$: Taylor expansion coefficients of $a_{\pm}$, $b_{\pm}$ in small $\nu_k$ limit. Defined in \eqref{req} and \eqref{req1}.
\item $\Delta$: UV (i.e. in asymptotic AdS$_4$ region) conformal dimension of scalar operator $\sO$. Defined in \eqref{uvdim}.
\item $\delta_k$: IR (i.e. in eCFT$_1$, or infrared AdS$_2$ region) conformal dimension of each Fourier mode of IR scalar operator. Defined in \eqref{nudef}. 
\item $\ga$: Dimensionless parameter describing overall scale of nonlinear condensed phase solution near bifurcating critical point. Defined in \eqref{finLsol}; see also \eqref{ABexpr12}. 
\item $G_R(\om, k)$: Full retarded correlator of UV operator $\sO$. Expression given in \eqref{roep21}.
\item $\sG_k(\om)$: IR (i.e. in eCFT$_1$) correlator of IR scalar operator. Expression at zero temperature given in \eqref{iiRc}; finite temperature generalization given in \eqref{finiteTch1}.
\item $\kappa_{\pm}$: Coefficients of various double trace-deformations that can be used to tune system through hybridized critical point; see \eqref{douir}, \eqref{douv} and \eqref{douir1}. 
\item $\kappa_c$: Critical value of $\kappa_+$ for hybridized critical point. Defined in \eqref{stare}.
\item $\ka_+^*$: Value of $\kappa_c$ for which hybridized phase transition line intersects bifurcating phase transition line, leading to marginal quantum phase transition. Defined in \eqref{kappstar}.
\item $m^2$: Bulk mass of scalar. See \eqref{uvdim} and \eqref{opep} for effect on UV and IR conformal dimension respectively.
\item $\mu_*$: Rescaled chemical potential. Defined in \eqref{gaugeexp}. 
\item $\nu_U$: Number related to UV conformal dimension by $\Delta = \frac{d}{2} + \nu_U$. Defined in \eqref{uvdim}.
\item $\nu_k, \nu$: Number related to IR conformal dimension by $\delta_k = \ha + \nu_k$. Defined in \eqref{nudef}; $\nu$ with no subscript is $\nu = \nu_{k=0}$.  
\item $q_*$: Rescaled charge $q$ of scalar field. Defined in \eqref{opep}. 
\item $R, R_2$: Curvature radii of asymptotic UV AdS$_4$ and IR AdS$_2$ regions respectively. Defined in \eqref{RNmetric} and \eqref{ads2M}. 
\item $u$: Control parameter describing distance from bifurcating quantum critical point, which is at $u = 0$, with condensed phase for $u < 0$. Defined in \eqref{uude}. 
\item $\chi_0$: Susceptibility approaching bifurcating quantum phase transition from uncondensed side. Defined in \eqref{chiz}.
\item $\chi_*$: Parameter characterizing non-analyticity in susceptibility across bifurcating quantum phase transition. Defined in \eqref{chiS}. 
\item $\psi(x)$: Digamma function, logarithmic derivative of gamma function $\psi(x) \equiv \frac{d}{dx} \log \Ga(x)$. Appears in thermal response near bifurcating critical point, e.g. \eqref{finTomsusc}. 

\een

\end{appendix}

\end{document}